\documentclass[aps,twocolumn,prd,showpacs,showkeys,preprintnumbers,superscriptaddress,nobibnotes,floatfix,longbibliography]{revtex4-1}
\usepackage{amsmath,amsfonts,amssymb,mathrsfs,color,longtable,hyperref,multirow,graphicx}

\newcommand{\Pinit}{\langle P_0 \rangle}
\newcommand{\Tmin}{T_{\rm min}}

\hypersetup{
    pdfnewwindow=true,    
    colorlinks=true,      
    linkcolor=blue,       
    citecolor=blue,       
    filecolor=blue,
    urlcolor=blue         
}

\begin{document}

\title{TeV Halos are Everywhere: Prospects for New Discoveries}

\author{Takahiro Sudoh}
\affiliation{Department of Astronomy, University of Tokyo, Hongo, Tokyo 113-0033, Japan}
\affiliation{Center for Cosmology and AstroParticle Physics (CCAPP), Ohio State University, Columbus, OH 43210, USA}

\author{Tim Linden}
\affiliation{Center for Cosmology and AstroParticle Physics (CCAPP), Ohio State University, Columbus, OH 43210, USA}
\affiliation{Department of Physics, Ohio State University, Columbus, OH 43210, USA}

\author{John F.~Beacom}
\affiliation{Center for Cosmology and AstroParticle Physics (CCAPP), Ohio State University, Columbus, OH 43210, USA}
\affiliation{Department of Physics, Ohio State University, Columbus, OH 43210, USA}
\affiliation{Department of Astronomy, Ohio State University, Columbus, OH 43210, USA
\\
{\tt \href{mailto:sudoh@astron.s.u-tokyo.ac.jp}{sudoh@astron.s.u-tokyo.ac.jp},  \href{mailto:linden.70@osu.edu}{linden.70@osu.edu}, \href{mailto:beacom.7@osu.edu}{beacom.7@osu.edu}}\\
{\tt \footnotesize \href{http://orcid.org/0000-0002-6884-1733}{0000-0002-6884-1733}, \href{http://orcid.org/0000-0001-9888-0971}{0000-0001-9888-0971}, \href{http://orcid.org/0000-0002-0005-2631}{0000-0002-0005-2631} \smallskip}}

\date{22 February, 2019}

\begin{abstract}
Milagro and HAWC have detected extended TeV gamma-ray emission around nearby pulsar wind nebulae (PWNe). Building on these discoveries, \citet{TeVhalo:Linden17} identified a new source class --- TeV halos --- powered by the interactions of high-energy electrons and positrons that have escaped from the PWN, but which remain trapped in a larger region where diffusion is inhibited compared to the interstellar medium. Many theoretical properties of TeV halos remain mysterious, but empirical arguments suggest that they are ubiquitous. The key to progress is finding more halos. We outline prospects for new discoveries and calculate their expectations and uncertainties. We predict, using models normalized to current data, that future HAWC and CTA observations will detect in total $\sim$50--240 TeV halos, though we note that multiple systematic uncertainties still exist. Further, the existing HESS source catalog could contain \mbox{$\sim$10--50} TeV halos that are presently classified as unidentified sources or PWN candidates. We quantify the importance of these detections for new probes of the evolution of TeV halos, pulsar properties, and the sources of high-energy gamma rays and cosmic rays.

\end{abstract}


\maketitle

\section{Introduction}
\label{sec:intro}

Milagro observations revealed extended TeV $\gamma$-ray emission surrounding the nearby Geminga pulsar, now confirmed by the High Altitude Water Cherenkov (HAWC) observatory~\cite{TeVhalo:MRGO09,TeVhalo:HAWC17,cat:2HWC}. Additionally, HAWC has detected similar emission surrounding another nearby pulsar, PSR B0656+14, commonly associated with the Monogem ring~\cite{Monogem:SNR}, and which we refer to as the ``Monogem pulsar.'' These sources are bright ($\sim10^{32}$ erg s$^{-1}$), have hard spectra ($\sim E^{-2.2}$), and are spatially extended ($\sim$ 25 pc). In addition, the High Energy Stereoscopic System (HESS) has detected a number of TeV $\gamma$-ray sources coincident with pulsars or pulsar wind nebulae (PWNe)~\cite{cat:HESSGP,cat:HESSPWN}. Though they refer to these as ``TeV PWN,'' they find that many are significantly larger than expected from PWN theory~\cite{TeVhalo:Linden17,HESS:largePWN,PWNreview:2006}. The sources noted above appear morphologically and dynamically distinct from PWNe detected in X-ray and radio observations. 

\begin{figure}
\includegraphics[width=0.783\columnwidth,height=0.75\columnwidth]{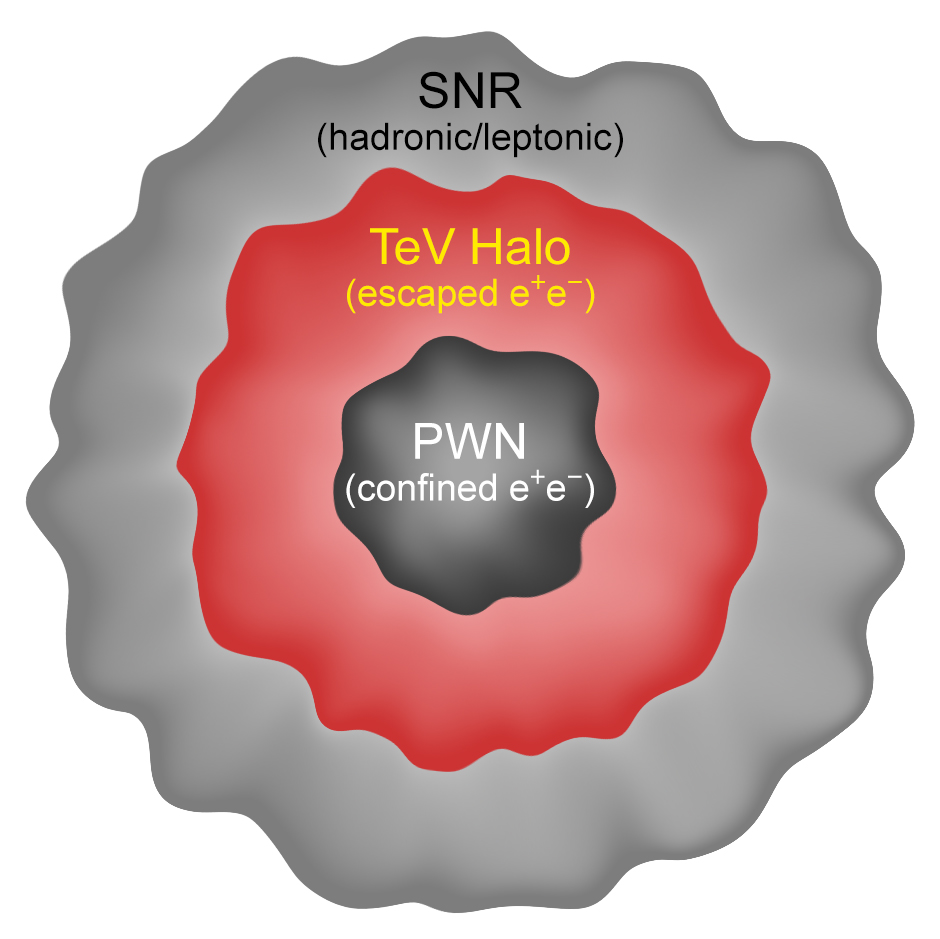}
\caption{\label{fig:pic}Schematic illustration of a TeV halo in relation to the more familiar PWN and supernova remnant (SNR). A TeV halo may not form early, and the SNR may be fading when the halo appears.}
\end{figure}

\citet{TeVhalo:Linden17} identified these sources as a new $\gamma$-ray source class (``TeV Halos'') and interpreted their emission as the result of electrons and positrons interacting with the ambient interstellar radiation field outside the PWN. The possibility of significantly extended leptonic emission was first predicted in Ref.~\cite{Aharonian1995}, and its importance was further discussed in Refs.~\cite{TeVhalo:AAK97,Aharonian_text,TeVhalo:Yuksel09,Aharonian2013}. Moreover, \citet{TeVhalo:Linden17} showed that a large fraction of 2HWC catalog sources are coincident with pulsars, and predicted that TeV halos are a generic feature of pulsar emission. 

In Fig.~\ref{fig:pic}, we show how a TeV halo compares to other features at the site of a past core-collapse supernova explosion. For a given source, it may be that not all components are detectable or even present at the same time. A PWN, powered by the rotational energy of the central pulsar, is delimited by the contact discontinuity between the shocked pulsar wind and the ejecta or interstellar matter. An SNR, powered by the energy of the supernova explosion, is delimited by its interaction with the interstellar medium.  A TeV halo is likely intermediate in size, is powered by cosmic rays diffusing away from the PWN, and does not have a well-defined boundary. The size of a PWN can be on the order of 0.1--1~pc, though some may range up to $\sim$10~pc \cite{PWNreview:2006,PWNmorph:Kargaltsev2008}, and the size of an SNR may span $\sim$1--100~pc \cite{SNRmorph:Badenes2010,SNRmorph:Stafford2018}, depending on their properties, evolutionary stages and environment. The typical size of a TeV halo is not known, but Geminga and Monogem observations indicate that it may be on the order of 10~pc for middle-aged pulsars. For the three types of object, differences in radii lead to larger differences in volumes that further support different physical origins.

The identification of TeV halos as a new source class is supported by the subsequent detection of two more TeV halos by HAWC~\cite{ATel:2017,ATel:2018}, one of which was predicted by Ref.~\cite{TeVhalo:Linden17}. However, many details about TeV halos remain unknown and further observations will have broad implications. Apart from shedding light on the properties of the TeV halos themselves, these observations will reveal new aspects of pulsar formation and evolution~\cite{TeVhalo:Linden17}, and will probe sources of high-energy $\gamma$-rays~\cite{TeVhalo:Hooper18,TeVhalo:Linden18,TeVhalo:MSP2018} and cosmic-ray electrons and positrons~\cite{TeVhalo:Hooper17,TeVhalo:Hooper18b,TeVhalo:Evoli18,TeVhalo:Fang18,TeVhalo:Profumo18,TeVhalo:Bucciantini18,Geminga:Tang18,Geminga:Xi18}.

Here, we outline a multifaceted strategy to discover more TeV halos and to constrain their evolution. We quantify the role of Galactic source searches and diffuse measurements using water Cherenkov telescopes like HAWC. We also consider Galactic and extragalactic source surveys by imaging air Cherenkov telescopes, focusing on the Cherenkov Telescope Array (CTA). Further, we show that follow-up studies of existing TeV $\gamma$-ray sources in the HESS catalog, especially those classified as PWN or unidentified sources, could find more TeV halos. For our overall approach, we use standard methods for pulsar population synthesis and treat the Geminga TeV halo as a prototype. Our results go significantly beyond those of prior work, yielding new insights into both the prospects for future TeV halo discoveries, and the implications of TeV halo observations for our understanding of astrophysics.

In Sec.~\ref{sec:review}, we briefly review the properties of TeV halos. In Sec.~\ref{sec:model}, we present our methods for modeling TeV halo populations. In Sec.~\ref{sec:now}, we compare predictions with current observations and constrain model parameters. In Sec.~\ref{sec:future}, we outline future directions to find more TeV halos. In Sec.~\ref{sec:conclusion}, we present our conclusions.

\section{What are TeV halos?}
\label{sec:review}
TeV halos are defined as the non-thermal emission produced in regions outside a PWN, but within a region where pulsar activity dominates cosmic-ray diffusion (\cite{TeVhalo:Linden17,Aharonian1995,TeVhalo:AAK97,Aharonian_text,TeVhalo:Yuksel09}). Within this region, multi-TeV $\gamma$-rays are produced by the inverse-Compton scattering of ambient photons by $\sim$10 TeV electrons and positrons accelerated by the pulsar wind termination shock. Observations indicate that the TeV halo produces bright $\gamma$-ray emission with a hard spectrum. 

We begin by examining the key features of the best-studied TeV halo, Geminga, which is about 340~kyr old \cite{ATNF} and believed to reside approximately 250~pc from Earth \cite{Geminga:distance}. HAWC detects TeV $\gamma$-ray emission extending to an angular size of $\sim$5$^\circ$, corresponding to $\sim$25~pc in physical extent \cite{TeVhalo:HAWC17}. The differential $\gamma$-ray luminosity at 7~TeV is 2.9$\times 10^{31}$$(d/250\ {\rm pc})^2$~erg s$^{-1}$, with a local spectral index of $-$2.2. 

The lack of gas-correlated emission indicates a leptonic origin. Within the context of an inverse-Compton model, several parameters regarding the electron population can be calculated \cite{TeVhalo:Hooper17,TeVhalo:HAWC17}. To produce the bright $\gamma$-ray luminosity, $\sim$10\% of the total pulsar spin-down power must be converted into e$^\pm$ pairs. Furthermore, to produce the hard $\gamma$-ray spectrum, the electron population should be injected with a hard power-law spectrum between $\sim$$-$1.5 and $-$2.2.

The most notable feature of TeV halos is their size. The Geminga TeV halo is significantly larger than its X-ray PWN, which is confined within $3^\prime$ of the central pulsar \cite{Geminga:XPWN17}. This indicates that the electrons and positrons responsible for TeV halo emission have already escaped the PWN and are interacting with the interstellar radiation field. The TeV halo morphology is consistent with cosmic-ray diffusion, rather than advection \cite{TeVhalo:HAWC17}. However, this diffusion must be inhibited. Assuming that the TeV halo medium is filled with the $\sim$1~eV~cm$^{-3}$ interstellar radiation field and the $\sim$3~$\mu$G magnetic field typical of the Galactic Plane, we would expect 10~TeV e$^\pm$ to cool in $\sim$40~kyr. In the interstellar medium, electrons and positrons that propagate for $\sim$40~kyr should diffuse over a distance of $\sim$700~pc \cite{crdiffusion:GALPROP11}. However, the TeV halo power appears to be confined within $\sim$25~pc of the pulsar center. 

TeV halo emission is not unique to Geminga. The HAWC collaboration has identified at least three other TeV halos with similar features: Monogem (111~kyr, 290~pc), PSR B0540+23 (253~kyr, 1.56~kpc), and PSR J0633+0632 (59~kyr, 1.35~kpc) \cite{ATNF,ATel:2017,ATel:2018}. In addition, \citet{TeVhalo:Linden17} listed 13 more TeV halo candidates in the 2HWC catalog. The 2HWC survey also provides a hint of TeV halo emission around millisecond pulsars \cite{TeVhalo:MSP2018}. 

In addition to HAWC, imaging air Cherenkov telescopes like HESS, MAGIC and VERITAS have detected a number of extended TeV $\gamma$-ray sources that are associated with pulsars or PWNe observed at other wavelengths. These systems are called ``TeV PWN," but many of them have an extension exceeding $\sim$10~pc \cite{cat:HESSGP,cat:HESSPWN}, while hydrodynamical simulations predict a typical PWN size on the order of 1~pc \cite{PWNmodel:vanderSwaluw2004,PWNreview:2006,PWNmodel:Gelfand2009,PWNmodel:Bucciantini2011,PWNmodel:Martin16,PWNmodel:Ishizaki18,PWNmodel:Zhu18,PWNmodel:vanRensburg18}. More pointedly, they are usually much more extended than the size of X-ray PWN observed from the same system \cite{TeVPWN:Kargaltsev10,TeVPWN:Kargaltsev13}. These observations suggest that some of these $\gamma$-ray sources may be interpreted as TeV halos, instead of emission from confined particles inside PWNe. In particular, HESS J1825-137 has the largest radius ($\sim$50~pc \cite{HESS:largePWN2019}) among ``TeV PWN" \cite{HESS:largePWN}. A TeV halo explanation for this source is already discussed in Refs.~\cite{Aharonian_text,Aharonian2013}.

Despite the significant number of TeV halos that have been (or are potentially) detected in current surveys, many of their properties remain mysterious. In particular, we do not understand the evolution of the key observable TeV halo properties: their luminosity, spectrum, and spatial morphology. In recent work \cite{TeVhalo:Linden17}, TeV halo predictions have been evaluated utilizing a ``Geminga-like" model, where the ratio of the $\gamma$-ray flux to $\dot{E}/d^2$ is constant for all systems with an efficiency set to the best-fit value of Geminga, and the physical size of all TeV halos is $\sim$10~pc. On one hand, this model appears reasonably consistent with the data --- choosing to instead normalize the TeV halo flux to the average $\gamma$-ray efficiency of all firmly identified TeV halos changes the normalization constant only by a factor of $\sim$2 compared to the ``Geminga-like" model. On the other hand, there is nearly an order of magnitude variation in the efficiencies of individual candidate sources, the origin of which is not understood.

A key question is when a TeV halo first forms. Several considerations indicate that the ``Geminga-like" model may not apply to young pulsars. Theoretically, high-energy cosmic rays are expected to be efficiently confined in young PWNe and quickly lose energy to adiabatic and synchrotron cooling in the strong PWN magnetic field \cite{PWNmodel:RG74,PWNmodel:KC84,PWNmodel:deJager09,PWNmodel:Tanaka10,PWNmodel:Torres2014,PWNmodel:Vorster13,PWNmodel:Olmi16}. This may imply that particles do not escape into young TeV halos. Moreover, the creation of a halo may require cosmic-ray self-generated turbulence, which is produced through the resonant interactions of Alfv{\'e}n waves with accelerated electrons and positrons. The growth-rate of self-generated turbulence is model dependent, but typically occurs on $>$~kyr timescales~\cite{TeVhalo:Evoli18}. 

Observationally, the Crab pulsar (964~yr, 2~kpc) does not appear to produce TeV halo emission \cite{crab:HESS,crab:MAGIC,crab:HAWC}, indicating that TeV halos may not be visible within the first kyr of pulsar evolution. An intriguing edge case is the Vela pulsar (11 kyr, 280~pc). Vela does not appear to produce a bright TeV halo (compared to the luminosity expected if the formation efficiency is Geminga-like). However, Vela does have dim, spatially-extended emission detected in radio and GeV-TeV $\gamma$-ray observations~\cite{Vela:Fermi10,Vela:HESS12,Vela:Grondin13,Vela:Fermi18}. This has historically been interpreted as a class of ``relic PWN'' that are left behind after the interaction of the expanding PWN and the SNR reverse shock, and which are powered by old electrons accumulated since the birth of the pulsar~\cite{Vela:Blondin01,Vela:deJager08,Vela:Hinton11,Vela:Slane18}. Interestingly, the size of this extended emission is $\sim$10~pc, comparable to that of observed TeV halos. Thus, Vela could be interpreted as a transition case, where inefficient TeV halos first form. Further TeV observations around $\sim$1--10~kyr pulsars are needed to study the properties of young systems. 

Because detailed examinations of young systems are beyond the scope of this study, in this paper we use a standard ``Geminga-like'' model, but introduce a new parameter $\Tmin$, before which pulsars are assumed to exhibit no TeV halo activity. Observations of the Crab and Vela suggest $\Tmin\gtrsim$~1--10~kyr, while Monogem and Geminga TeV halos suggest $\Tmin\lesssim$~100--300~kyr. 

Another key question is whether TeV halo activity is ubiquitous to all pulsars. Theoretically, the creation of halos requires strongly inhabited diffusion around pulsars, which might be expected for all pulsars if a steep cosmic-ray gradient around them efficiently excites self-generated turbulence \cite{TeVhalo:Evoli18}. Observationally, \citet{TeVhalo:Linden17} listed seven middle-aged pulsars that should be detected by HAWC, under the assumption that every pulsar has a Geminga-like TeV halo, and find that five are in fact associated with the 2HWC sources. These are consistent with the expectation that a significant fraction of pulsars have TeV halos.

We operate under the assumption that all pulsars older than $\Tmin$ produce TeV halos. This can be tested in future surveys. We do not consider the maximum age of TeV halos, because late-time sources are not important due to their small spindown power. 

\section{TeV Halo Population Models}
\label{sec:model}

To model the population of TeV halos, we generate an ensemble of pulsars with randomly assigned initial spin periods ($P_0$) and magnetic fields ($B_0$). We assume $M$~=~1.4~$M_\odot$ and $R$~=~12~km for all pulsars \cite{NS:review2007}. We then assign each pulsar an age ($T_{\rm age}$) drawn from a uniform distribution spanning from 0 to 1 Gyr, and calculate the pulsar spindown power as: 
\begin{equation}
\label{eq:Edot}
        \dot{E}(t) = \frac{8\pi^4B_0^2R^6}{3c^3P_0^4}\left(1+\frac{t}{\tau_{\rm sd}}\right)^{-2},
\end{equation}
where $\tau_{\rm sd} = 3Ic^3P_0^2/4\pi^2B_0^2R^6$ is the spindown timescale \cite{Shapiro1983,TeVhalo:Hooper18}. We associate each pulsar with a randomly distributed position within the Milky Way, based on the pulsar distributions determined by Refs.~\cite{psr_d:YK04,GALPROP2017,psr_d:L06}. Specifically, we adopt the radial distribution of Ref.~\cite{psr_d:YK04}, and a scale height of 200~pc \cite{GALPROP2017}, and calculate the pulsar position relative to Earth assuming a galactocentric distance of 8.5~kpc. Our results are only slightly affected if we use the alternative spatial distributions defined in Ref.~\cite{psr_d:L06}. We have verified that our models are reasonably consistent with the observations of nearby neutron stars, i.e., the seven isolated neutron stars and several pulsars younger than 1~Myr within around 500~pc \cite{INS2005,INS2009}. Further, we calculate the probability that the pulsed radio emission from each pulsar is beamed towards Earth following the empirical relation defined in Ref.~\cite{psr:beam},

\begin{equation}
    f_{\rm beam} = \left[9\left(\log_{10}\frac{P}{10\ {\rm s}}\right)^2 + 3\right]\ \%.
    \label{eq:beam}
\end{equation}
Of all choices in our calculation, the most significant are those of $P_0$ and $B_0$, due to their strong dependence in Eq.~\ref{eq:Edot}: $B_0^2/P_0^4$ in the pre-factor and $P_0^2/B_0^2$ in $\tau_{\rm sd}$.

The $P_0$ distribution is poorly constrained \cite{psrP0:FK06,psrP0:PWN,psrP0:WR11,psrP0:Popov12,psrP0:Noutsos13,psrP0:Igoshev13,psrP0:2018,psrP0:G2004,psrB0:G14}, because population statistics are not sensitive to it \cite{psrP0:G2004,psrB0:G14}. Conventionally, pulsar population models adopt a Gaussian distribution with $\Pinit$~=~300~ms and $\sigma_{P_0}$~=~150~ms, based on radio observations \cite{psrP0:FK06}. However, studies of the $\gamma$-rays pulsar population hint at much smaller values $\Pinit$~=~50~ms and $\sigma_{P_0}$~=~50/$\sqrt{2}$~ms \cite{psrP0:WR11}. In what follows, we present results for both $P_0$ distributions. We also test an intermediate case of $\Pinit$~=~120~ms and $\sigma_{P_0}$~=~60~ms. Finally, in Appendix~\ref{app:uniform_distribution}, we examine models that utilize a uniform, rather than a Gaussian distribution, for the initial spin period. In all cases, we set a minimum spin period at the Newtonian centrifugal breakup limit, $P_{0,\min}$~=~$0.85$~$(M/1.4\ M_\odot)^{1/2}$~$(R/12\  {\rm km})^{3/2}$~ms \cite{NS:review2007}.

For the $B_0$ distribution, we adopt a log-normal magnetic field distribution with mean $\langle\log_{10}B_0\rangle$~=~12.65 and 
standard deviation $\sigma_{\log_{10}B_0}$~=~0.55, which is derived from population studies of radio pulsars \cite{psrP0:FK06}. Other studies predict magnetic fields that are larger by a factor of 2--4 \cite{psrB0:P10,psrB0:G14}. This uncertainty is discussed in Sec.~\ref{subsec:constraints}. We do not include any term to account for the decay of the magnetic field strength, because it occurs on timescales of $>$Myr, much greater than the age of the majority of detectable TeV halos.

\begin{figure}[t]
\includegraphics[width=\columnwidth]{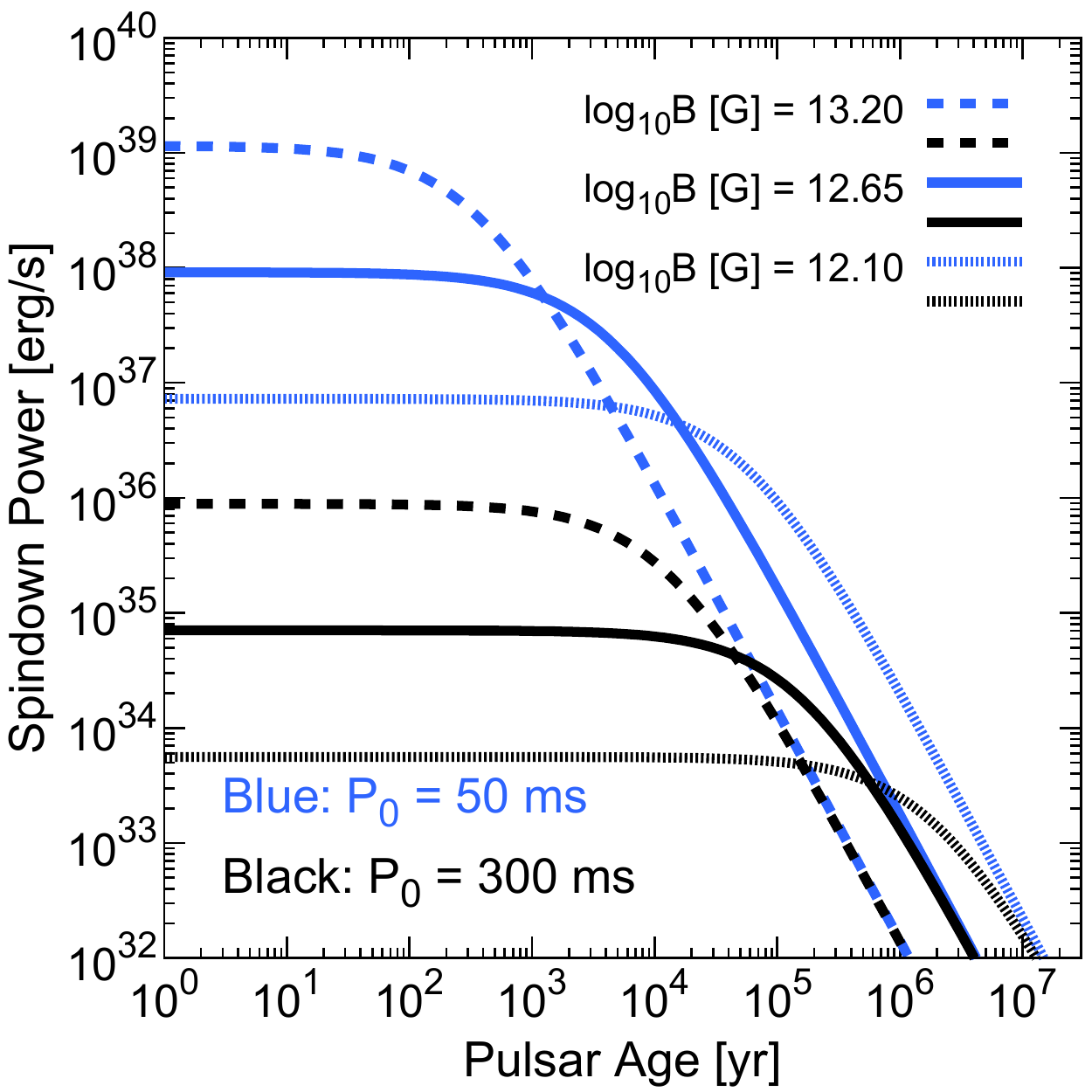}
\caption{\label{fig:pulsar_evol}Evolution of the pulsar spindown power for six representative cases, as labeled.}
\end{figure}

In Fig.~\ref{fig:pulsar_evol}, we show the evolution of the spindown power for six representative pulsars. Most of the integrated spindown power is spent before $\sim \tau_{\rm sd}$, which is 4~kyr for $P_0$~=~50~ms and 160~kyr for 300~ms in the fiducial case of $B_0$~=~$10^{12.65}$~G (this would be more evident if we had plotted the power per log time, which would include multiplying by a factor $t$).

Using the ensemble of pulsars generated above, we assign a ``Geminga-like" TeV halo to each pulsar that has an age older than $\Tmin$. We normalize the differential $\gamma$-ray flux at 7 TeV ($\Phi_{\rm 7TeV}$) for each TeV halo relative to Geminga, using the spindown power ($\dot{E}$) and distance ($d$) as

\begin{equation}
\label{eq:flux_Geminga_like}
        \Phi_{\rm 7TeV} = \Phi_{\rm 7TeV}^{\mathcal{G}}\left(\frac{\dot{E}}{\dot{E}^{\mathcal{G}}}\right)\left(\frac{d^{\mathcal{G}}}{d}\right)^{2}.
\end{equation}
We adopt physical quantities for Geminga (superscript ``$\mathcal{G}$") as summarized in Table~\ref{tab:Geminga}.

\begin{table}[b]
    \centering
    \begin{tabular}{|c|c|c|c|}
    \hline
       \multirow{3}{*}{\bf{Observed}} & 
        $\dot{E}^{\mathcal{G}}$ [erg s$^{-1}$]& 3.2$\times 10^{34}$ & \cite{ATNF}\\
         & $d^{\mathcal{G}}$ [pc]& 250 & \cite{Geminga:distance} \\
         & $\Phi_{\rm 7TeV}^{\mathcal{G}}$ [TeV$^{-1}$cm$^{-2}$s$^{-1}$]&  4.87$\times$10$^{-14}$ & \cite{cat:2HWC} \\
         \hline
        \multirow{2}{*}{\bf{Calculated}} & $F_{\rm TeV}^{\mathcal{G}}$ [cm$^{-2}$s$^{-1}$] & $3.5\times 10^{-12}$ & \\
        & $L_{\rm TeV}^{\mathcal{G}}$ [erg s$^{-1}$] & 1.1$\times$10$^{32}$ & \\
    \hline
    \end{tabular}
    \caption{Physical quantities for Geminga and its TeV halo.}
    \label{tab:Geminga}
\end{table}

In addition to directly observable parameters such as the spindown energy and the 7-TeV $\gamma$-ray flux, our models also require us to derive parameters such as the integrated $\gamma$-ray flux ($F_{\rm TeV}$) and luminosity ($L_{\rm TeV}$) for each pulsar; we calculate these above 1 TeV. To normalize these parameters to Geminga, we follow the theoretical treatment of Ref.~\cite{TeVhalo:Linden18}, which calculates the inverse-Compton scattering $\gamma$-ray spectrum from Ref.~\cite{BG70}. We model the electron spectrum following Ref.~\cite{TeVhalo:Hooper17}, which derives the best-fit $\gamma$-ray spectrum from a combination of HAWC (7 TeV) and Milagro (35 TeV) observations of the Geminga TeV halo \cite{cat:2HWC,TeVhalo:MRGO09}. Specifically, we assume that electrons are injected with a power-law index of \mbox{$\alpha$ = 1.9} that cuts off exponentially at $E_{\rm cut}$ = 49 TeV. We adopt an energy-independent escape time of 1.8$\times$10$^4$ yr from the TeV halo emission region \cite{TeVhalo:Hooper17}. The total e$^\pm$ luminosity is normalized to be $\eta\dot{E}$. We find the best-fit value of $\eta = 0.12$ from the observed $\gamma$-ray flux. The derived values of $F_{\rm TeV}^{\mathcal{G}}$ and $L_{\rm TeV}^{\mathcal{G}}$ are provided in Table~\ref{tab:Geminga}. We again calculate $F_{\rm TeV}$ and $L_{\rm TeV}$ for every other pulsar by scaling the best-fit Geminga values with $\dot{E}$ and $d$ as shown in Eq.~(\ref{eq:flux_Geminga_like}). 

The 2HWC catalog reports the photon index at 7 TeV (2.23 for Geminga). If we extrapolate this spectral index down to 1~TeV and use the 7-TeV differential flux, we derive values of $F_{\rm TeV}^{\mathcal{G}}$ that fall within $\sim$20$\%$ of the theoretically derived photon flux reported in Table~\ref{tab:Geminga}. On the other hand, the values for $L_{\rm TeV}^{\mathcal{G}}$ are increased nearly by a factor of 2, and hence our calculated luminosity in Table~\ref{tab:Geminga} may be pessimistic.

\begin{figure}[t]
\includegraphics[width=\columnwidth]{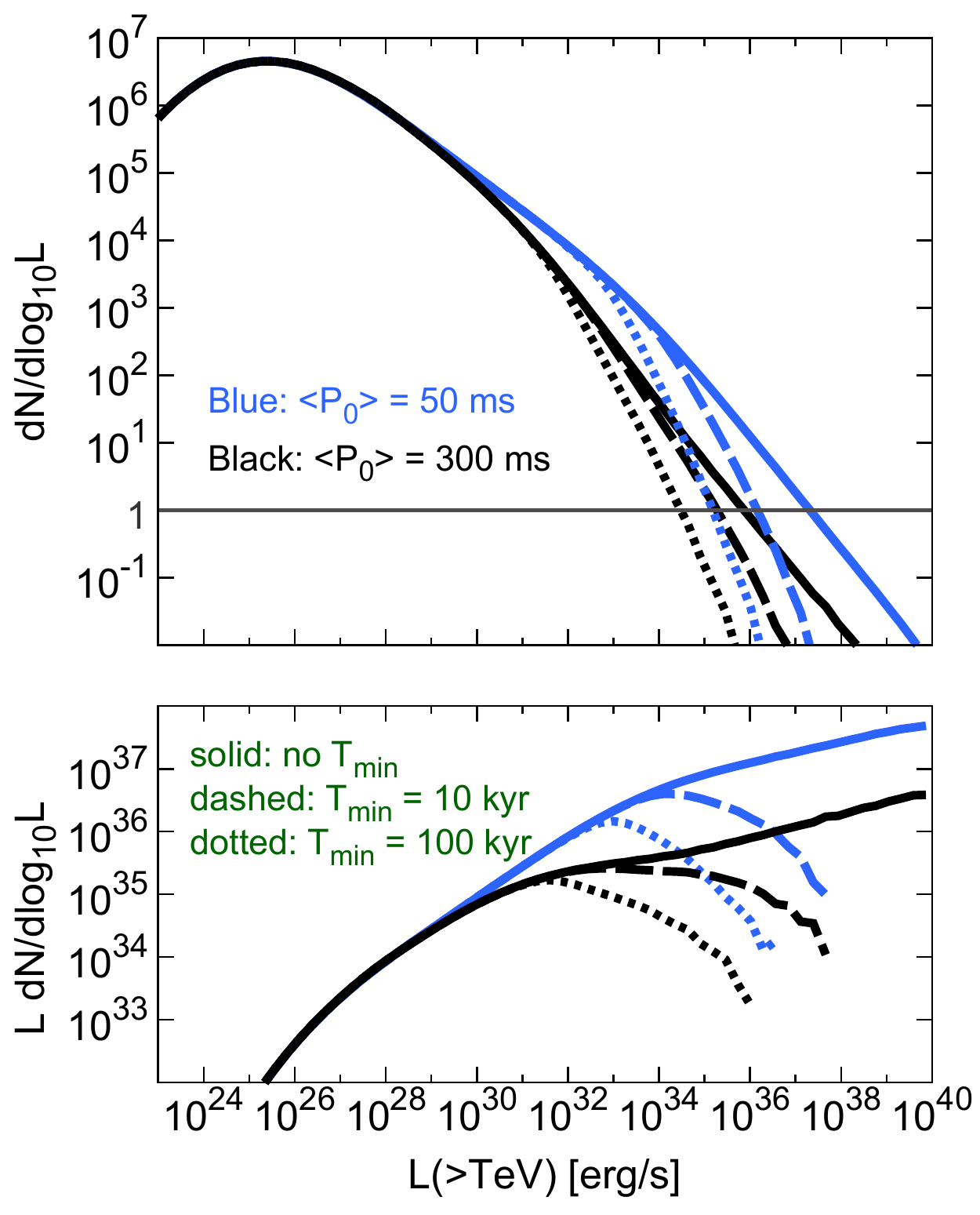}
\caption{\label{fig:LF}TeV $\gamma$-ray luminosity functions of TeV halos for two choices of $P_0$ distribution. The upper panel shows source counts and the lower panel shows the contributions to the total luminosity.}
\end{figure}

In Fig.~\ref{fig:LF}, we show the $\gamma$-ray luminosity function of Milky Way TeV halos for two different $P_0$ distributions and three different values of $\Tmin$. We normalize the total number of Milky Way pulsars using a pulsar birth rate of 0.015 yr$^{-1}$ \cite{psr_d:L06}. The upper panel shows the number weighting only, while the lower panel also includes the luminosity weighting.

If we do not set $\Tmin$, the bright end of the number count (upper panel) has a slope of $\sim L^{-0.8}$, which is driven by the distribution of $P_0$. As pulsars get older (above $\tau_{\rm sd}$), they lose spindown power following $\dot{E}\propto t^{-2}$ and move to the left in this plot, producing a shallower slope of $\sim L^{-0.5}$ before the peak, where pulsars with average properties ($P_0,B_0,T_{\rm age}$) reside. We do not include the effect that old pulsars may terminate their activities below the radio death line \cite{deathline1993}, because late-time sources have small $\gamma$-ray luminosities and contribute negligibly to the source count. Furthermore, due to the shallow slopes of the number count, dim sources contribute negligibly to the total Galactic emission, as shown in Fig~\ref{fig:LF} (bottom).

\section{Existing Model Constraints}
\label{sec:now}
The range of model parameters used in our predictions below can be constrained by current data. In Sec.~\ref{subsec:2HWC}, we predict the number of TeV halos that should be detected in the 2HWC source catalog. In Sec.~\ref{subsec:MRGO}, we estimate the contribution of unresolved TeV halos to the diffuse TeV $\gamma$-ray emission across the Galactic Plane, comparing our predictions with Milagro observations. In Sec.~\ref{subsec:constraints}, we summarize model constraints and briefly discuss uncertainties.

\subsection{Sources in the 2HWC Catalog}
\label{subsec:2HWC}

The 2HWC catalog utilizes 507 days of HAWC data and identifies 39 high-significance sources within the field of view of $-20^\circ <$ decl. $< 60^\circ$. The sensitivity depends on the photon spectral index and the source declination. We adopt the average of quoted values for spectral indices of $-2.5$ and $-2.0$ and the declination dependence given in Ref.~\cite{cat:2HWC}. The best sensitivity of 4.3$\times 10^{-15}$ TeV$^{-1}$ cm$^{-2}$ s$^{-1}$ (9\% of the Geminga TeV halo flux) occurs at a declination of 20$^\circ$, and is degraded by a factor of $\sim$2 for declinations that differ by 30$^\circ$. We take into account the degradation of the flux sensitivity for sources that are larger than the size of the PSF, utilizing a model where the sensitivity decreases by a factor of $\theta_{\rm size}/\theta_{\rm PSF}$ compared to the point source sensitivity~\cite{ACT:Hinton2009}. We assume a PSF size of $\theta_{\rm PSF} = 0.2^\circ$ for HAWC~\cite{HAWC2013}. To determine the source size, we again utilize a Geminga-like model, assuming that all TeV halos have the same physical size as that of the Geminga halo ($\theta_{\rm size}=2^\circ$ at a distance of 250~pc). We ignore source confusion, where HAWC may identify neighboring or overlapping sources as one source, because our calculations show it to be unimportant.

We constrain our TeV halo models by requiring that they do not produce too many or too few systems that would be detected in the 2HWC catalog search. We set the maximum number of potential TeV halos in the 2HWC catalog at 36, because three sources (the Crab, Mrk501, and Mrk421) are associated with objects that are definitively not TeV halos. For the minimum number, we choose 2, because Geminga and Monogem were detected while the two other sources announced by Astronomer's Telegrams \cite{ATel:2017,ATel:2018} did not meet the flux threshold to be included in the 2HWC catalog. Both of these choices are conservative, as they do not take into account additional information concerning individual 2HWC objects.

We can additionally constrain the number of detectable TeV halos that would have radio beams oriented towards Earth. Such sources are especially compelling because the spatial coincidence points towards a TeV halo origin. We note that while the Monogem pulsar is a firmly detected radio pulsar, the Geminga pulsar has extremely dim radio emission and would not have been detected in blind radio searches~\cite{Geminga:radio97}. Hence, we conservatively assume that at least 1~TeV halo (Monogem) has been detected in the 2HWC catalog with a radio beam oriented towards Earth.

The lower limits on the number of beamed and unbeamed TeV halos would become much stronger if the TeV halo candidates that were first identified by Ref.~\citep{TeVhalo:Linden17} are confirmed by subsequent observations. Ref.~\citep{TeVhalo:Linden17} finds three additional 2HWC sources that are consistent with the position of middle-aged radio pulsars, and twelve additional 2HWC sources that are consistent with the positions of younger pulsars. They estimate that only 2.6 chance coincidences would be expected if the 2HWC sources were not associated with pulsar activity. We compare our model predictions with these candidate sources.

\begin{figure}[t]
\centering
\includegraphics[width=1.0\columnwidth]{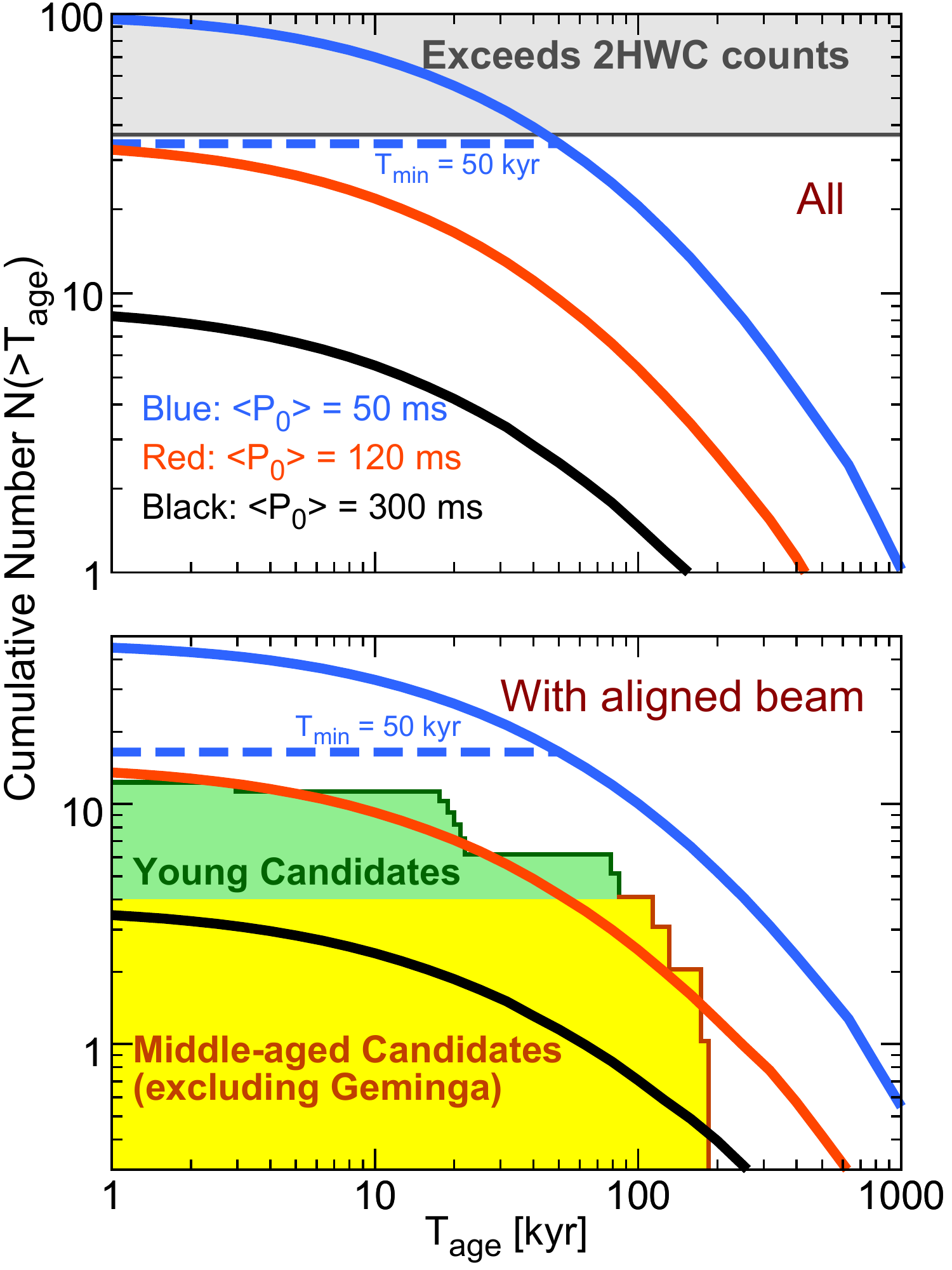}
\caption{\label{fig:2HWC}Predicted numbers of TeV halos in the 2HWC catalog for all sources (top) and sources with radio beams aligned towards Earth (bottom), for different choices of $\Pinit$ and $\Tmin$. Unmarked curves have $\Tmin=0$. In the bottom panel, the cumulative histogram for TeV halo candidates is also shown, separated into young and middle-aged candidates following Ref.~\cite{TeVhalo:Linden17}.}
\end{figure}

In Fig.~\ref{fig:2HWC} (top), we show the total number of detectable TeV halos produced by our model, regardless of whether the system has a radio beam that is oriented towards Earth. This prediction should thus be compared to the total number of detected 2HWC sources. We plot results for models with initial pulsar spin periods of $\Pinit$~=~50~ms, 120~ms, and 300~ms, and find that our parameters allow us to vary the predicted number of detected halos by about one order of magnitude. 

This variation translates into a constraint on the value of $\Tmin$. If typical pulsars are born with relatively large spin periods (e.g., $\Pinit$ = 300 ms), TeV halos produce about 10 sources in models where $\Tmin$~=~0. In this case, the lower bound of $\Tmin$ is not strongly constrained by 2HWC data. On the other hand, if we utilize our minimum value of $\Pinit$~=~50 ms, TeV halos produce about 100 sources in models where $\Tmin$~=~0. Because this exceeds the total number of 2HWC sources, this would require a simultaneous constraint of $\Tmin$~$\gtrsim$~50 kyr. In the remainder of the section, we adopt $\Tmin$ = 50 kyr for the case of $\Pinit$~=~50~ms as the most optimistic case, which predicts that most of the 39 sources in the 2HWC catalog are TeV halos. Models with $\Pinit$~=~120~ms provide a critical case, approximately saturating the number of detectable TeV halos in models with $\Tmin$~=0. Thus, in this case, the value of $\Tmin$ is not strongly constrained at this point, but may be better constrained if future observations indicate that several 2HWC sources are not TeV halos.

To produce at least two detectable TeV halos, we need to set $\Tmin\lesssim$ 300~kyr for $\Pinit$~=~120ms. This constraint is not strong, because such a large value of $\Tmin$ is already disfavored by the observations of Monogem (110~kyr) and Geminga (340~kyr). On the other hand, for $\Pinit$~=~300ms, we can constrain $\Tmin\lesssim$ 70~kyr.

In Fig.~\ref{fig:2HWC} (bottom), we show model predictions for the expected number of TeV halos in the 2HWC catalog that have radio beams aligned with Earth, compared with the age distribution of these TeV halo candidates. We first focus on middle-aged pulsars ($>$100~kyr). The $\Pinit$~=~50~ms model predicts $\sim$9 sources, which slightly exceeds the number of TeV halo candidate systems identified in Ref.~\citep{TeVhalo:Linden17}. This model is allowed, but if future observations rule out the TeV halo nature of several of these systems, it would be in tension with the data.

On the other hand, models with $\Pinit$~=~300~ms produce $\lesssim$~1 middle-aged TeV halo with a radio beam directed towards Earth, which approximately saturates the lower limit produced by the identification of Monogem. This model is allowed, but if future observations confirm the TeV halo origin of candidate sources, it would be disfavored. Intriguingly, though we adopted models with $\Pinit$~=~50~ms and $\Pinit$~=~300~ms based on previous pulsar studies~\cite{psrP0:FK06,psrP0:WR11}, they coincidentally also serve as reasonable estimates for the largest and smallest values allowed by the 2HWC data. The firm interpretation of existing 2HWC observations could potentially rule out either model.

We additionally show an intermediate case, with $\Pinit$~=~120~ms, which predicts the observation of $\sim2$ middle-aged TeV halos with radio beams oriented towards Earth. This model matches current observations well, and is likely remain consistent regardless of the interpretation of the 2HWC candidate sources.

\begin{figure}[t]
\includegraphics[width=1.0\columnwidth]{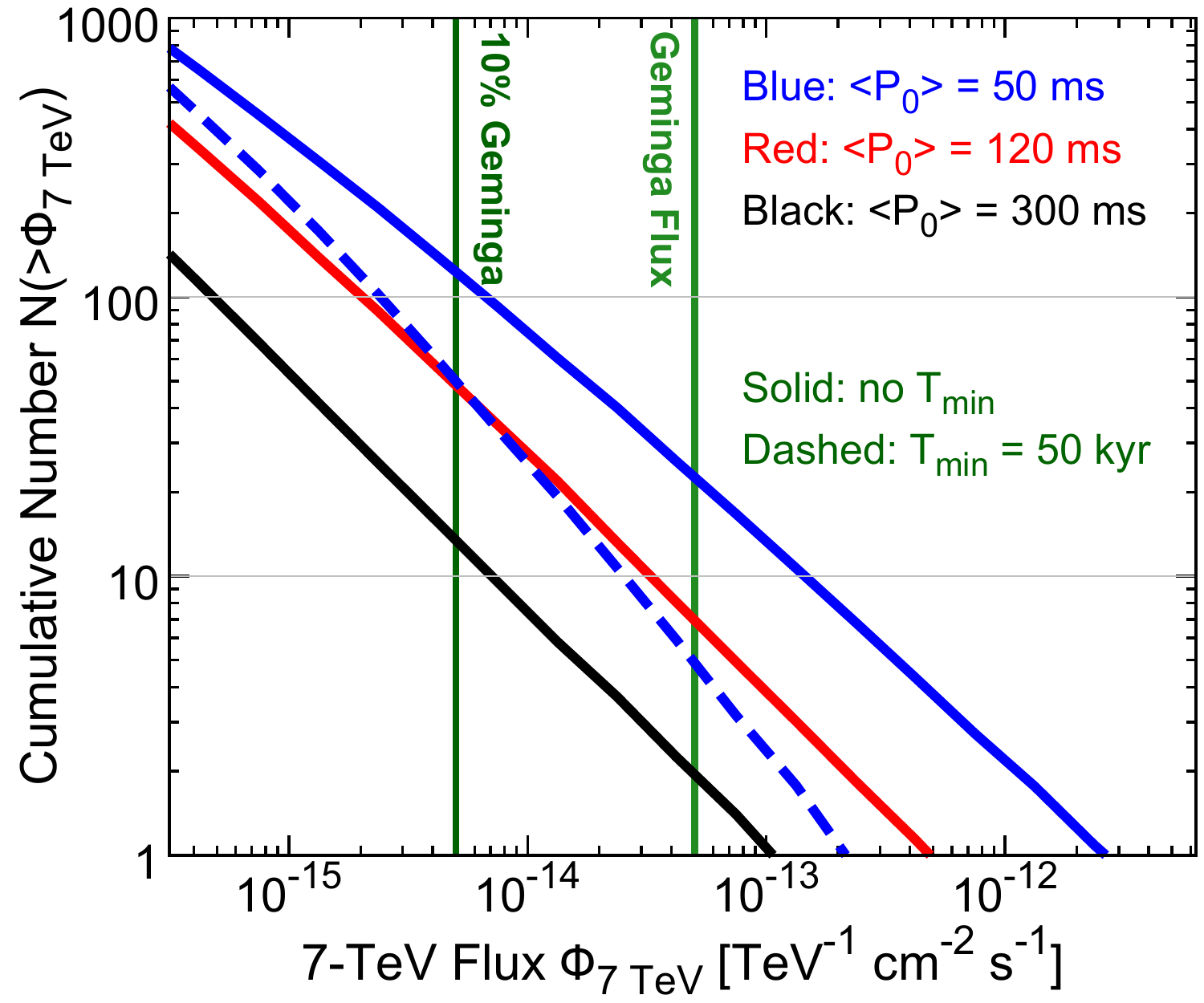}
\caption{\label{fig:flux}Cumulative number distribution of the differential $\gamma$-ray flux at an energy of 7~TeV for all TeV halos within the HAWC field of view, as labeled. The 2HWC sensitivity is approximately equal to 10$\%$ of the Geminga flux.}
\end{figure}

Expanding our analysis to include all TeV halos with aligned radio beams regardless of the TeV halo age (including young sources), we find that the interpretations become trickier because models predict fairly similar TeV halo number counts. Models with $\Pinit$~=~300~ms produce $\sim$~3 TeV halos in case that $\Tmin$~=~0. Meanwhile, models with $\Pinit$~=~50 ms predict approximately $\sim$15 sources for $\Tmin$~=~50~kyr. Our intermediate model with $\Pinit$~=~120~ms also predicts the observation of $\sim$15 sources for $\Tmin$~=~0. However, the age distribution of observed TeV halos differs markedly between models with and without a firm value of $\Tmin$. Thus, future observations that correlate TeV halo activity with pulsars of known ages can more clearly distinguish between models of TeV halo formation, even in light of degeneracies between $\Pinit$ and $\Tmin$.

In Fig.~\ref{fig:flux}, we show the cumulative flux distribution of all TeV halos within the HAWC field of view. The 50-ms model with no $\Tmin$ produces $\sim$20 sources that have $\gamma$-ray fluxes larger than that of Geminga, while the 2HWC catalogue contains 5--12 such potential sources (depending on source extension), so the prediction is somewhat too high. Furthermore, this model predicts a few sources that are at least an order of magnitude brighter than Geminga, while no such source is reported, so the prediction is again somewhat too high, though consistent with Poisson fluctuations. Therefore, while this model is not ruled out by the flux distribution, it is in slight tension. All of the other models are consistent with data.

Due to the steep slope at the bright end of the luminosity function (Fig.~\ref{fig:LF}), nearby sources are expected to dominate the source count. Indeed, in our estimate, about 50$\%$ of observable sources are located within $\simeq$~3~kpc from Earth. This suggests that many observed TeV halos might have large angular sizes, indicating the importance of HAWC, which is suited for detecting extended sources.

\subsection{Diffuse TeV Gamma-Ray Emission Measurements with Milagro}
\label{subsec:MRGO}
\begin{figure}[t]
\includegraphics[width=1.0\columnwidth]{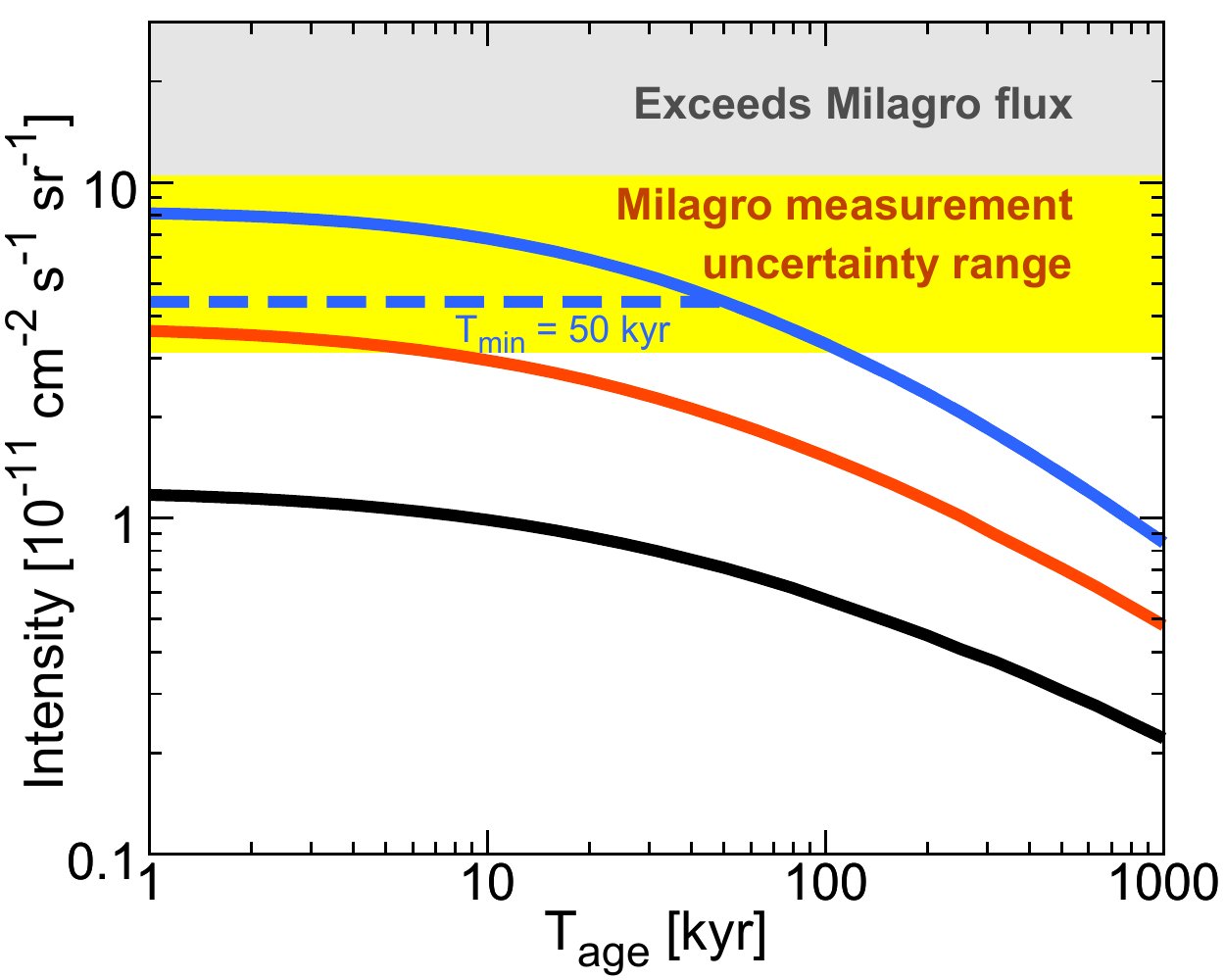}
\caption{\label{fig:MRGO} Cumulative contribution to the diffuse Galactic $\gamma$-ray flux from unresolved TeV halos with ages above $T_{\rm age}$, compared to the measurement by Milagro above 3.5 TeV. Solid unmarked curves have $\Tmin=0$. Line colors have same meaning as Fig.~\ref{fig:2HWC}.}
\end{figure}

Milagro measured the diffuse Galactic $\gamma$-ray flux above 3.5 TeV, finding $\phi(>$3.5 TeV) = (6.8$\pm$1.5$\pm$2.2) $\times 10^{-11}$ cm$^{-2}$ s$^{-1}$ sr$^{-1}$ within a region spanning $40^\circ<l < 100^\circ$ and $|b| < 5^\circ$ \cite{diffuse:MRGO05}.  We constrain our model by requiring that unresolved TeV halos do not overproduce this flux. Due to the hard TeV halo spectrum, alternative diffuse emission measurements at lower energies by ARGO-YBJ \cite{diffuse:ARGO15} or a higher energy in a smaller region analyzed by Milagro \cite{diffuse:MRGO08} give comparable constraints.

To estimate the contribution from TeV halos to this emission, we include contributions from unresolved TeV halos with fluxes below that of Geminga. We also include the contribution from electrons and positrons that escape from unresolved TeV halos and provide a diffuse emission component that fills the interstellar medium. We remove contributions from any individual halo with a $\gamma$-ray flux exceeding Geminga because such a source would be detected by Milagro \cite{TeVhalo:Linden18}. Because the number of such sources is small ($\sim$1 or fewer) and particles lose a significant fraction of their energy in the halo region, their contribution to the diffuse emission is not important. This treatment also allows us to remove unrealistically bright individual halos predicted for the $\Tmin=0$ model, as seen in Fig.~\ref{fig:LF}. Note that we show our model predictions in cumulative contributions from pulsars above 1~kyr. Thus, the results for $\Tmin=0$ are identical for any model $\Tmin\leq$~1~kyr, and are not affected by these unrealistically bright halos that only occur for $\Tmin\simeq0$. We have verified that the number of sources that contribute is large enough that the result is not subject to statistical fluctuations.

In Fig.~\ref{fig:MRGO}, we show that the current diffuse measurement does not strongly constrain our models. However, we stress that having a more precise measurement in the future could provide complementary constraints to future source surveys. 

The diffuse TeV $\gamma$-rays are particularly important, because, as first shown in Ref.~\cite{diffuse:TeVexcess07}, the Milagro measurements \cite{diffuse:MRGO05} of the diffuse flux from the Milky Way plane are significantly higher than expected from extrapolations of the GeV data (the ``TeV excess"). In Ref.~\cite{TeVhalo:Linden18}, it was shown that TeV halos could provide an explanation of this long-standing mystery. Our results also show that unresolved TeV halos could significantly contribute to the diffuse TeV $\gamma$-ray flux. We note that the diffuse emission is dominated by bright sources (Fig.~\ref{fig:LF}). The predicted contribution for $\Pinit$~=~300~ms is smaller than that estimated by Ref.~\cite{TeVhalo:Linden18}, which adopted a harder electron spectrum of $\alpha$~=~1.7 and $E_{\rm cut}$~=~100~TeV. In other words, a better determination of the average electron injection spectrum could increase the predicted flux from unresolved TeV halos, producing tighter constraints on $\Pinit$ and $\Tmin$. Interestingly, in the case of $\Pinit$~=~50~ms or 120~ms, unresolved TeV halos can explain a significant fraction of the Milagro diffuse data without changing the spectral shape from that of our best-fit Geminga model.

\subsection{Summary of Allowed Models and Uncertainties}
\label{subsec:constraints}
Our results are primarily affected by the $P_0$ distribution, and the 2HWC source count allows us to constrain 50~ms~$\lesssim\Pinit\lesssim$~300~ms. Both the 50-ms and 300-ms models are barely allowed, and further investigations of TeV halo candidates will place stronger constraints on $\Pinit$. This indicates that TeV halo observations can provide an important new probe of the $P_0$ distribution, which is difficult to constrain by pulsar statistics.

2HWC data require $\Tmin\gtrsim50$~kyr for $\Pinit$~=~50~ms and $\Tmin\lesssim70$~kyr for $\Pinit$~=~300~ms. The value of $\Tmin$ is not well constrained for $\Pinit\gtrsim$120~ms, but the firm identification of TeV halos around Geminga and Monogem suggests $\Tmin\lesssim$100--300~kyr. Further observations are needed to better constrain this parameter. We stress that 50 kyr is not a strict minimum age for a TeV halo. Rather, we found that, operating under the assumption that the initial spin period of pulsars has an average $\Pinit$~=~50~ms, this minimum age was required to ensure that the TeV halo number was consistent with data. However, the true initial period may have a larger mean, which would eliminate the need for such a cutoff. Alternatively, there may be significant variations between individual objects that are not taken into account in this model. Furthermore, our calculations have several uncertainties noted below, which could relieve the constraint on  $\Tmin$ for the $\Pinit$~=~50~ms model.

We have fixed the distribution of $B_0$ to follow a lognormal distribution with $\langle\log_{10}B_0\rangle$~=~12.65 and $\sigma_{\log_{10}B_0}$~=~0.55. Other studies that examined the magnetic field evolution of pulsars find best-fit mean values that are about 2--4 times larger \cite{psrB0:P10,psrB0:G14}. In these models, the larger magnetic field causes pulsars to spin down faster, producing a smaller spin-down power for pulsars with ages exceeding $\sim$1~kyr. Adopting an alternative model with $\langle\log_{10}B_0\rangle$~=~13.10 and $\sigma_{\log_{10}B_0}$~=~0.65 as derived in Ref.~\cite{psrB0:G14}, we find the predicted number of detectable TeV halos are reduced by a factor of $\sim$ 2. This increases the tension between $\Pinit$~=~300~ms and current HAWC observations, but relieves some tension between $\Pinit$~=~50~ms and the HAWC data. In particular, for these stronger magnetic fields, $\Pinit$~=~50~ms models become consistent with HAWC upper limits for much smaller values of $\Tmin\gtrsim$10~kyr. Further examinations of the $B_0$ distribution will also be important for the study of TeV halo populations.

We also note that throughout this section we focus on ``Geminga-like" TeV halos. We can also adopt different models to take into account deviations from this assumption. We first study the effect of variations in the $\gamma$-ray efficiencies. There might be nearly an order of magnitude variation among individual sources, as noted in Sec.~\ref{sec:review}. We examine alternative models where $\gamma$-ray fluxes are multiplied by a factor of $10^x$. If $x$ is fixed to 0.5 (i.e., all TeV halos are about 3 times brighter than the Geminga halo), then the number of detectable sources is increased by a factor of $\sim$3. Similarly, if $x$ is fixed to -0.5, the source count decreases by a factor of $\sim$3. We then examine the case where $x$ is a random variable drawn from a normal distribution with mean 0 and standard deviation 0.5. The primary effect of such a dispersion would be to smooth out the falling number-count distribution (Fig~\ref{fig:LF}), and increase the number of detectable sources. We find that the number of detectable sources is increased by a factor of $\sim$2 in the case of $\Pinit$~=~120~ms.

Finally, we study the possibility that pulsars younger than $\Tmin$ produce TeV halos with different properties. In particular, at early ages, TeV halos may have smaller $\gamma$-ray efficiencies, because most of the injected energy should be lost to synchrotron emission and there may be less particle energy escaping into the TeV halos, as discussed in Sec.~\ref{sec:review}. Throughout this paper, this effect is simply treated by sharply cutting off contributions from pulsars younger than $\Tmin$, but one could alternatively assume a smooth changes in the $\gamma$-ray efficiencies. This could lead to a detectable population that does not exceed 2HWC constraints. To be more quantitative on this point, we examine alternative models where the $\gamma$-ray fluxes are smoothly reduced by a factor of $(T_{\rm age}/{\rm 340\ kyr})^\beta$ for pulsars younger than Geminga. This replaces the sharp cutoff ($\Tmin$) in our standard formalism. We find that the $\Pinit$~=~50~ms model does not produce too many TeV halos for $\beta\gtrsim$0.7. Further studies are needed to more thoroughly examine this parameter space. 

In Table~\ref{tab:uncertainties}, we show each major uncertainty, mention an alternative model, and roughly indicate the net effect of this model on the predicted number of TeV halo sources. In addition to models mentioned above, we further test several different scenarios, which are explained in Appendix~\ref{app:model_uncertainty}. The exact effect of different uncertainties depends on the standard model that we use for comparison, so we adopt a constant default model of $\langle P_0\rangle$~=~120 ms and $T_{\rm min}$~=~10~kyr in all cases, and show their age dependence in Fig.~\ref{fig:uncertainty} in Appendix~\ref{app:model_uncertainty}.

\begin{table*}
\caption{\label{tab:uncertainties} The most important uncertainties in the number of TeV halos that are discussed in this work. For each uncertainty, we note an alternative model, and roughly indicate the effect that such a model would have on the predicted TeV halo source count.}
\begin{ruledtabular}
\begin{tabular}{cccc}
Name of Uncertainty & Default & Alternative & Effect  \\ \hline
\multicolumn{4}{c}{pulsar population} \\ \hline
$P_0$ distribution & Gaussian & Uniform & {\bf Increase}, $\times2$\\ \hline
$B_0$ distribution & $\langle\log_{10}B_0\rangle$ = 12.65 & $\langle\log_{10}B_0\rangle$ = 13.10 & {\bf Decrease}, $\times0.5$\\ \hline
\multicolumn{4}{c}{$\gamma$-ray efficiency} \\ \hline
\multirow{2}{*}{$\dot{E}$ dependence}  & 
\multirow{2}{*}{$L_\gamma\propto\dot{E}$} & 
$L_\gamma\propto\dot{E}^{0.8}$ & {\bf Decrease}, $\times0.5$ \\  \cline{3-4}
 & & $L_\gamma\propto\dot{E}^{1.2}$ & {\bf Increase}, $\times2$ \\ \hline
\multirow{2}{*}{Age dependence}  &
\multirow{2}{*}{$L_\gamma/\dot{E}$ = const.} &
$L_\gamma/\dot{E} \propto (T_{\rm age})^{0.5}$ & {\bf Decrease}, $\times0.3$ \\  \cline{3-4}
 & & $L_\gamma/\dot{E} \propto (T_{\rm age})^{-0.5}$& {\bf Increase}, $\times3$ \\ \hline
Source-to-source scatter &
None &
$\log_{10}(L_\gamma/\dot{E}) \sim N(1,0.5^2)$ (lognormal, $\sigma=0.5$) & {\bf Increase}, $\times2$\\
\end{tabular}
\end{ruledtabular}
\end{table*}

All of the uncertainties noted above could change the number of detectable sources by a factor of $\sim$2. While these changes are important, they are subdominant to the effect of variations in the $P_0$ distribution, and support our assertion that the $P_0$ distribution dominates the uncertainty in our models. Note that different $B_0$ distributions may lead to smaller number counts, while source variations may increase the number of detectable systems, implying that our default case occupies a reasonable middle value. More TeV halo observations would allow us to better examine these models, and place stronger constraints on pulsar properties. 

\section{Future Directions}
\label{sec:future}
Upcoming surveys have great power to detect TeV halos. In Sec.~\ref{subsec:10HWC}, we quantitatively assess the prospects for Galactic source searches with HAWC and CTA. In Sec.~\ref{subsec:CTA}, we do the same for extragalactic searches with CTA. In Sec.~\ref{subsec:HESS}, we show that detailed morphological studies of existing HESS sources could potentially identify many TeV halos.

\subsection{Extended Source Survey with HAWC and CTA}
\label{subsec:10HWC}

We begin by outlining methods to identify TeV halos. One straightforward way to claim that TeV emission is powered by a pulsar is to detect the radio beam from the pulsed emission or to find a compact PWN at the center of the $\gamma$-ray source. We may also detect a TeV halo component in a composite (TeV halo + PWN) system by examining if its $\gamma$-ray emission can be fit by two morphological components rather than one. In the case of bow-shock pulsars, we can more clearly discriminate TeV halos from PWN, whose size is clearly determined by the stand-off radius \cite{PWN:bowshock2017}. 

In some cases we may be able to detect extended emission around PWNe in other wavelengths, from synchrotron radiation produced by the same electrons and positrons that escape the compact PWN and produce TeV halo emission. Interestingly, Refs.~\cite{XrayHalo:Uchiyama2009,XrayHalo:Bamba2010} potentially detected such emission in X-rays, suggesting the potential for identifying TeV halos in multi-wavelength observations.

\begin{figure}[t]
\includegraphics[width=\columnwidth]{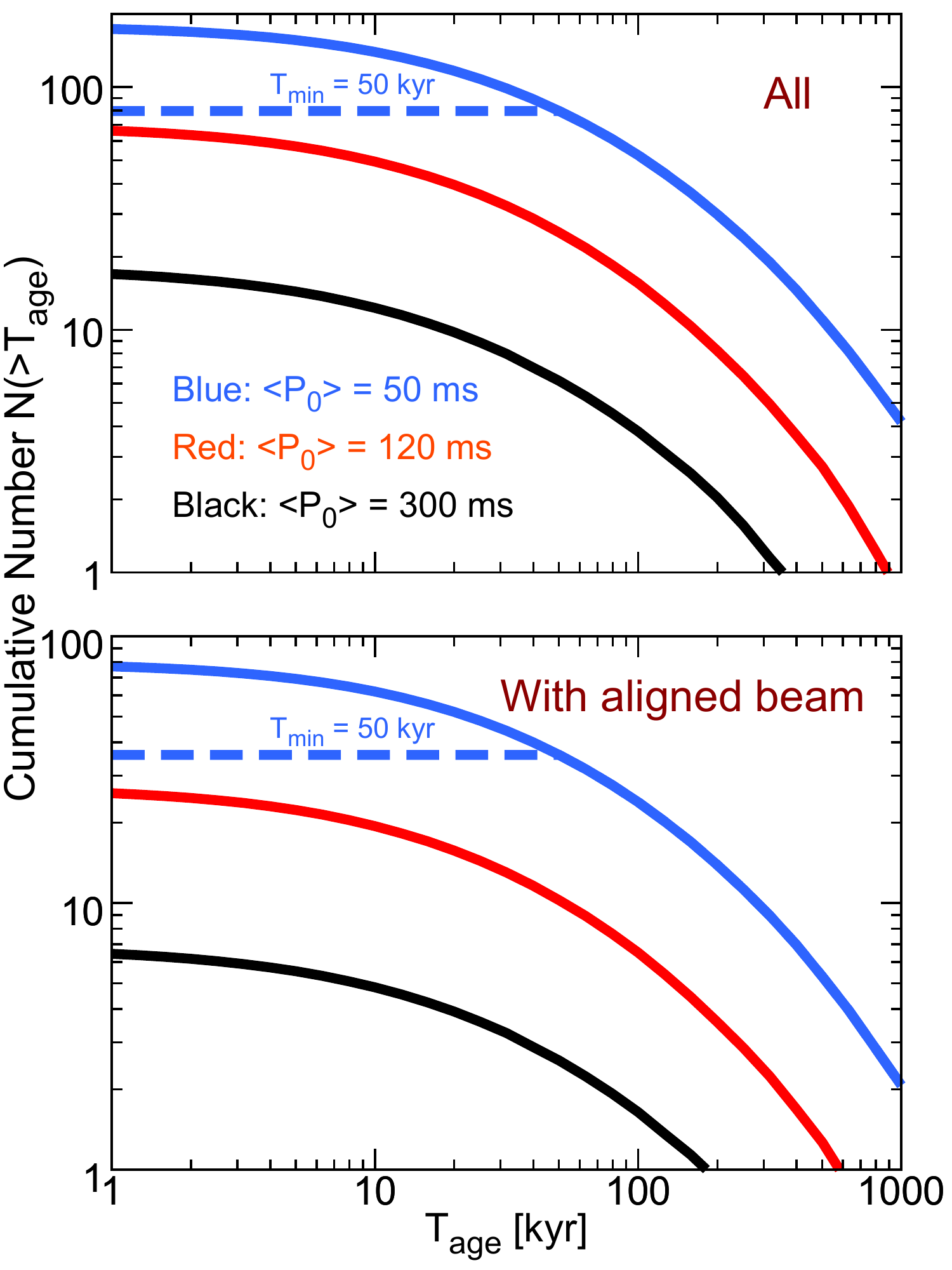}
\caption{\label{fig:10hwc}Same as Fig.~\ref{fig:2HWC}, but using 10 years of HAWC observations. Solid unmarked curves have $\Tmin=0$.}
\end{figure}

In Fig.~\ref{fig:10hwc}, we show expectations for the TeV halo population that could be uncovered by HAWC observations. We assume a 10-yr sensitivity that is improved by a factor of $\sqrt{5}$ compared to the quoted sensitivity of the 2HWC catalog, following the same declination dependence. This corresponds to a sensitivity that is approximately 4\% of the Geminga flux for sources residing in optimal sky positions. These predictions are pessimistic, because HAWC has recently installed an upgrade and increased the instrumented area \cite{HAWCupgrade2017}, an effect which is not included in our calculation.

The 10-yr HAWC survey promises to discover a significant TeV halo population. Even in the pessimistic case of $\Pinit$~=~300~ms, we expect that $\sim$20~sources (including 4 already found) will be detected for $\Tmin = 0$. In the most optimistic case (e.g., $\Pinit$~=~50~ms and $\Tmin$ = 50 kyr), HAWC would be expected to detect $\sim$80 TeV halos. We note that these cases are nearly ruled out by existing TeV halo observations (Sec.~\ref{subsec:2HWC}). Our intermediate case ($\Pinit$~=~120~ms) predicts $\sim$70 sources if $\Tmin = 0$, though this model prediction is only barely allowed from the 2HWC source count (see Fig~\ref{fig:2HWC}). Such a large number of sources would allow us to significantly improve our constraints on the spectral, morphological, and evolutionary properties of TeV halos. 

We stress that our predictions are based on the Geminga-like assumption provided in Eq.~(\ref{eq:flux_Geminga_like}), combined with standard models of pulsar population synthesis established by previous studies (e.g., \cite{psrP0:FK06,psrpoppy}). If HAWC detects a significantly smaller number of TeV halos, it would indicate that Eq.~(\ref{eq:flux_Geminga_like}) cannot be applied to all pulsars, and that the observed Geminga-like halos must have unusually large TeV $\gamma$-ray efficiencies or may have specific properties or environment that generate halos. Conversely, if significantly more sources are observed than predicted, it would indicate that observed sources have relatively dim halos, compared to the average population. Either result would substantially enhance our understanding of these systems. 

\begin{figure*}[t]
\resizebox{18cm}{!}{\includegraphics{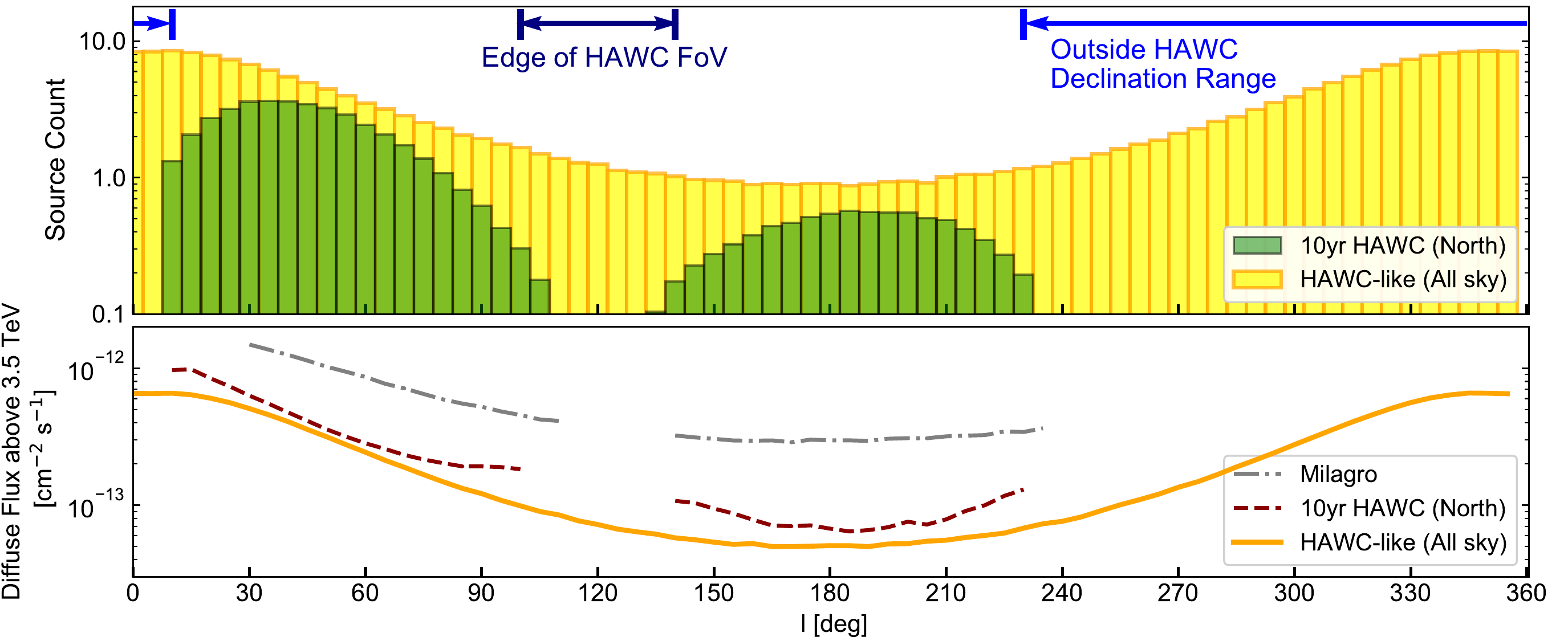}}
\caption{\label{fig:south}The prediction for the Galactic longitude distribution of the TeV halo number (top) and the diffuse $\gamma$-ray flux from unresolved TeV halos (bottom) from $|b|<5^\circ$ in bins of $\Delta l=5^\circ$. All results utilize a  model with $\Pinit$~=~120~ms and $\Tmin$~=~1~kyr. A ``HAWC-like" sensitivity is modeled as a theoretical instrument with a sensitivity of 3\% of the Geminga flux across the entire sky.}
\end{figure*}

In Fig.~\ref{fig:south}, we show our prediction for the Galactic longitude distribution of detected TeV halos (top) and the diffuse flux from unresolved TeV halos (bottom) for our intermediate case of $\Pinit$~=~120~ms. In addition to 10-yr HAWC observations, we make a prediction for hypothetical HAWC-like telescope that uniformly observes the sky with a sensitivity that is 3\% of the Geminga flux. We also show the predicted contribution from TeV halos to the Milagro diffuse measurement, which probed the region within $30^\circ < l < 110^\circ$ and $136^\circ < l < 216^\circ$ \cite{diffuse:MRGO05,diffuse:MRGO08}, and also to the HAWC diffuse measurement, for the region that falls fairly within the field of view. The source count (top panel) shows the sizeable impact of a HAWC-like water Cherenkov telescope at the Southern hemisphere, like the Southern Gamma-Ray Survey Observatory \cite{South2017a,South2017b,South2017c}; it would allow us to probe a region at the edge or outside the HAWC field of view, where we expect a significant number of detectable TeV halos. It also demonstrates that the effect of source confusion is not large; we expect at most $\sim$4~sources over $\Delta l=5^\circ$, which means that the typical intersource spacing is large compared to the angular resolution ($\sim 0.1^\circ$ in radius) and the typical source size (also $\sim 0.1^\circ$ with significant variations).

In Fig.~\ref{fig:south} (bottom), we show that HAWC measurements will greatly improve our understanding of the diffuse TeV flux. In particular, in the region of $40^\circ < l < 100^\circ$, where the diffuse TeV excess was first identified, our predictions indicate that more than half of the diffuse emission from TeV halos will be resolved into individual sources by future HAWC surveys. Apart from shedding light on the nature of the diffuse TeV emission, this would also put constraints on the population of unresolved sources that contribute to the remaining diffuse flux.

We also make a prediction for the future Galactic Plane survey with CTA \cite{CTA:17}. We assume that CTA will observe from $l=0^\circ$ to $360^\circ$ and $|b|<3^\circ$ with a sensitivity of 3 mCrab, where 1 Crab is defined as a $\gamma$-ray flux above 1 TeV of 2.26$\times10^{-11}$ cm$^{-2}$s$^{-1}$, and 3 mCrab corresponds to 2$\%$ of the Geminga flux (defined as $F_{\rm TeV}^{\mathcal{G}}$ in Sec.~\ref{sec:model}). We assume a PSF size of $\theta_{\rm PSF} = 0.05^\circ$ to take take into account the degradation of the sensitivity for extended sources. Because the PSF of CTA is smaller, this effect is more important compared to HAWC observations.

Even in the pessimistic case of $\Pinit$~=~300~ms, we predict that $\sim$30 TeV halos could be detected. In the case of $\Pinit$~=~50~ms with $\Tmin$~=~50~kyr, we predict that about 160 TeV halos could be detected. Our intermediate model, $\Pinit$~=~120~ms, also predicts $\sim$150 for $\Tmin$~=~0. These detections will be highly complementary to HAWC observations. HAWC is located in the Northern hemisphere, while CTA is expected to have better sensitivity in the Southern hemisphere. Moreover, while HAWC is suited for spatially extended sources, CTA can find more distant and dimmer sources. In our prediction for HAWC, the 10--90\% containment fraction of TeV halo distances corresponds to roughly 1--10 kpc. In contrast, for CTA, the 10--90\% containment window spans from 3--15 kpc. Together, these observations can map out much of the Galaxy in TeV halos. This also indicates that another water Cherenkov telescope at the Southern hemisphere would be critical for detecting nearby sources throughout the Galactic Plane. 

Finally, we study the effect of source confusion for CTA. First, we find that CTA is expected to detect $\sim7$ sources in a 5$^\circ$ wide bin even in the densest regions (near the Galactic Center). Second, 90\% of CTA sources has distances of above 3~kpc, which translates to an angular extension less than 0.5$^\circ$ assuming the source size of 25~pc. These indicate that the source confusion has marginal effect on our results, but further understanding of luminosity function of other sources and CTA properties are needed to better quantify this effect.

\subsection{Extragalactic Survey with CTA}
\label{subsec:CTA}
Milky Way TeV halo searches must deal with large angular source size and distance uncertainties. One way to avoid these issues is to search for extragalactic sources. A good target is the Large Magellanic Cloud (LMC), which is nearby and face-on, and which will also be extensively observed by the CTA as part of its Key Scientific Program \cite{CTA:17}. CTA observations are expected to achieve an integrated energy flux sensitivity of 3 $\times$~10$^{-14}$ erg cm$^{-2}$ s$^{-1}$ above 1~TeV. The collaboration expects to uncover $\sim$10~sources that are primarily SNRs, without including TeV halos.

We estimate the number of TeV halos that can be detected in the LMC by the CTA survey. We adopt a standard distance of 50~kpc, and count the number of halos with luminosities exceeding 9~$\times~10^{33}$~erg s$^{-1}$ above 1~TeV. The birth rate of pulsars in the LMC is normalized to be 0.005 yr$^{-1}$, which is the lower value obtained by a previous study of LMC pulsar population modelling \cite{LMC:Ridley10}. Though the interstellar infrared radiation field in the LMC is weaker compared to the Milky Way \cite{LMCrad}, the predicted TeV halo flux is reduced only by a factor of 1.3 even if we set $\rho_{\rm IR}$~=~0, due to the contribution from the cosmic microwave background photons. Since this modification is degenerate with a number of uncertainties, we do not take this into account in what follows.

\begin{figure}[t]
\includegraphics[width=\columnwidth]{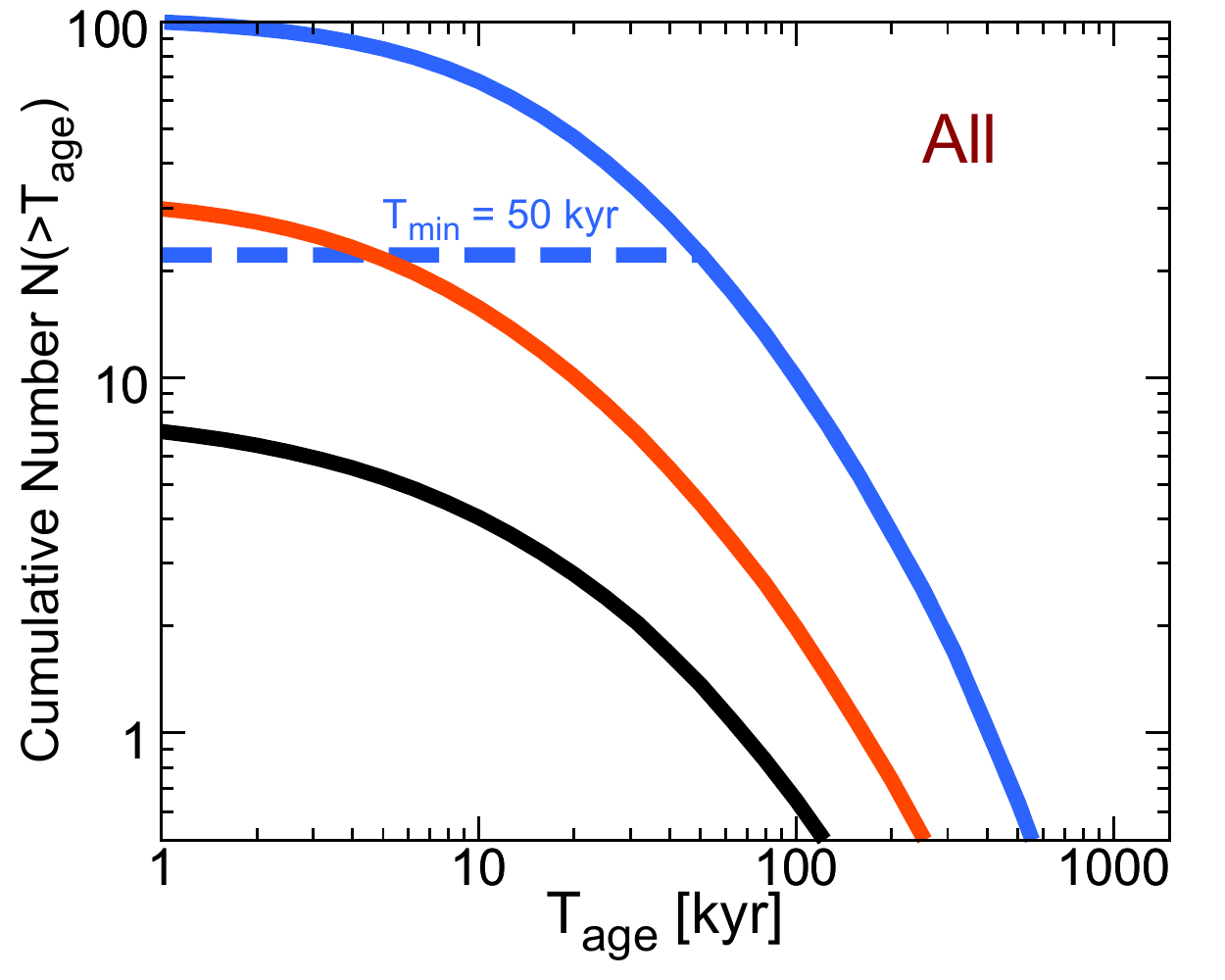}
\caption{\label{fig:lmc}Same as the upper panel of Fig.~\ref{fig:10hwc}, but for the LMC survey with CTA. Curves have $\Tmin=0$ unless marked.}
\end{figure}

In Fig.~\ref{fig:lmc}, we show that CTA will likely detect at least $\sim$1, and potentially $\sim$30, extragalactic TeV halos in the LMC, substantially increasing the total number of sources detected with this survey. These observations will provide more important information than the source count alone, shedding insight into the brightest TeV halos in a region without significant distance uncertainties. Thus, CTA observations will provide complementary constraints to HAWC Milky Way observations. The differentiation of extragalactic TeV halos from PWNe will be challenging, because both will appear pointlike even with the unparalleled angular resolution of CTA. The best path forward will be to employ followup radio observations of synchrotron counterparts to further examine the emission morphology. Because the size of halos is poorly understood, there may also be a possibility that we could observe TeV halos extended beyond a radius of $\sim$ 100~pc, which could be detected as extended sources even in the LMC.

We note that the observed radio pulsars in the LMC appear to have a strikingly different distribution of spindown powers compared to expectations from pulsar evolution (Fig.~\ref{fig:LF}). In particular, current observations detect two very bright ($\dot{E}>10^{38}$ erg s$^{-1}$) pulsars, while the 11 other detected pulsars have low $\dot{E}$ ($\dot{E}< 6 \times 10^{34}$ erg s$^{-1}$). There are no pulsars in between these ranges~\cite{ATNF}. This is most likely due to the combination of selection effects, weak correlations between $\dot{E}$ and radio luminosity \cite{psrL:Szary14}, and the randomness of pulse radiation beamed toward us. Future pulsar surveys may enable us to better examine the pulsar population in the LMC. Since TeV halo emissions are expected to be more isotropic and may better correlate with $\dot{E}$ than radio pulse emissions, they could provide complementary information regarding the population of bright pulsars, potentially resolving this tension, or confirming it. In the latter case, it would demand significant modifications to the theory of pulsar formation and evolution.

Finally, we note that we might also be able to observe a similar number of TeV halos in the Small Magellanic Cloud, because it has a distance and pulsar formation rate comparable to the LMC \cite{LMC:Ridley10}. This would potentially provide information regarding the evolution of TeV halos in low metallicity environments. 

\subsection{Followup Study for HESS Sources}
\label{subsec:HESS}

So far, we have focused on the existing survey catalog by HAWC, which is suited for extended source surveys. However, existing source catalogs from imaging air Cherenkov telescopes should also contain as-yet identified TeV halos.

\begin{figure}[t]
\includegraphics[width=\columnwidth]{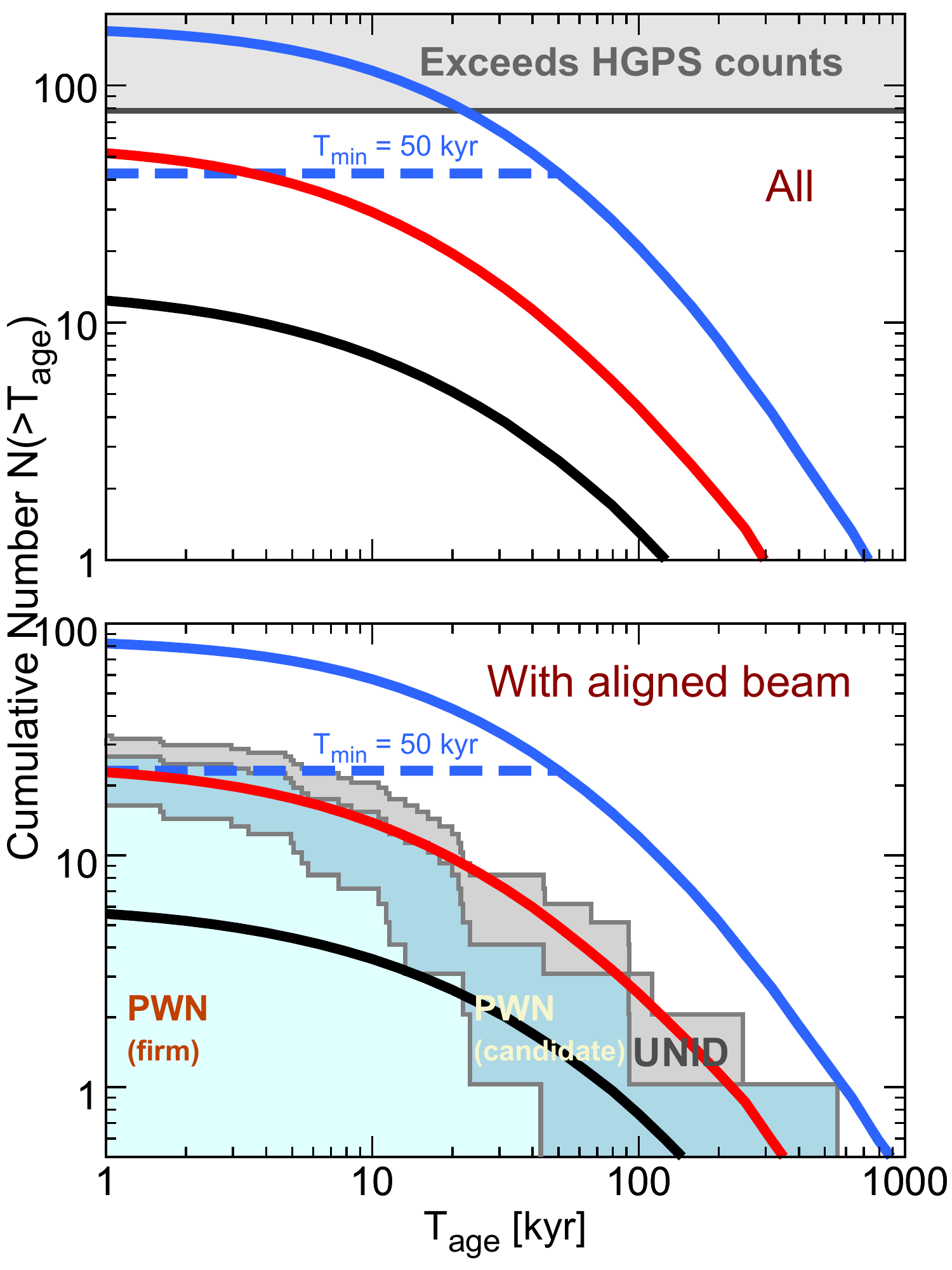}
\caption{\label{fig:hess} Same as Fig.~\ref{fig:2HWC}, but for the HESS Galactic Plane Survey (HGPS). In the bottom panel, the histogram of HGPS sources associated with ATNF pulsars is also shown, divided into three classes following Refs.~\cite{cat:HESSGP,cat:HESSPWN}. Note that from all unidentified sources in HGPS we only plot those associated with radio pulsars.}
\end{figure}

Here we focus on the HESS Galactic Plane Survey (HGPS) catalog, which has detected 78 sources in total, 42 of which are associated with ATNF pulsars \cite{cat:HESSGP}. Five of these tentative pulsar associations are known to be either an SNR or a binary, as well as the Arc and Galactic Center, while the remaining 37 sources are categorized as firmly identified PWN (including composite system), candidate PWN, or unidentified sources. We examine how many of these sources could be interpreted as TeV halos.

We make a prediction following the methodology for the CTA Galactic Plane survey in Sec.~\ref{subsec:10HWC}. We include sources between Galactic longitudes spanning from \mbox{$l=250^\circ$} to $65^\circ$ and latitude $|b|<3^\circ$. The sensitivity of HGPS is non-uniform across the Galactic Plane. Since our goal is not to make precise estimates of the HESS sensitivity, we simply assume a sensitivity of 1$\%$ Crab for point sources and utilize a PSF of $\theta_{\rm PSF} = 0.08^\circ$. We further assume that any source of $\theta_{\rm size} > 0.7^\circ$ is not observed, because HGPS is not able to detect such an extended source. This removes the contribution from any halos within about 700~pc of Earth.

In Fig.~\ref{fig:hess} (top), we show that 10--50 sources in the HGPS catalog could be TeV halos. This indicates that detailed morphological studies of HGPS sources could uncover many TeV halos in this catalog. The definitive identification of these sources would be important in constraining particle transport due to the unparalleled angular resolution of HESS.

In Fig.~\ref{fig:hess} (bottom), we compare the predicted number of TeV halos to the observed number of radio pulsar associations. Our model predicts that, among the 37 sources associated with ATNF pulsars, $\sim$6--20 sources could be TeV halos. Interestingly, the shape of our predicted age distribution matches observed data well. Because our predictions are based on the assumption that the $\gamma$-ray flux is proportional to $\dot{E}$, this agreement suggests that these HGPS sources are powered by pulsar activity, either PWNe or TeV halos, rather than SNR.

\section{Conclusions}
\label{sec:conclusion}

TeV halos are a new class of $\gamma$-ray sources \cite{TeVhalo:Linden17,TeVhalo:MRGO09,TeVhalo:HAWC17,TeVhalo:AAK97,TeVhalo:Yuksel09,TeVhalo:Linden18,TeVhalo:Hooper18,TeVhalo:Evoli18,TeVhalo:MSP2018,ATel:2017,ATel:2018}. They are bright, have hard spectra, and are spatially extended. They are powered by electrons and positrons that escaped from the PWNe, but which remain confined in a region where diffusion is strongly suppressed. Empirical arguments suggest TeV halos are common around pulsars. However, among many candidate sources, only four TeV halos have been confirmed so far. The rest are likely undetected due to the diminishing sensitivity of TeV instruments to extended $\gamma$-ray sources.

In this work, for the first time, we theoretically quantify the role of future surveys to detect more TeV halos. We also study new implications for pulsar physics and existing $\gamma$-ray sources. We use standard methods for pulsar population synthesis and focus on a model where the TeV halo luminosity is calculated based on Geminga observations. Our analysis produced three main results.

\begin{itemize}

\item {\bf TeV halos could be the most important source class in future TeV $\gamma$-ray surveys.} Within the context of our standard Geminga-like model, and utilizing the range of $P_0$ and $T_{\rm min}$ that are consistent with current datasets, we predict that HAWC will eventually detect $\sim$20--80 TeV halos, and future Galactic surveys by CTA will also find $\sim$30--160 halos, for a total of $\sim$50--240 halos. Further, CTA can potentially detect $\sim$10 TeV halos in the LMC and SMC. This indicates that TeV halos could be the dominant source class in the TeV $\gamma$-ray sky. Such a large number of sources would allow us to examine their properties and evolution in great detail.

\item {\bf Further studies of unidentified TeV sources and ``TeV PWN" are needed.} We find that the HGPS catalog might contain $\sim$10--50 TeV halos, which are currently classified as either unidentified sources or PWNe. These results have three implications. First, imaging air Cherenkov telescopes like HESS, MAGIC and VERITAS can also play an important role in studying TeV halos. In particular, their high angular resolution would be critical in examining particle transport inside the halo. Second, it might be important for modelling of ``TeV PWN" to take into account the emission from TeV halo regions. Third, X-ray and radio observations of ``TeV PWN" may find many extended halos around compact PWNe. This synchrotron emission counterpart could help to identify TeV halos.

\item {\bf TeV halos observations can constrain pulsar properties.} Our predictions are primarily affected by the distribution of the initial spindown period, which is not well constrained by pulsar population studies. In other words, TeV halo observations can provide complementary constraints to existing radio surveys. Current 2HWC data allow 50~ms~$\lesssim$~$\Pinit$~$\lesssim$~300~ms, and further studies will place tighter constraints.

\end{itemize}

We finally note that TeV halo observations may unlock new opportunities to study astrophysics. 

\begin{itemize}
    \item TeV halo observations would allow us to detect pulsars with radio emission not aligned toward Earth and hence which have been missed in previous blind searches \cite{TeVhalo:Linden17}. Further, the angular size of halos could provide useful distance estimations for Galactic pulsars. Thus, many observations of TeV halos could allow us to map out pulsars in the Galaxy, including misaligned systems. This new method would work as an independent and complementary method compared to radio observations. In this regard, the Southern Gamma-Ray Survey Observatory would play an important role in detecting TeV halos across the Galactic Plane, especially in the inner Galaxy. In addition, next-generation telescopes like LHASSO \cite{LHAASO:2016} can find more TeV halos.
    \item TeV halo observations would substantially improve our understanding of total galactic $\gamma$-ray emission. Usually, galactic $\gamma$-ray emission is assumed to be dominated by hadronic processes induced by diffusing protons and nuclei, especially in the GeV energy range. The bright and hard-spectrum emission from TeV halos suggests that leptonic emission mechanisms may be important for the diffuse emission in the TeV energy range. This has two implications for the cosmic background emission. First, the hard spectrum of TeV halo emission might make ordinary galaxies more important for TeV $\gamma$-ray background than expected only from hadronic emission, which falls off steeply. Second, the leptonic nature of TeV halo emission
    may make star-forming galaxies less important for the TeV neutrino background than expected from the assumption that all TeV galactic $\gamma$-ray emission is hadronic. In particular, predicting neutrino flux from galaxies by simply extrapolating their $\gamma$-ray flux could result in a substantial overestimate.
    \item TeV halos could help pinpoint the sources of IceCube neutrinos in our Galaxy. A promising way to search for neutrino emitters is to look into $\gamma$-ray source catalogs. However, if most $\gamma$-ray sources are TeV halos, there is less room for hadronic sources. This has both positive and negative implications for neutrino astronomy. It is unfortunate, since we can only expect high-energy neutrinos from a small fraction of identifiable $\gamma$-ray sources. On the other hand, if we can identify TeV halos in $\gamma$-ray source catalogs, we can ignore their neutrino contributions and reduce the trials factor in IceCube neutrino cross-correlations.
    \item Existing TeV halo observations indicate that pulsars contribute to the cosmic-ray electron and positron flux. However, future observations are necessary to understand the exact degree they contribute, which classes of systems are important, and the constraints which can be put on residual contributions by exotic physics such as dark matter annihilation. In particular, follow up observations of TeV halos with GeV $\gamma$-rays and other wavelengths can provide important complementary information capable of constraining the pulsar contribution to the positron excess, which is seen in GeV energy range.
\end{itemize}
Finally, it is remarkable that such an important source class escaped identification until the development of TeV $\gamma$-ray instruments, especially those that can detect extended sources. Although TeV halos may have corresponding emission in radio, X-ray, and GeV photons, this was not sufficiently obvious to recognize this source class. The significance of these observations thus predicts the importance of future TeV surveys in understanding the multi-wavelength sky.

\section*{Acknowledgements}

We are grateful to Felix Aharonian, Aya Bamba, Chad Brisbois, Mattia Di Mauro, Fiorenza Donato, Henrike Fleischhack, Dan Hooper, Petra Huentemeyer, Wataru Ishizaki, Kelly Malone, Miguel Mostafa, Kohta Murase, Pat Slane, Andrew Smith, Terry Walker, and especially Katie Auchettl for discussions and helpful comments on the manuscript. We are grateful to Sophie Beacom for help with Fig.~\ref{fig:pic}. We also thank two anonymous referees for helpful comments. TS is supported by Research Fellowship of Japan Society for the Promotion of Science (JSPS), and also supported by JSPS KAKENHI Grant Number JP 18J20943. JFB is supported by NSF grant PHY-1714479.

\newpage
\clearpage


\appendix

\section{Effect of Uniform Pulsar Period Distributions}
\label{app:uniform_distribution}

One of the most important modeling assumptions in this paper is the spin-period of the pulsar at birth. In the main paper, we utilized several models that utilized a Gaussian distribution to describe the initial pulsar period distribution. We chose three average periods of 50~ms, 120~ms, or 300~ms with this Gaussian, along with variances $\sigma_{P_0}$~=~50/$\sqrt{2}$~ms, $\sigma_{P_0}$~=~60~ms and $\sigma_{P_0}$~=~150~ms, respectively. However, recent modelling of pulsar population indicates that a uniform period distribution of $0<P_0<500$~ms may be consistent with the pulsar statistics~\cite{Gullon:2015zca}. 

\begin{figure}[b]
\includegraphics[width=\columnwidth]{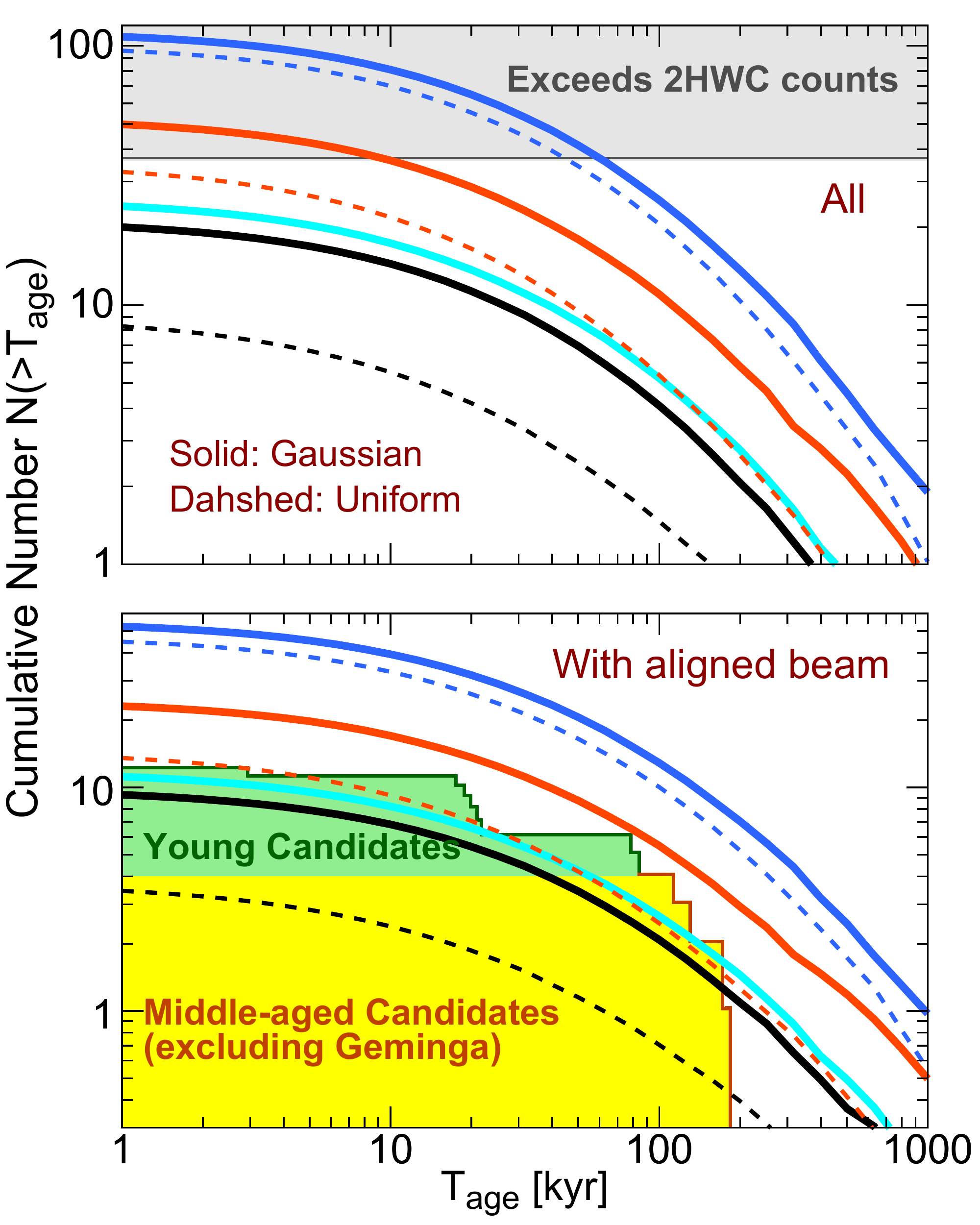}
\caption{\label{fig:uniformP0} Same as Fig.~\ref{fig:2HWC}, but for different models for the $P_0$ distribution with Gaussian (dotted) and uniform (solid). The cyan curve shows the case of a uniform distribution from 0 to 500 ms.}
\end{figure}

In Fig.~\ref{fig:uniformP0}, we show a version of Fig.~\ref{fig:2HWC} that shows the expected number of TeV halos in the 2HWC catalog as a function of the pulsar age for models with uniform initial spin-period distributions between the breakup limit and $2\langle P_0\rangle$. In addition to three cases of $\Pinit$~=50, 120, and 300~ms, we show the case of a uniform $P_0$ distribution up to 500~ms, as suggested in Ref.~\cite{Gullon:2015zca}. 

We find that, in general, models with a uniform pulsar distribution of $P_0$ predict a greater number of pulsar than Gaussian models. This is because the uniform distribution produces a larger number of sources with very small $P_0$. In general, the effect of these models on the number of pulsars as a function of $T_{\rm age}$ is about a factor of $\sim$ 2, compared to models with a Gaussian distribution centered at $\Pinit$.

\section{Model Uncertainties}
\label{app:model_uncertainty}

\begin{figure*}
\resizebox{16cm}{!}{\includegraphics{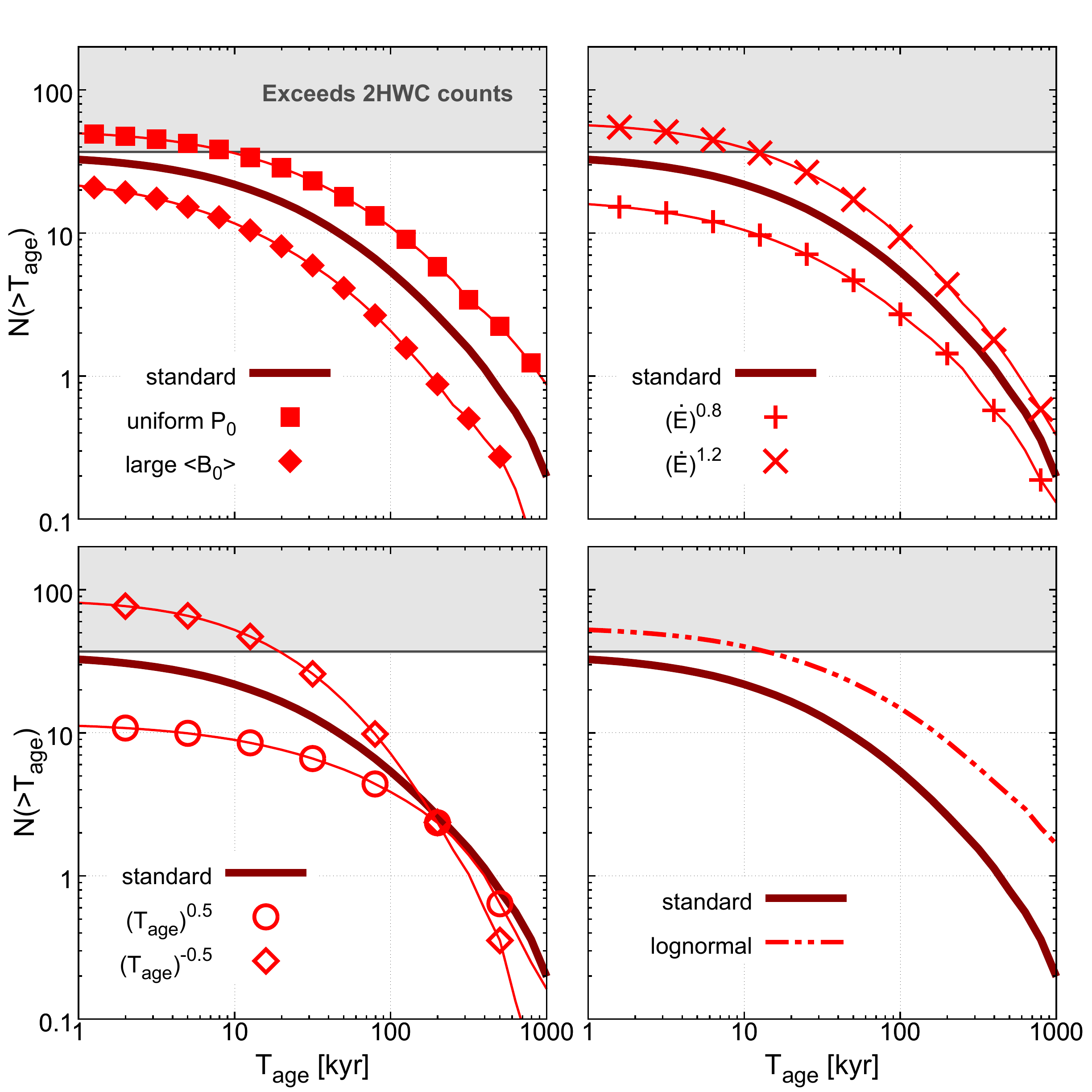}}
\caption{\label{fig:uncertainty}Same as the upper panel of Fig.~\ref{fig:2HWC}, but for alternative models in the case of $\Pinit$~=~120~ms.}
\end{figure*}

Here, we describe all uncertainties in Table~\ref{tab:uncertainties}. In Fig.~\ref{fig:uncertainty}, we show how our results are changed for alternative models for the case of $\langle P_0\rangle$~=~120 ms. 

\begin{itemize}
    \item {\bf $P_0$ distribution}: In the main text, we adopt a Gaussian distribution following Refs.~\cite{psrP0:FK06,psrP0:WR11}. In Fig.~\ref{fig:uncertainty} (top left), we compare this with a uniform distribution, motivated by Ref.~\cite{Gullon:2015zca}.
    \item {\bf $B_0$ distribution}: In the main text, we adopt a log-normal distribution with mean $\langle\log_{10}B_0\rangle$~=~12.65 and  $\sigma_{\log_{10}B_0}$~=~0.55 following Ref.~\cite{psrP0:FK06}. In  Fig.~\ref{fig:uncertainty} (top left), we compare this with another lognormal distribution with $\langle\log_{10}B_0\rangle$~=~13.10 and $\sigma_{\log_{10}B_0}$~=~0.65, as defined in Ref.~\cite{psrB0:G14}.
    \item {\bf $\dot{E}$ dependence}: In the main text, we adopt $L_\gamma\propto\dot{E}$. In Fig.~\ref{fig:uncertainty} (top right), we compare this with different models which adopt $L_\gamma\propto\dot{E}^{0.8}$ and $L_\gamma\propto\dot{E}^{1.2}$, normalizing the $\gamma$-ray flux with the Geminga halo.
    \item {\bf age dependence}: In the main text, we assume that $\gamma$-ray efficiency $L_\gamma/\dot{E}$ is constant for all pulsars. In Fig.~\ref{fig:uncertainty} (bottom left), we compare this with different models where $L_\gamma/\dot{E}$ depends on the age of pulsars as $L_\gamma/\dot{E}\propto (T_{\rm age})^{0.5}$ and $L_\gamma/\dot{E}\propto (T_{\rm age})^{-0.5}$, normalizing the $\gamma$-ray flux with the Geminga halo and adopting 340~kyr as the age of Geminga. In the case of $L_\gamma/\dot{E}\propto (T_{\rm age})^{0.5}$, we do not assume age dependence for sources older than 340~kyr, in order to avoid producing unrealistically bright late-time sources.
    \item{\bf source-to-source scatter}: In the main text, we assume that $L_\gamma/\dot{E}$ is constant for all systems. In Fig.~\ref{fig:uncertainty} (bottom right), we compare this with different models where $\log_{10}(L_\gamma/\dot{E})$ is a random variable drawn from a normal distribution with mean 1 and standard deviation 0.5.
\end{itemize}

\newpage
\clearpage

\twocolumngrid
\bibliography{tevhalomodels.bib}

\begin{thebibliography}{103}%
\makeatletter
\providecommand \@ifxundefined [1]{%
 \@ifx{#1\undefined}
}%
\providecommand \@ifnum [1]{%
 \ifnum #1\expandafter \@firstoftwo
 \else \expandafter \@secondoftwo
 \fi
}%
\providecommand \@ifx [1]{%
 \ifx #1\expandafter \@firstoftwo
 \else \expandafter \@secondoftwo
 \fi
}%
\providecommand \natexlab [1]{#1}%
\providecommand \enquote  [1]{``#1''}%
\providecommand \bibnamefont  [1]{#1}%
\providecommand \bibfnamefont [1]{#1}%
\providecommand \citenamefont [1]{#1}%
\providecommand \href@noop [0]{\@secondoftwo}%
\providecommand \href [0]{\begingroup \@sanitize@url \@href}%
\providecommand \@href[1]{\@@startlink{#1}\@@href}%
\providecommand \@@href[1]{\endgroup#1\@@endlink}%
\providecommand \@sanitize@url [0]{\catcode `\\12\catcode `\$12\catcode
  `\&12\catcode `\#12\catcode `\^12\catcode `\_12\catcode `\%12\relax}%
\providecommand \@@startlink[1]{}%
\providecommand \@@endlink[0]{}%
\providecommand \url  [0]{\begingroup\@sanitize@url \@url }%
\providecommand \@url [1]{\endgroup\@href {#1}{\urlprefix }}%
\providecommand \urlprefix  [0]{URL }%
\providecommand \Eprint [0]{\href }%
\providecommand \doibase [0]{http://dx.doi.org/}%
\providecommand \selectlanguage [0]{\@gobble}%
\providecommand \bibinfo  [0]{\@secondoftwo}%
\providecommand \bibfield  [0]{\@secondoftwo}%
\providecommand \translation [1]{[#1]}%
\providecommand \BibitemOpen [0]{}%
\providecommand \bibitemStop [0]{}%
\providecommand \bibitemNoStop [0]{.\EOS\space}%
\providecommand \EOS [0]{\spacefactor3000\relax}%
\providecommand \BibitemShut  [1]{\csname bibitem#1\endcsname}%
\let\auto@bib@innerbib\@empty
\bibitem [{\citenamefont {{Linden}}\ \emph {et~al.}(2017)\citenamefont
  {{Linden}}, \citenamefont {{Auchettl}}, \citenamefont {{Bramante}},
  \citenamefont {{Cholis}}, \citenamefont {{Fang}}, \citenamefont {{Hooper}},
  \citenamefont {{Karwal}},\ and\ \citenamefont {{Li}}}]{TeVhalo:Linden17}%
  \BibitemOpen
  \bibfield  {author} {\bibinfo {author} {\bibfnamefont {T.}~\bibnamefont
  {{Linden}}}, \bibinfo {author} {\bibfnamefont {K.}~\bibnamefont
  {{Auchettl}}}, \bibinfo {author} {\bibfnamefont {J.}~\bibnamefont
  {{Bramante}}}, \bibinfo {author} {\bibfnamefont {I.}~\bibnamefont
  {{Cholis}}}, \bibinfo {author} {\bibfnamefont {K.}~\bibnamefont {{Fang}}},
  \bibinfo {author} {\bibfnamefont {D.}~\bibnamefont {{Hooper}}}, \bibinfo
  {author} {\bibfnamefont {T.}~\bibnamefont {{Karwal}}}, \ and\ \bibinfo
  {author} {\bibfnamefont {S.~W.}\ \bibnamefont {{Li}}},\ }\bibfield  {title}
  {\enquote {\bibinfo {title} {{Using HAWC to discover invisible pulsars}},}\
  }\href {\doibase 10.1103/PhysRevD.96.103016} {\bibfield  {journal} {\bibinfo
  {journal} {\prd}\ }\textbf {\bibinfo {volume} {96}},\ \bibinfo {eid} {103016}
  (\bibinfo {year} {2017})},\ \Eprint {http://arxiv.org/abs/1703.09704}
  {arXiv:1703.09704 [astro-ph.HE]} \BibitemShut {NoStop}%
\bibitem [{\citenamefont {{Abdo}}\ \emph {et~al.}(2009)\citenamefont {{Abdo}}
  \emph {et~al.}}]{TeVhalo:MRGO09}%
  \BibitemOpen
  \bibfield  {author} {\bibinfo {author} {\bibfnamefont {A.~A.}\ \bibnamefont
  {{Abdo}}} \emph {et~al.},\ }\bibfield  {title} {\enquote {\bibinfo {title}
  {{Milagro Observations of Multi-TeV Emission from Galactic Sources in the
  Fermi Bright Source List}},}\ }\href {\doibase 10.1088/0004-637X/700/2/L127}
  {\bibfield  {journal} {\bibinfo  {journal} {Astrophys. J. Lett.}\ }\textbf
  {\bibinfo {volume} {700}},\ \bibinfo {pages} {L127--L131} (\bibinfo {year}
  {2009})},\ \Eprint {http://arxiv.org/abs/0904.1018} {arXiv:0904.1018
  [astro-ph.HE]} \BibitemShut {NoStop}%
\bibitem [{\citenamefont {{Abeysekara}}\ \emph
  {et~al.}(2017{\natexlab{a}})\citenamefont {{Abeysekara}} \emph
  {et~al.}}]{TeVhalo:HAWC17}%
  \BibitemOpen
  \bibfield  {author} {\bibinfo {author} {\bibfnamefont {A.~U.}\ \bibnamefont
  {{Abeysekara}}} \emph {et~al.},\ }\bibfield  {title} {\enquote {\bibinfo
  {title} {{Extended gamma-ray sources around pulsars constrain the origin of
  the positron flux at Earth}},}\ }\href {\doibase 10.1126/science.aan4880}
  {\bibfield  {journal} {\bibinfo  {journal} {Science}\ }\textbf {\bibinfo
  {volume} {358}},\ \bibinfo {pages} {911--914} (\bibinfo {year}
  {2017}{\natexlab{a}})},\ \Eprint {http://arxiv.org/abs/1711.06223}
  {arXiv:1711.06223 [astro-ph.HE]} \BibitemShut {NoStop}%
\bibitem [{\citenamefont {{Abeysekara}}\ \emph
  {et~al.}(2017{\natexlab{b}})\citenamefont {{Abeysekara}} \emph
  {et~al.}}]{cat:2HWC}%
  \BibitemOpen
  \bibfield  {author} {\bibinfo {author} {\bibfnamefont {A.~U.}\ \bibnamefont
  {{Abeysekara}}} \emph {et~al.},\ }\bibfield  {title} {\enquote {\bibinfo
  {title} {{The 2HWC HAWC Observatory Gamma-Ray Catalog}},}\ }\href {\doibase
  10.3847/1538-4357/aa7556} {\bibfield  {journal} {\bibinfo  {journal} {\apj}\
  }\textbf {\bibinfo {volume} {843}},\ \bibinfo {eid} {40} (\bibinfo {year}
  {2017}{\natexlab{b}})},\ \Eprint {http://arxiv.org/abs/1702.02992}
  {arXiv:1702.02992 [astro-ph.HE]} \BibitemShut {NoStop}%
\bibitem [{\citenamefont {{Thorsett}}\ \emph {et~al.}(2003)\citenamefont
  {{Thorsett}}, \citenamefont {{Benjamin}}, \citenamefont {{Brisken}},
  \citenamefont {{Golden}},\ and\ \citenamefont {{Goss}}}]{Monogem:SNR}%
  \BibitemOpen
  \bibfield  {author} {\bibinfo {author} {\bibfnamefont {S.~E.}\ \bibnamefont
  {{Thorsett}}}, \bibinfo {author} {\bibfnamefont {R.~A.}\ \bibnamefont
  {{Benjamin}}}, \bibinfo {author} {\bibfnamefont {W.~F.}\ \bibnamefont
  {{Brisken}}}, \bibinfo {author} {\bibfnamefont {A.}~\bibnamefont {{Golden}}},
  \ and\ \bibinfo {author} {\bibfnamefont {W.~M.}\ \bibnamefont {{Goss}}},\
  }\bibfield  {title} {\enquote {\bibinfo {title} {{Pulsar PSR B0656+14, the
  Monogem Ring, and the Origin of the ``Knee'' in the Primary Cosmic-Ray
  Spectrum}},}\ }\href {\doibase 10.1086/377682} {\bibfield  {journal}
  {\bibinfo  {journal} {Astrophys. J. Lett.}\ }\textbf {\bibinfo {volume}
  {592}},\ \bibinfo {pages} {L71--L73} (\bibinfo {year} {2003})},\ \Eprint
  {http://arxiv.org/abs/astro-ph/0306462} {astro-ph/0306462} \BibitemShut
  {NoStop}%
\bibitem [{\citenamefont {{Abdalla}}\ \emph
  {et~al.}(2018{\natexlab{a}})\citenamefont {{Abdalla}} \emph
  {et~al.}}]{cat:HESSGP}%
  \BibitemOpen
  \bibfield  {author} {\bibinfo {author} {\bibfnamefont {H.}~\bibnamefont
  {{Abdalla}}} \emph {et~al.} (\bibinfo {collaboration} {H.E.S.S.
  Collaboration}),\ }\bibfield  {title} {\enquote {\bibinfo {title} {{The
  H.E.S.S. Galactic plane survey}},}\ }\href {\doibase
  10.1051/0004-6361/201732098} {\bibfield  {journal} {\bibinfo  {journal}
  {Astron. and Astrophys.}\ }\textbf {\bibinfo {volume} {612}},\ \bibinfo {eid}
  {A1} (\bibinfo {year} {2018}{\natexlab{a}})},\ \Eprint
  {http://arxiv.org/abs/1804.02432} {arXiv:1804.02432 [astro-ph.HE]}
  \BibitemShut {NoStop}%
\bibitem [{\citenamefont {{Abdalla}}\ \emph
  {et~al.}(2018{\natexlab{b}})\citenamefont {{Abdalla}} \emph
  {et~al.}}]{cat:HESSPWN}%
  \BibitemOpen
  \bibfield  {author} {\bibinfo {author} {\bibfnamefont {H.}~\bibnamefont
  {{Abdalla}}} \emph {et~al.} (\bibinfo {collaboration} {H.E.S.S.
  Collaboration}),\ }\bibfield  {title} {\enquote {\bibinfo {title} {{The
  population of TeV pulsar wind nebulae in the H.E.S.S. Galactic Plane
  Survey}},}\ }\href {\doibase 10.1051/0004-6361/201629377} {\bibfield
  {journal} {\bibinfo  {journal} {Astron. and Astrophys.}\ }\textbf {\bibinfo
  {volume} {612}},\ \bibinfo {eid} {A2} (\bibinfo {year}
  {2018}{\natexlab{b}})},\ \Eprint {http://arxiv.org/abs/1702.08280}
  {arXiv:1702.08280 [astro-ph.HE]} \BibitemShut {NoStop}%
\bibitem [{\citenamefont {{Khangulyan}}\ \emph {et~al.}(2018)\citenamefont
  {{Khangulyan}}, \citenamefont {{Koldoba}}, \citenamefont {{Ustyugova}},
  \citenamefont {{Bogovalov}},\ and\ \citenamefont
  {{Aharonian}}}]{HESS:largePWN}%
  \BibitemOpen
  \bibfield  {author} {\bibinfo {author} {\bibfnamefont {D.}~\bibnamefont
  {{Khangulyan}}}, \bibinfo {author} {\bibfnamefont {A.~V.}\ \bibnamefont
  {{Koldoba}}}, \bibinfo {author} {\bibfnamefont {G.~V.}\ \bibnamefont
  {{Ustyugova}}}, \bibinfo {author} {\bibfnamefont {S.~V.}\ \bibnamefont
  {{Bogovalov}}}, \ and\ \bibinfo {author} {\bibfnamefont {F.}~\bibnamefont
  {{Aharonian}}},\ }\bibfield  {title} {\enquote {\bibinfo {title} {{On the
  Anomalously Large Extension of the Pulsar Wind Nebula HESS J1825-137}},}\
  }\href {\doibase 10.3847/1538-4357/aac20f} {\bibfield  {journal} {\bibinfo
  {journal} {\apj}\ }\textbf {\bibinfo {volume} {860}},\ \bibinfo {eid} {59}
  (\bibinfo {year} {2018})},\ \Eprint {http://arxiv.org/abs/1712.10161}
  {arXiv:1712.10161 [astro-ph.HE]} \BibitemShut {NoStop}%
\bibitem [{\citenamefont {{Gaensler}}\ and\ \citenamefont
  {{Slane}}(2006)}]{PWNreview:2006}%
  \BibitemOpen
  \bibfield  {author} {\bibinfo {author} {\bibfnamefont {B.~M.}\ \bibnamefont
  {{Gaensler}}}\ and\ \bibinfo {author} {\bibfnamefont {P.~O.}\ \bibnamefont
  {{Slane}}},\ }\bibfield  {title} {\enquote {\bibinfo {title} {{The Evolution
  and Structure of Pulsar Wind Nebulae}},}\ }\href {\doibase
  10.1146/annurev.astro.44.051905.092528} {\bibfield  {journal} {\bibinfo
  {journal} {Ann. Rev. of Astron. and Astrophys.}\ }\textbf {\bibinfo {volume}
  {44}},\ \bibinfo {pages} {17--47} (\bibinfo {year} {2006})},\ \Eprint
  {http://arxiv.org/abs/astro-ph/0601081} {astro-ph/0601081} \BibitemShut
  {NoStop}%
\bibitem [{\citenamefont {{Aharonian}}(1995)}]{Aharonian1995}%
  \BibitemOpen
  \bibfield  {author} {\bibinfo {author} {\bibfnamefont {F.~A.}\ \bibnamefont
  {{Aharonian}}},\ }\bibfield  {title} {\enquote {\bibinfo {title} {{Very high
  energy gamma-ray astronomy and the origin of cosmic rays.}}}\ }\href
  {\doibase 10.1016/0920-5632(95)00022-2} {\bibfield  {journal} {\bibinfo
  {journal} {Nuclear Physics B Proceedings Supplements}\ }\textbf {\bibinfo
  {volume} {39}},\ \bibinfo {pages} {193--206} (\bibinfo {year}
  {1995})}\BibitemShut {NoStop}%
\bibitem [{\citenamefont {{Aharonian}}\ \emph {et~al.}(1997)\citenamefont
  {{Aharonian}}, \citenamefont {{Atoyan}},\ and\ \citenamefont
  {{Kifune}}}]{TeVhalo:AAK97}%
  \BibitemOpen
  \bibfield  {author} {\bibinfo {author} {\bibfnamefont {F.~A.}\ \bibnamefont
  {{Aharonian}}}, \bibinfo {author} {\bibfnamefont {A.~M.}\ \bibnamefont
  {{Atoyan}}}, \ and\ \bibinfo {author} {\bibfnamefont {T.}~\bibnamefont
  {{Kifune}}},\ }\bibfield  {title} {\enquote {\bibinfo {title} {{Inverse
  Compton gamma radiation of faint synchrotron X-ray nebulae around
  pulsars}},}\ }\href {\doibase 10.1093/mnras/291.1.162} {\bibfield  {journal}
  {\bibinfo  {journal} {Mon. Not. R. Astron. Soc.}\ }\textbf {\bibinfo {volume}
  {291}},\ \bibinfo {pages} {162--176} (\bibinfo {year} {1997})}\BibitemShut
  {NoStop}%
\bibitem [{\citenamefont {{Aharonian}}(2004)}]{Aharonian_text}%
  \BibitemOpen
  \bibfield  {author} {\bibinfo {author} {\bibfnamefont {F.~A.}\ \bibnamefont
  {{Aharonian}}},\ }\href {\doibase 10.1142/4657} {\emph {\bibinfo {title}
  {Very High Energy Cosmic Gamma Radiation: A Crucial Window on the Extreme
  Universe.}}}\ (\bibinfo  {publisher} {World Scientific Publishing Co},\
  \bibinfo {year} {2004})\BibitemShut {NoStop}%
\bibitem [{\citenamefont {{Y{\"u}ksel}}\ \emph {et~al.}(2009)\citenamefont
  {{Y{\"u}ksel}}, \citenamefont {{Kistler}},\ and\ \citenamefont
  {{Stanev}}}]{TeVhalo:Yuksel09}%
  \BibitemOpen
  \bibfield  {author} {\bibinfo {author} {\bibfnamefont {H.}~\bibnamefont
  {{Y{\"u}ksel}}}, \bibinfo {author} {\bibfnamefont {M.~D.}\ \bibnamefont
  {{Kistler}}}, \ and\ \bibinfo {author} {\bibfnamefont {T.}~\bibnamefont
  {{Stanev}}},\ }\bibfield  {title} {\enquote {\bibinfo {title} {{TeV Gamma
  Rays from Geminga and the Origin of the GeV Positron Excess}},}\ }\href
  {\doibase 10.1103/PhysRevLett.103.051101} {\bibfield  {journal} {\bibinfo
  {journal} {Physical Review Letters}\ }\textbf {\bibinfo {volume} {103}},\
  \bibinfo {eid} {051101} (\bibinfo {year} {2009})},\ \Eprint
  {http://arxiv.org/abs/0810.2784} {arXiv:0810.2784} \BibitemShut {NoStop}%
\bibitem [{\citenamefont {{Aharonian}}(2013)}]{Aharonian2013}%
  \BibitemOpen
  \bibfield  {author} {\bibinfo {author} {\bibfnamefont {F.}~\bibnamefont
  {{Aharonian}}},\ }\bibfield  {title} {\enquote {\bibinfo {title} {{Gamma Rays
  at Very High Energies}},}\ }\href {\doibase 10.1007/978-3-642-36134-0_1}
  {\bibfield  {journal} {\bibinfo  {journal} {Astrophysics at Very High
  Energies, Saas-Fee Advanced Course, Volume 40.~ISBN
  978-3-642-36133-3.~Springer-Verlag Berlin Heidelberg, 2013, p.~1}\ }\textbf
  {\bibinfo {volume} {40}},\ \bibinfo {pages} {1} (\bibinfo {year}
  {2013})}\BibitemShut {NoStop}%
\bibitem [{\citenamefont {{Kargaltsev}}\ and\ \citenamefont
  {{Pavlov}}(2008)}]{PWNmorph:Kargaltsev2008}%
  \BibitemOpen
  \bibfield  {author} {\bibinfo {author} {\bibfnamefont {O.}~\bibnamefont
  {{Kargaltsev}}}\ and\ \bibinfo {author} {\bibfnamefont {G.~G.}\ \bibnamefont
  {{Pavlov}}},\ }\bibfield  {title} {\enquote {\bibinfo {title} {{Pulsar Wind
  Nebulae in the Chandra Era}},}\ }in\ \href {\doibase 10.1063/1.2900138}
  {\emph {\bibinfo {booktitle} {40 Years of Pulsars: Millisecond Pulsars,
  Magnetars and More}}},\ \bibinfo {series} {American Institute of Physics
  Conference Series}, Vol.\ \bibinfo {volume} {983},\ \bibinfo {editor} {edited
  by\ \bibinfo {editor} {\bibfnamefont {C.}~\bibnamefont {{Bassa}}}, \bibinfo
  {editor} {\bibfnamefont {Z.}~\bibnamefont {{Wang}}}, \bibinfo {editor}
  {\bibfnamefont {A.}~\bibnamefont {{Cumming}}}, \ and\ \bibinfo {editor}
  {\bibfnamefont {V.~M.}\ \bibnamefont {{Kaspi}}}}\ (\bibinfo {year} {2008})\
  pp.\ \bibinfo {pages} {171--185},\ \Eprint {http://arxiv.org/abs/0801.2602}
  {arXiv:0801.2602 [astro-ph]} \BibitemShut {NoStop}%
\bibitem [{\citenamefont {{Badenes}}\ \emph {et~al.}(2010)\citenamefont
  {{Badenes}}, \citenamefont {{Maoz}},\ and\ \citenamefont
  {{Draine}}}]{SNRmorph:Badenes2010}%
  \BibitemOpen
  \bibfield  {author} {\bibinfo {author} {\bibfnamefont {Carles}\ \bibnamefont
  {{Badenes}}}, \bibinfo {author} {\bibfnamefont {Dan}\ \bibnamefont {{Maoz}}},
  \ and\ \bibinfo {author} {\bibfnamefont {Bruce~T.}\ \bibnamefont
  {{Draine}}},\ }\bibfield  {title} {\enquote {\bibinfo {title} {{On the size
  distribution of supernova remnants in the Magellanic Clouds}},}\ }\href
  {\doibase 10.1111/j.1365-2966.2010.17023.x} {\bibfield  {journal} {\bibinfo
  {journal} {Mon. Not. R. Astron. Soc.}\ }\textbf {\bibinfo {volume} {407}},\
  \bibinfo {pages} {1301--1313} (\bibinfo {year} {2010})},\ \Eprint
  {http://arxiv.org/abs/1003.3030} {arXiv:1003.3030 [astro-ph.GA]} \BibitemShut
  {NoStop}%
\bibitem [{\citenamefont {{Stafford}}\ \emph {et~al.}(2018)\citenamefont
  {{Stafford}}, \citenamefont {{Lopez}}, \citenamefont {{Auchettl}},\ and\
  \citenamefont {{Holland-Ashford}}}]{SNRmorph:Stafford2018}%
  \BibitemOpen
  \bibfield  {author} {\bibinfo {author} {\bibfnamefont {Jennifer~N.}\
  \bibnamefont {{Stafford}}}, \bibinfo {author} {\bibfnamefont {Laura~A.}\
  \bibnamefont {{Lopez}}}, \bibinfo {author} {\bibfnamefont {Katie}\
  \bibnamefont {{Auchettl}}}, \ and\ \bibinfo {author} {\bibfnamefont {Tyler}\
  \bibnamefont {{Holland-Ashford}}},\ }\bibfield  {title} {\enquote {\bibinfo
  {title} {{The Age Evolution of the Radio Morphology of Supernova
  Remnants}},}\ }\href@noop {} {\bibfield  {journal} {\bibinfo  {journal}
  {arXiv e-prints}\ ,\ \bibinfo {eid} {arXiv:1808.08234}} (\bibinfo {year}
  {2018})},\ \Eprint {http://arxiv.org/abs/1808.08234} {arXiv:1808.08234
  [astro-ph.HE]} \BibitemShut {NoStop}%
\bibitem [{\citenamefont {{Riviere}}\ \emph {et~al.}(2017)\citenamefont
  {{Riviere}}, \citenamefont {{Fleischhack}},\ and\ \citenamefont
  {{Sandoval}}}]{ATel:2017}%
  \BibitemOpen
  \bibfield  {author} {\bibinfo {author} {\bibfnamefont {C.}~\bibnamefont
  {{Riviere}}}, \bibinfo {author} {\bibfnamefont {H.}~\bibnamefont
  {{Fleischhack}}}, \ and\ \bibinfo {author} {\bibfnamefont {A.}~\bibnamefont
  {{Sandoval}}},\ }\bibfield  {title} {\enquote {\bibinfo {title} {{HAWC
  detection of TeV emission near PSR B0540+23}},}\ }\href@noop {} {\bibfield
  {journal} {\bibinfo  {journal} {The Astronomer's Telegram}\ }\textbf
  {\bibinfo {volume} {10941}} (\bibinfo {year} {2017})}\BibitemShut {NoStop}%
\bibitem [{\citenamefont {{Brisbois}}\ \emph {et~al.}(2018)\citenamefont
  {{Brisbois}}, \citenamefont {{Riviere}}, \citenamefont {{Fleischhack}},\ and\
  \citenamefont {{Smith}}}]{ATel:2018}%
  \BibitemOpen
  \bibfield  {author} {\bibinfo {author} {\bibfnamefont {C.}~\bibnamefont
  {{Brisbois}}}, \bibinfo {author} {\bibfnamefont {C.}~\bibnamefont
  {{Riviere}}}, \bibinfo {author} {\bibfnamefont {H.}~\bibnamefont
  {{Fleischhack}}}, \ and\ \bibinfo {author} {\bibfnamefont {A.}~\bibnamefont
  {{Smith}}},\ }\bibfield  {title} {\enquote {\bibinfo {title} {{HAWC detection
  of TeV source HAWC J0635+070}},}\ }\href@noop {} {\bibfield  {journal}
  {\bibinfo  {journal} {The Astronomer's Telegram}\ }\textbf {\bibinfo {volume}
  {12013}} (\bibinfo {year} {2018})}\BibitemShut {NoStop}%
\bibitem [{\citenamefont {{Hooper}}\ \emph {et~al.}(2018)\citenamefont
  {{Hooper}}, \citenamefont {{Cholis}},\ and\ \citenamefont
  {{Linden}}}]{TeVhalo:Hooper18}%
  \BibitemOpen
  \bibfield  {author} {\bibinfo {author} {\bibfnamefont {D.}~\bibnamefont
  {{Hooper}}}, \bibinfo {author} {\bibfnamefont {I.}~\bibnamefont {{Cholis}}},
  \ and\ \bibinfo {author} {\bibfnamefont {T.}~\bibnamefont {{Linden}}},\
  }\bibfield  {title} {\enquote {\bibinfo {title} {{TeV gamma rays from
  Galactic Center pulsars}},}\ }\href {\doibase 10.1016/j.dark.2018.05.004}
  {\bibfield  {journal} {\bibinfo  {journal} {Physics of the Dark Universe}\
  }\textbf {\bibinfo {volume} {21}},\ \bibinfo {pages} {40--46} (\bibinfo
  {year} {2018})},\ \Eprint {http://arxiv.org/abs/1705.09293} {arXiv:1705.09293
  [astro-ph.HE]} \BibitemShut {NoStop}%
\bibitem [{\citenamefont {{Linden}}\ and\ \citenamefont
  {{Buckman}}(2018)}]{TeVhalo:Linden18}%
  \BibitemOpen
  \bibfield  {author} {\bibinfo {author} {\bibfnamefont {T.}~\bibnamefont
  {{Linden}}}\ and\ \bibinfo {author} {\bibfnamefont {B.~J.}\ \bibnamefont
  {{Buckman}}},\ }\bibfield  {title} {\enquote {\bibinfo {title} {{Pulsar TeV
  Halos Explain the Diffuse TeV Excess Observed by Milagro}},}\ }\href
  {\doibase 10.1103/PhysRevLett.120.121101} {\bibfield  {journal} {\bibinfo
  {journal} {Physical Review Letters}\ }\textbf {\bibinfo {volume} {120}},\
  \bibinfo {eid} {121101} (\bibinfo {year} {2018})},\ \Eprint
  {http://arxiv.org/abs/1707.01905} {arXiv:1707.01905 [astro-ph.HE]}
  \BibitemShut {NoStop}%
\bibitem [{\citenamefont {{Hooper}}\ and\ \citenamefont
  {{Linden}}(2018{\natexlab{a}})}]{TeVhalo:MSP2018}%
  \BibitemOpen
  \bibfield  {author} {\bibinfo {author} {\bibfnamefont {D.}~\bibnamefont
  {{Hooper}}}\ and\ \bibinfo {author} {\bibfnamefont {T.}~\bibnamefont
  {{Linden}}},\ }\bibfield  {title} {\enquote {\bibinfo {title} {{Millisecond
  pulsars, TeV halos, and implications for the Galactic Center gamma-ray
  excess}},}\ }\href {\doibase 10.1103/PhysRevD.98.043005} {\bibfield
  {journal} {\bibinfo  {journal} {\prd}\ }\textbf {\bibinfo {volume} {98}},\
  \bibinfo {eid} {043005} (\bibinfo {year} {2018}{\natexlab{a}})},\ \Eprint
  {http://arxiv.org/abs/1803.08046} {arXiv:1803.08046 [astro-ph.HE]}
  \BibitemShut {NoStop}%
\bibitem [{\citenamefont {{Hooper}}\ \emph {et~al.}(2017)\citenamefont
  {{Hooper}}, \citenamefont {{Cholis}}, \citenamefont {{Linden}},\ and\
  \citenamefont {{Fang}}}]{TeVhalo:Hooper17}%
  \BibitemOpen
  \bibfield  {author} {\bibinfo {author} {\bibfnamefont {D.}~\bibnamefont
  {{Hooper}}}, \bibinfo {author} {\bibfnamefont {I.}~\bibnamefont {{Cholis}}},
  \bibinfo {author} {\bibfnamefont {T.}~\bibnamefont {{Linden}}}, \ and\
  \bibinfo {author} {\bibfnamefont {K.}~\bibnamefont {{Fang}}},\ }\bibfield
  {title} {\enquote {\bibinfo {title} {{HAWC observations strongly favor pulsar
  interpretations of the cosmic-ray positron excess}},}\ }\href {\doibase
  10.1103/PhysRevD.96.103013} {\bibfield  {journal} {\bibinfo  {journal}
  {\prd}\ }\textbf {\bibinfo {volume} {96}},\ \bibinfo {eid} {103013} (\bibinfo
  {year} {2017})},\ \Eprint {http://arxiv.org/abs/1702.08436} {arXiv:1702.08436
  [astro-ph.HE]} \BibitemShut {NoStop}%
\bibitem [{\citenamefont {{Hooper}}\ and\ \citenamefont
  {{Linden}}(2018{\natexlab{b}})}]{TeVhalo:Hooper18b}%
  \BibitemOpen
  \bibfield  {author} {\bibinfo {author} {\bibfnamefont {Dan}\ \bibnamefont
  {{Hooper}}}\ and\ \bibinfo {author} {\bibfnamefont {Tim}\ \bibnamefont
  {{Linden}}},\ }\bibfield  {title} {\enquote {\bibinfo {title} {{Measuring the
  local diffusion coefficient with H.E.S.S. observations of very high-energy
  electrons}},}\ }\href {\doibase 10.1103/PhysRevD.98.083009} {\bibfield
  {journal} {\bibinfo  {journal} {\prd}\ }\textbf {\bibinfo {volume} {98}},\
  \bibinfo {eid} {083009} (\bibinfo {year} {2018}{\natexlab{b}})},\ \Eprint
  {http://arxiv.org/abs/1711.07482} {arXiv:1711.07482 [astro-ph.HE]}
  \BibitemShut {NoStop}%
\bibitem [{\citenamefont {{Evoli}}\ \emph {et~al.}(2018)\citenamefont
  {{Evoli}}, \citenamefont {{Linden}},\ and\ \citenamefont
  {{Morlino}}}]{TeVhalo:Evoli18}%
  \BibitemOpen
  \bibfield  {author} {\bibinfo {author} {\bibfnamefont {C.}~\bibnamefont
  {{Evoli}}}, \bibinfo {author} {\bibfnamefont {T.}~\bibnamefont {{Linden}}}, \
  and\ \bibinfo {author} {\bibfnamefont {G.}~\bibnamefont {{Morlino}}},\
  }\bibfield  {title} {\enquote {\bibinfo {title} {{Self-generated cosmic-ray
  confinement in TeV halos: Implications for TeV {$\gamma$} -ray emission and
  the positron excess}},}\ }\href {\doibase 10.1103/PhysRevD.98.063017}
  {\bibfield  {journal} {\bibinfo  {journal} {\prd}\ }\textbf {\bibinfo
  {volume} {98}},\ \bibinfo {eid} {063017} (\bibinfo {year} {2018})},\ \Eprint
  {http://arxiv.org/abs/1807.09263} {arXiv:1807.09263 [astro-ph.HE]}
  \BibitemShut {NoStop}%
\bibitem [{\citenamefont {{Fang}}\ \emph {et~al.}(2018)\citenamefont {{Fang}},
  \citenamefont {{Bi}}, \citenamefont {{Yin}},\ and\ \citenamefont
  {{Yuan}}}]{TeVhalo:Fang18}%
  \BibitemOpen
  \bibfield  {author} {\bibinfo {author} {\bibfnamefont {K.}~\bibnamefont
  {{Fang}}}, \bibinfo {author} {\bibfnamefont {X.-J.}\ \bibnamefont {{Bi}}},
  \bibinfo {author} {\bibfnamefont {P.-F.}\ \bibnamefont {{Yin}}}, \ and\
  \bibinfo {author} {\bibfnamefont {Q.}~\bibnamefont {{Yuan}}},\ }\bibfield
  {title} {\enquote {\bibinfo {title} {{Two-zone Diffusion of Electrons and
  Positrons from Geminga Explains the Positron Anomaly}},}\ }\href {\doibase
  10.3847/1538-4357/aad092} {\bibfield  {journal} {\bibinfo  {journal} {\apj}\
  }\textbf {\bibinfo {volume} {863}},\ \bibinfo {eid} {30} (\bibinfo {year}
  {2018})},\ \Eprint {http://arxiv.org/abs/1803.02640} {arXiv:1803.02640
  [astro-ph.HE]} \BibitemShut {NoStop}%
\bibitem [{\citenamefont {{Profumo}}\ \emph {et~al.}(2018)\citenamefont
  {{Profumo}}, \citenamefont {{Reynoso-Cordova}}, \citenamefont {{Kaaz}},\ and\
  \citenamefont {{Silverman}}}]{TeVhalo:Profumo18}%
  \BibitemOpen
  \bibfield  {author} {\bibinfo {author} {\bibfnamefont {S.}~\bibnamefont
  {{Profumo}}}, \bibinfo {author} {\bibfnamefont {J.}~\bibnamefont
  {{Reynoso-Cordova}}}, \bibinfo {author} {\bibfnamefont {N.}~\bibnamefont
  {{Kaaz}}}, \ and\ \bibinfo {author} {\bibfnamefont {M.}~\bibnamefont
  {{Silverman}}},\ }\bibfield  {title} {\enquote {\bibinfo {title} {{Lessons
  from HAWC pulsar wind nebulae observations: The diffusion constant is not a
  constant; pulsars remain the likeliest sources of the anomalous positron
  fraction; cosmic rays are trapped for long periods of time in pockets of
  inefficient diffusion}},}\ }\href {\doibase 10.1103/PhysRevD.97.123008}
  {\bibfield  {journal} {\bibinfo  {journal} {\prd}\ }\textbf {\bibinfo
  {volume} {97}},\ \bibinfo {eid} {123008} (\bibinfo {year} {2018})},\ \Eprint
  {http://arxiv.org/abs/1803.09731} {arXiv:1803.09731 [astro-ph.HE]}
  \BibitemShut {NoStop}%
\bibitem [{\citenamefont {{Bucciantini}}(2018)}]{TeVhalo:Bucciantini18}%
  \BibitemOpen
  \bibfield  {author} {\bibinfo {author} {\bibfnamefont {N.}~\bibnamefont
  {{Bucciantini}}},\ }\bibfield  {title} {\enquote {\bibinfo {title} {{Escape
  of high-energy particles from bow-shock pulsar wind nebulae}},}\ }\href
  {\doibase 10.1093/mnras/sty2237} {\bibfield  {journal} {\bibinfo  {journal}
  {Mon. Not. R. Astron. Soc.}\ }\textbf {\bibinfo {volume} {480}},\ \bibinfo
  {pages} {5419--5426} (\bibinfo {year} {2018})},\ \Eprint
  {http://arxiv.org/abs/1808.08757} {arXiv:1808.08757 [astro-ph.HE]}
  \BibitemShut {NoStop}%
\bibitem [{\citenamefont {{Tang}}\ and\ \citenamefont
  {{Piran}}(2018)}]{Geminga:Tang18}%
  \BibitemOpen
  \bibfield  {author} {\bibinfo {author} {\bibfnamefont {Xiaping}\ \bibnamefont
  {{Tang}}}\ and\ \bibinfo {author} {\bibfnamefont {Tsvi}\ \bibnamefont
  {{Piran}}},\ }\bibfield  {title} {\enquote {\bibinfo {title} {{Positron flux
  and gamma-ray emission from Geminga pulsar and pulsar wind nebula}},}\
  }\href@noop {} {\bibfield  {journal} {\bibinfo  {journal} {arXiv e-prints}\
  ,\ \bibinfo {eid} {arXiv:1808.02445}} (\bibinfo {year} {2018})},\ \Eprint
  {http://arxiv.org/abs/1808.02445} {arXiv:1808.02445 [astro-ph.HE]}
  \BibitemShut {NoStop}%
\bibitem [{\citenamefont {{Xi}}\ \emph {et~al.}(2018)\citenamefont {{Xi}},
  \citenamefont {{Liu}}, \citenamefont {{Huang}}, \citenamefont {{Fang}},
  \citenamefont {{Yan}},\ and\ \citenamefont {{Wang}}}]{Geminga:Xi18}%
  \BibitemOpen
  \bibfield  {author} {\bibinfo {author} {\bibfnamefont {Shao-Qiang}\
  \bibnamefont {{Xi}}}, \bibinfo {author} {\bibfnamefont {Ruo-Yu}\ \bibnamefont
  {{Liu}}}, \bibinfo {author} {\bibfnamefont {Zhi-Qiu}\ \bibnamefont
  {{Huang}}}, \bibinfo {author} {\bibfnamefont {Kun}\ \bibnamefont {{Fang}}},
  \bibinfo {author} {\bibfnamefont {Huirong}\ \bibnamefont {{Yan}}}, \ and\
  \bibinfo {author} {\bibfnamefont {Xiang-Yu}\ \bibnamefont {{Wang}}},\
  }\bibfield  {title} {\enquote {\bibinfo {title} {{GeV observations of the
  extended pulsar wind nebulae challenge the pulsar interpretations of the
  cosmic-ray positron excess}},}\ }\href@noop {} {\bibfield  {journal}
  {\bibinfo  {journal} {arXiv e-prints}\ ,\ \bibinfo {eid} {arXiv:1810.10928}}
  (\bibinfo {year} {2018})},\ \Eprint {http://arxiv.org/abs/1810.10928}
  {arXiv:1810.10928 [astro-ph.HE]} \BibitemShut {NoStop}%
\bibitem [{\citenamefont {{Manchester}}\ \emph {et~al.}(2005)\citenamefont
  {{Manchester}}, \citenamefont {{Hobbs}}, \citenamefont {{Teoh}},\ and\
  \citenamefont {{Hobbs}}}]{ATNF}%
  \BibitemOpen
  \bibfield  {author} {\bibinfo {author} {\bibfnamefont {R.~N.}\ \bibnamefont
  {{Manchester}}}, \bibinfo {author} {\bibfnamefont {G.~B.}\ \bibnamefont
  {{Hobbs}}}, \bibinfo {author} {\bibfnamefont {A.}~\bibnamefont {{Teoh}}}, \
  and\ \bibinfo {author} {\bibfnamefont {M.}~\bibnamefont {{Hobbs}}},\
  }\bibfield  {title} {\enquote {\bibinfo {title} {{VizieR Online Data Catalog:
  ATNF Pulsar Catalog (Manchester+, 2005)}},}\ }\href@noop {} {\bibfield
  {journal} {\bibinfo  {journal} {VizieR Online Data Catalog}\ }\textbf
  {\bibinfo {volume} {7245}} (\bibinfo {year} {2005})}\BibitemShut {NoStop}%
\bibitem [{\citenamefont {{Verbiest}}\ \emph {et~al.}(2012)\citenamefont
  {{Verbiest}}, \citenamefont {{Weisberg}}, \citenamefont {{Chael}},
  \citenamefont {{Lee}},\ and\ \citenamefont {{Lorimer}}}]{Geminga:distance}%
  \BibitemOpen
  \bibfield  {author} {\bibinfo {author} {\bibfnamefont {J.~P.~W.}\
  \bibnamefont {{Verbiest}}}, \bibinfo {author} {\bibfnamefont {J.~M.}\
  \bibnamefont {{Weisberg}}}, \bibinfo {author} {\bibfnamefont {A.~A.}\
  \bibnamefont {{Chael}}}, \bibinfo {author} {\bibfnamefont {K.~J.}\
  \bibnamefont {{Lee}}}, \ and\ \bibinfo {author} {\bibfnamefont {D.~R.}\
  \bibnamefont {{Lorimer}}},\ }\bibfield  {title} {\enquote {\bibinfo {title}
  {{On Pulsar Distance Measurements and Their Uncertainties}},}\ }\href
  {\doibase 10.1088/0004-637X/755/1/39} {\bibfield  {journal} {\bibinfo
  {journal} {\apj}\ }\textbf {\bibinfo {volume} {755}},\ \bibinfo {eid} {39}
  (\bibinfo {year} {2012})},\ \Eprint {http://arxiv.org/abs/1206.0428}
  {arXiv:1206.0428 [astro-ph.GA]} \BibitemShut {NoStop}%
\bibitem [{\citenamefont {{Posselt}}\ \emph {et~al.}(2017)\citenamefont
  {{Posselt}}, \citenamefont {{Pavlov}}, \citenamefont {{Slane}}, \citenamefont
  {{Romani}}, \citenamefont {{Bucciantini}}, \citenamefont {{Bykov}},
  \citenamefont {{Kargaltsev}}, \citenamefont {{Weisskopf}},\ and\
  \citenamefont {{Ng}}}]{Geminga:XPWN17}%
  \BibitemOpen
  \bibfield  {author} {\bibinfo {author} {\bibfnamefont {B.}~\bibnamefont
  {{Posselt}}}, \bibinfo {author} {\bibfnamefont {G.~G.}\ \bibnamefont
  {{Pavlov}}}, \bibinfo {author} {\bibfnamefont {P.~O.}\ \bibnamefont
  {{Slane}}}, \bibinfo {author} {\bibfnamefont {R.}~\bibnamefont {{Romani}}},
  \bibinfo {author} {\bibfnamefont {N.}~\bibnamefont {{Bucciantini}}}, \bibinfo
  {author} {\bibfnamefont {A.~M.}\ \bibnamefont {{Bykov}}}, \bibinfo {author}
  {\bibfnamefont {O.}~\bibnamefont {{Kargaltsev}}}, \bibinfo {author}
  {\bibfnamefont {M.~C.}\ \bibnamefont {{Weisskopf}}}, \ and\ \bibinfo {author}
  {\bibfnamefont {C.-Y.}\ \bibnamefont {{Ng}}},\ }\bibfield  {title} {\enquote
  {\bibinfo {title} {{Geminga's Puzzling Pulsar Wind Nebula}},}\ }\href
  {\doibase 10.3847/1538-4357/835/1/66} {\bibfield  {journal} {\bibinfo
  {journal} {\apj}\ }\textbf {\bibinfo {volume} {835}},\ \bibinfo {eid} {66}
  (\bibinfo {year} {2017})},\ \Eprint {http://arxiv.org/abs/1611.03496}
  {arXiv:1611.03496 [astro-ph.HE]} \BibitemShut {NoStop}%
\bibitem [{\citenamefont {{Trotta}}\ \emph {et~al.}(2011)\citenamefont
  {{Trotta}}, \citenamefont {{J{\'o}hannesson}}, \citenamefont {{Moskalenko}},
  \citenamefont {{Porter}}, \citenamefont {{Ruiz de Austri}},\ and\
  \citenamefont {{Strong}}}]{crdiffusion:GALPROP11}%
  \BibitemOpen
  \bibfield  {author} {\bibinfo {author} {\bibfnamefont {R.}~\bibnamefont
  {{Trotta}}}, \bibinfo {author} {\bibfnamefont {G.}~\bibnamefont
  {{J{\'o}hannesson}}}, \bibinfo {author} {\bibfnamefont {I.~V.}\ \bibnamefont
  {{Moskalenko}}}, \bibinfo {author} {\bibfnamefont {T.~A.}\ \bibnamefont
  {{Porter}}}, \bibinfo {author} {\bibfnamefont {R.}~\bibnamefont {{Ruiz de
  Austri}}}, \ and\ \bibinfo {author} {\bibfnamefont {A.~W.}\ \bibnamefont
  {{Strong}}},\ }\bibfield  {title} {\enquote {\bibinfo {title} {{Constraints
  on Cosmic-ray Propagation Models from A Global Bayesian Analysis}},}\ }\href
  {\doibase 10.1088/0004-637X/729/2/106} {\bibfield  {journal} {\bibinfo
  {journal} {\apj}\ }\textbf {\bibinfo {volume} {729}},\ \bibinfo {eid} {106}
  (\bibinfo {year} {2011})},\ \Eprint {http://arxiv.org/abs/1011.0037}
  {arXiv:1011.0037 [astro-ph.HE]} \BibitemShut {NoStop}%
\bibitem [{\citenamefont {{van der Swaluw}}\ \emph {et~al.}(2004)\citenamefont
  {{van der Swaluw}}, \citenamefont {{Downes}},\ and\ \citenamefont
  {{Keegan}}}]{PWNmodel:vanderSwaluw2004}%
  \BibitemOpen
  \bibfield  {author} {\bibinfo {author} {\bibfnamefont {E.}~\bibnamefont {{van
  der Swaluw}}}, \bibinfo {author} {\bibfnamefont {T.~P.}\ \bibnamefont
  {{Downes}}}, \ and\ \bibinfo {author} {\bibfnamefont {R.}~\bibnamefont
  {{Keegan}}},\ }\bibfield  {title} {\enquote {\bibinfo {title} {{An
  evolutionary model for pulsar-driven supernova remnants. A hydrodynamical
  model}},}\ }\href {\doibase 10.1051/0004-6361:20035700} {\bibfield  {journal}
  {\bibinfo  {journal} {Astron. and Astrophys.}\ }\textbf {\bibinfo {volume}
  {420}},\ \bibinfo {pages} {937--944} (\bibinfo {year} {2004})},\ \Eprint
  {http://arxiv.org/abs/astro-ph/0311388} {arXiv:astro-ph/0311388 [astro-ph]}
  \BibitemShut {NoStop}%
\bibitem [{\citenamefont {{Gelfand}}\ \emph {et~al.}(2009)\citenamefont
  {{Gelfand}}, \citenamefont {{Slane}},\ and\ \citenamefont
  {{Zhang}}}]{PWNmodel:Gelfand2009}%
  \BibitemOpen
  \bibfield  {author} {\bibinfo {author} {\bibfnamefont {J.~D.}\ \bibnamefont
  {{Gelfand}}}, \bibinfo {author} {\bibfnamefont {P.~O.}\ \bibnamefont
  {{Slane}}}, \ and\ \bibinfo {author} {\bibfnamefont {W.}~\bibnamefont
  {{Zhang}}},\ }\bibfield  {title} {\enquote {\bibinfo {title} {{A Dynamical
  Model for the Evolution of a Pulsar Wind Nebula Inside a Nonradiative
  Supernova Remnant}},}\ }\href {\doibase 10.1088/0004-637X/703/2/2051}
  {\bibfield  {journal} {\bibinfo  {journal} {\apj}\ }\textbf {\bibinfo
  {volume} {703}},\ \bibinfo {pages} {2051--2067} (\bibinfo {year} {2009})},\
  \Eprint {http://arxiv.org/abs/0904.4053} {arXiv:0904.4053 [astro-ph.HE]}
  \BibitemShut {NoStop}%
\bibitem [{\citenamefont {{Bucciantini}}(2011)}]{PWNmodel:Bucciantini2011}%
  \BibitemOpen
  \bibfield  {author} {\bibinfo {author} {\bibfnamefont {N.}~\bibnamefont
  {{Bucciantini}}},\ }\bibfield  {title} {\enquote {\bibinfo {title} {{MHD
  models of Pulsar Wind Nebulae}},}\ }\href {\doibase
  10.1007/978-3-642-17251-9_39} {\bibfield  {journal} {\bibinfo  {journal}
  {Astrophysics and Space Science Proceedings}\ }\textbf {\bibinfo {volume}
  {21}},\ \bibinfo {pages} {473} (\bibinfo {year} {2011})},\ \Eprint
  {http://arxiv.org/abs/1005.4781} {arXiv:1005.4781 [astro-ph.HE]} \BibitemShut
  {NoStop}%
\bibitem [{\citenamefont {{Mart{\'{\i}}n}}\ \emph {et~al.}(2016)\citenamefont
  {{Mart{\'{\i}}n}}, \citenamefont {{Torres}},\ and\ \citenamefont
  {{Pedaletti}}}]{PWNmodel:Martin16}%
  \BibitemOpen
  \bibfield  {author} {\bibinfo {author} {\bibfnamefont {J.}~\bibnamefont
  {{Mart{\'{\i}}n}}}, \bibinfo {author} {\bibfnamefont {D.~F.}\ \bibnamefont
  {{Torres}}}, \ and\ \bibinfo {author} {\bibfnamefont {G.}~\bibnamefont
  {{Pedaletti}}},\ }\bibfield  {title} {\enquote {\bibinfo {title} {{Molecular
  environment, reverberation, and radiation from the pulsar wind nebula in CTA
  1}},}\ }\href {\doibase 10.1093/mnras/stw684} {\bibfield  {journal} {\bibinfo
   {journal} {Mon. Not. R. Astron. Soc.}\ }\textbf {\bibinfo {volume} {459}},\
  \bibinfo {pages} {3868--3879} (\bibinfo {year} {2016})},\ \Eprint
  {http://arxiv.org/abs/1603.09328} {arXiv:1603.09328 [astro-ph.HE]}
  \BibitemShut {NoStop}%
\bibitem [{\citenamefont {{Ishizaki}}\ \emph {et~al.}(2018)\citenamefont
  {{Ishizaki}}, \citenamefont {{Asano}},\ and\ \citenamefont
  {{Kawaguchi}}}]{PWNmodel:Ishizaki18}%
  \BibitemOpen
  \bibfield  {author} {\bibinfo {author} {\bibfnamefont {W.}~\bibnamefont
  {{Ishizaki}}}, \bibinfo {author} {\bibfnamefont {K.}~\bibnamefont {{Asano}}},
  \ and\ \bibinfo {author} {\bibfnamefont {K.}~\bibnamefont {{Kawaguchi}}},\
  }\bibfield  {title} {\enquote {\bibinfo {title} {{Outflow and Emission Model
  of Pulsar Wind Nebulae with the Back Reaction of Particle Diffusion}},}\
  }\href {\doibase 10.3847/1538-4357/aae389} {\bibfield  {journal} {\bibinfo
  {journal} {\apj}\ }\textbf {\bibinfo {volume} {867}},\ \bibinfo {eid} {141}
  (\bibinfo {year} {2018})},\ \Eprint {http://arxiv.org/abs/1809.09054}
  {arXiv:1809.09054 [astro-ph.HE]} \BibitemShut {NoStop}%
\bibitem [{\citenamefont {{Zhu}}\ \emph {et~al.}(2018)\citenamefont {{Zhu}},
  \citenamefont {{Zhang}},\ and\ \citenamefont {{Fang}}}]{PWNmodel:Zhu18}%
  \BibitemOpen
  \bibfield  {author} {\bibinfo {author} {\bibfnamefont {Bo-Tao}\ \bibnamefont
  {{Zhu}}}, \bibinfo {author} {\bibfnamefont {Li}~\bibnamefont {{Zhang}}}, \
  and\ \bibinfo {author} {\bibfnamefont {Jun}\ \bibnamefont {{Fang}}},\
  }\bibfield  {title} {\enquote {\bibinfo {title} {{Multiband nonthermal
  radiative properties of pulsar wind nebulae}},}\ }\href {\doibase
  10.1051/0004-6361/201629108} {\bibfield  {journal} {\bibinfo  {journal}
  {Astron. and Astrophys.}\ }\textbf {\bibinfo {volume} {609}},\ \bibinfo {eid}
  {A110} (\bibinfo {year} {2018})}\BibitemShut {NoStop}%
\bibitem [{\citenamefont {{van Rensburg}}\ \emph {et~al.}(2018)\citenamefont
  {{van Rensburg}}, \citenamefont {{Kr{\"u}ger}},\ and\ \citenamefont
  {{Venter}}}]{PWNmodel:vanRensburg18}%
  \BibitemOpen
  \bibfield  {author} {\bibinfo {author} {\bibfnamefont {C.}~\bibnamefont {{van
  Rensburg}}}, \bibinfo {author} {\bibfnamefont {P.~P.}\ \bibnamefont
  {{Kr{\"u}ger}}}, \ and\ \bibinfo {author} {\bibfnamefont {C.}~\bibnamefont
  {{Venter}}},\ }\bibfield  {title} {\enquote {\bibinfo {title} {{Spatially
  dependent modelling of pulsar wind nebula G0.9+0.1}},}\ }\href {\doibase
  10.1093/mnras/sty826} {\bibfield  {journal} {\bibinfo  {journal} {Mon. Not.
  R. Astron. Soc.}\ }\textbf {\bibinfo {volume} {477}},\ \bibinfo {pages}
  {3853--3868} (\bibinfo {year} {2018})},\ \Eprint
  {http://arxiv.org/abs/1803.10625} {arXiv:1803.10625 [astro-ph.HE]}
  \BibitemShut {NoStop}%
\bibitem [{\citenamefont {{Kargaltsev}}\ and\ \citenamefont
  {{Pavlov}}(2010)}]{TeVPWN:Kargaltsev10}%
  \BibitemOpen
  \bibfield  {author} {\bibinfo {author} {\bibfnamefont {O.}~\bibnamefont
  {{Kargaltsev}}}\ and\ \bibinfo {author} {\bibfnamefont {G.~G.}\ \bibnamefont
  {{Pavlov}}},\ }\bibfield  {title} {\enquote {\bibinfo {title} {{Pulsar-wind
  nebulae in X-rays and TeV {$\gamma$}-rays}},}\ }\href {\doibase
  10.1063/1.3475228} {\bibfield  {journal} {\bibinfo  {journal} {X-ray
  Astronomy 2009; Present Status, Multi-Wavelength Approach and Future
  Perspectives}\ }\textbf {\bibinfo {volume} {1248}},\ \bibinfo {pages}
  {25--28} (\bibinfo {year} {2010})},\ \Eprint {http://arxiv.org/abs/1002.0885}
  {arXiv:1002.0885 [astro-ph.HE]} \BibitemShut {NoStop}%
\bibitem [{\citenamefont {{Kargaltsev}}\ \emph {et~al.}(2013)\citenamefont
  {{Kargaltsev}}, \citenamefont {{Rangelov}},\ and\ \citenamefont
  {{Pavlov}}}]{TeVPWN:Kargaltsev13}%
  \BibitemOpen
  \bibfield  {author} {\bibinfo {author} {\bibfnamefont {O.}~\bibnamefont
  {{Kargaltsev}}}, \bibinfo {author} {\bibfnamefont {B.}~\bibnamefont
  {{Rangelov}}}, \ and\ \bibinfo {author} {\bibfnamefont {G.~G.}\ \bibnamefont
  {{Pavlov}}},\ }\bibfield  {title} {\enquote {\bibinfo {title} {{Gamma-ray and
  X-ray Properties of Pulsar Wind Nebulae and Unidentified Galactic TeV
  Sources}},}\ }\href@noop {} {\bibfield  {journal} {\bibinfo  {journal} {arXiv
  e-prints}\ } (\bibinfo {year} {2013})},\ \Eprint
  {http://arxiv.org/abs/1305.2552} {arXiv:1305.2552 [astro-ph.HE]} \BibitemShut
  {NoStop}%
\bibitem [{\citenamefont {{Abdalla}}\ \emph {et~al.}(2019)\citenamefont
  {{Abdalla}} \emph {et~al.}}]{HESS:largePWN2019}%
  \BibitemOpen
  \bibfield  {author} {\bibinfo {author} {\bibfnamefont {H.}~\bibnamefont
  {{Abdalla}}} \emph {et~al.} (\bibinfo {collaboration} {H.~E.~S.~S.
  Collaboration}),\ }\bibfield  {title} {\enquote {\bibinfo {title} {{Particle
  transport within the pulsar wind nebula HESS J1825-137}},}\ }\href {\doibase
  10.1051/0004-6361/201834335} {\bibfield  {journal} {\bibinfo  {journal}
  {Astron. and Astrophys.}\ }\textbf {\bibinfo {volume} {621}},\ \bibinfo {eid}
  {A116} (\bibinfo {year} {2019})},\ \Eprint {http://arxiv.org/abs/1810.12676}
  {arXiv:1810.12676 [astro-ph.HE]} \BibitemShut {NoStop}%
\bibitem [{\citenamefont {{Rees}}\ and\ \citenamefont
  {{Gunn}}(1974)}]{PWNmodel:RG74}%
  \BibitemOpen
  \bibfield  {author} {\bibinfo {author} {\bibfnamefont {M.~J.}\ \bibnamefont
  {{Rees}}}\ and\ \bibinfo {author} {\bibfnamefont {J.~E.}\ \bibnamefont
  {{Gunn}}},\ }\bibfield  {title} {\enquote {\bibinfo {title} {{The origin of
  the magnetic field and relativistic particles in the Crab Nebula}},}\ }\href
  {\doibase 10.1093/mnras/167.1.1} {\bibfield  {journal} {\bibinfo  {journal}
  {Mon. Not. R. Astron. Soc.}\ }\textbf {\bibinfo {volume} {167}},\ \bibinfo
  {pages} {1--12} (\bibinfo {year} {1974})}\BibitemShut {NoStop}%
\bibitem [{\citenamefont {{Kennel}}\ and\ \citenamefont
  {{Coroniti}}(1984)}]{PWNmodel:KC84}%
  \BibitemOpen
  \bibfield  {author} {\bibinfo {author} {\bibfnamefont {C.~F.}\ \bibnamefont
  {{Kennel}}}\ and\ \bibinfo {author} {\bibfnamefont {F.~V.}\ \bibnamefont
  {{Coroniti}}},\ }\bibfield  {title} {\enquote {\bibinfo {title}
  {{Magnetohydrodynamic model of Crab nebula radiation}},}\ }\href {\doibase
  10.1086/162357} {\bibfield  {journal} {\bibinfo  {journal} {\apj}\ }\textbf
  {\bibinfo {volume} {283}},\ \bibinfo {pages} {710--730} (\bibinfo {year}
  {1984})}\BibitemShut {NoStop}%
\bibitem [{\citenamefont {{de Jager}}\ \emph {et~al.}(2009)\citenamefont {{de
  Jager}}, \citenamefont {{Ferreira}}, \citenamefont {{Djannati-Ata{\"i}}},
  \citenamefont {{Dalton}}, \citenamefont {{Deil}}, \citenamefont {{Kosack}},
  \citenamefont {{Renaud}}, \citenamefont {{Schwanke}},\ and\ \citenamefont
  {{Tibolla}}}]{PWNmodel:deJager09}%
  \BibitemOpen
  \bibfield  {author} {\bibinfo {author} {\bibfnamefont {O.~C.}\ \bibnamefont
  {{de Jager}}}, \bibinfo {author} {\bibfnamefont {S.~E.~S.}\ \bibnamefont
  {{Ferreira}}}, \bibinfo {author} {\bibfnamefont {A.}~\bibnamefont
  {{Djannati-Ata{\"i}}}}, \bibinfo {author} {\bibfnamefont {M.}~\bibnamefont
  {{Dalton}}}, \bibinfo {author} {\bibfnamefont {C.}~\bibnamefont {{Deil}}},
  \bibinfo {author} {\bibfnamefont {K.}~\bibnamefont {{Kosack}}}, \bibinfo
  {author} {\bibfnamefont {M.}~\bibnamefont {{Renaud}}}, \bibinfo {author}
  {\bibfnamefont {U.}~\bibnamefont {{Schwanke}}}, \ and\ \bibinfo {author}
  {\bibfnamefont {O.}~\bibnamefont {{Tibolla}}},\ }\bibfield  {title} {\enquote
  {\bibinfo {title} {{Unidentified Gamma-Ray Sources as Ancient Pulsar Wind
  Nebulae}},}\ }\href@noop {} {\bibfield  {journal} {\bibinfo  {journal} {ArXiv
  e-prints}\ } (\bibinfo {year} {2009})},\ \Eprint
  {http://arxiv.org/abs/0906.2644} {arXiv:0906.2644 [astro-ph.HE]} \BibitemShut
  {NoStop}%
\bibitem [{\citenamefont {{Tanaka}}\ and\ \citenamefont
  {{Takahara}}(2010)}]{PWNmodel:Tanaka10}%
  \BibitemOpen
  \bibfield  {author} {\bibinfo {author} {\bibfnamefont {S.~J.}\ \bibnamefont
  {{Tanaka}}}\ and\ \bibinfo {author} {\bibfnamefont {F.}~\bibnamefont
  {{Takahara}}},\ }\bibfield  {title} {\enquote {\bibinfo {title} {{A Model of
  the Spectral Evolution of Pulsar Wind Nebulae}},}\ }\href {\doibase
  10.1088/0004-637X/715/2/1248} {\bibfield  {journal} {\bibinfo  {journal}
  {\apj}\ }\textbf {\bibinfo {volume} {715}},\ \bibinfo {pages} {1248--1257}
  (\bibinfo {year} {2010})},\ \Eprint {http://arxiv.org/abs/1004.3098}
  {arXiv:1004.3098 [astro-ph.HE]} \BibitemShut {NoStop}%
\bibitem [{\citenamefont {{Torres}}\ \emph {et~al.}(2014)\citenamefont
  {{Torres}}, \citenamefont {{Cillis}}, \citenamefont {{Mart{\'{\i}}n}},\ and\
  \citenamefont {{de O{\~n}a Wilhelmi}}}]{PWNmodel:Torres2014}%
  \BibitemOpen
  \bibfield  {author} {\bibinfo {author} {\bibfnamefont {D.~F.}\ \bibnamefont
  {{Torres}}}, \bibinfo {author} {\bibfnamefont {A.}~\bibnamefont {{Cillis}}},
  \bibinfo {author} {\bibfnamefont {J.}~\bibnamefont {{Mart{\'{\i}}n}}}, \ and\
  \bibinfo {author} {\bibfnamefont {E.}~\bibnamefont {{de O{\~n}a Wilhelmi}}},\
  }\bibfield  {title} {\enquote {\bibinfo {title} {{Time-dependent modeling of
  TeV-detected, young pulsar wind nebulae}},}\ }\href {\doibase
  10.1016/j.jheap.2014.02.001} {\bibfield  {journal} {\bibinfo  {journal}
  {Journal of High Energy Astrophysics}\ }\textbf {\bibinfo {volume} {1}},\
  \bibinfo {pages} {31--62} (\bibinfo {year} {2014})},\ \Eprint
  {http://arxiv.org/abs/1402.5485} {arXiv:1402.5485 [astro-ph.HE]} \BibitemShut
  {NoStop}%
\bibitem [{\citenamefont {{Vorster}}\ \emph {et~al.}(2013)\citenamefont
  {{Vorster}}, \citenamefont {{Tibolla}}, \citenamefont {{Ferreira}},\ and\
  \citenamefont {{Kaufmann}}}]{PWNmodel:Vorster13}%
  \BibitemOpen
  \bibfield  {author} {\bibinfo {author} {\bibfnamefont {M.~J.}\ \bibnamefont
  {{Vorster}}}, \bibinfo {author} {\bibfnamefont {O.}~\bibnamefont
  {{Tibolla}}}, \bibinfo {author} {\bibfnamefont {S.~E.~S.}\ \bibnamefont
  {{Ferreira}}}, \ and\ \bibinfo {author} {\bibfnamefont {S.}~\bibnamefont
  {{Kaufmann}}},\ }\bibfield  {title} {\enquote {\bibinfo {title}
  {{Time-dependent Modeling of Pulsar Wind Nebulae}},}\ }\href {\doibase
  10.1088/0004-637X/773/2/139} {\bibfield  {journal} {\bibinfo  {journal}
  {\apj}\ }\textbf {\bibinfo {volume} {773}},\ \bibinfo {eid} {139} (\bibinfo
  {year} {2013})},\ \Eprint {http://arxiv.org/abs/1309.7137} {arXiv:1309.7137
  [astro-ph.HE]} \BibitemShut {NoStop}%
\bibitem [{\citenamefont {{Olmi}}\ \emph {et~al.}(2016)\citenamefont {{Olmi}},
  \citenamefont {{Del Zanna}}, \citenamefont {{Amato}}, \citenamefont
  {{Bucciantini}},\ and\ \citenamefont {{Mignone}}}]{PWNmodel:Olmi16}%
  \BibitemOpen
  \bibfield  {author} {\bibinfo {author} {\bibfnamefont {B.}~\bibnamefont
  {{Olmi}}}, \bibinfo {author} {\bibfnamefont {L.}~\bibnamefont {{Del Zanna}}},
  \bibinfo {author} {\bibfnamefont {E.}~\bibnamefont {{Amato}}}, \bibinfo
  {author} {\bibfnamefont {N.}~\bibnamefont {{Bucciantini}}}, \ and\ \bibinfo
  {author} {\bibfnamefont {A.}~\bibnamefont {{Mignone}}},\ }\bibfield  {title}
  {\enquote {\bibinfo {title} {{Multi-D magnetohydrodynamic modelling of pulsar
  wind nebulae: recent progress and open questions}},}\ }\href {\doibase
  10.1017/S0022377816000957} {\bibfield  {journal} {\bibinfo  {journal}
  {Journal of Plasma Physics}\ }\textbf {\bibinfo {volume} {82}},\ \bibinfo
  {eid} {635820601} (\bibinfo {year} {2016})},\ \Eprint
  {http://arxiv.org/abs/1610.07956} {arXiv:1610.07956 [astro-ph.HE]}
  \BibitemShut {NoStop}%
\bibitem [{\citenamefont {{Aharonian}}\ \emph {et~al.}(2006)\citenamefont
  {{Aharonian}} \emph {et~al.}}]{crab:HESS}%
  \BibitemOpen
  \bibfield  {author} {\bibinfo {author} {\bibfnamefont {F.}~\bibnamefont
  {{Aharonian}}} \emph {et~al.} (\bibinfo {collaboration} {H.E.S.S.
  Collaboration}),\ }\bibfield  {title} {\enquote {\bibinfo {title}
  {{Observations of the Crab nebula with HESS}},}\ }\href {\doibase
  10.1051/0004-6361:20065351} {\bibfield  {journal} {\bibinfo  {journal}
  {Astron. and Astrophys.}\ }\textbf {\bibinfo {volume} {457}},\ \bibinfo
  {pages} {899--915} (\bibinfo {year} {2006})},\ \Eprint
  {http://arxiv.org/abs/astro-ph/0607333} {astro-ph/0607333} \BibitemShut
  {NoStop}%
\bibitem [{\citenamefont {{Albert}}\ \emph {et~al.}(2008)\citenamefont
  {{Albert}} \emph {et~al.}}]{crab:MAGIC}%
  \BibitemOpen
  \bibfield  {author} {\bibinfo {author} {\bibfnamefont {J.}~\bibnamefont
  {{Albert}}} \emph {et~al.},\ }\bibfield  {title} {\enquote {\bibinfo {title}
  {{VHE {$\gamma$}-Ray Observation of the Crab Nebula and its Pulsar with the
  MAGIC Telescope}},}\ }\href {\doibase 10.1086/525270} {\bibfield  {journal}
  {\bibinfo  {journal} {\apj}\ }\textbf {\bibinfo {volume} {674}},\ \bibinfo
  {pages} {1037--1055} (\bibinfo {year} {2008})},\ \Eprint
  {http://arxiv.org/abs/0705.3244} {arXiv:0705.3244} \BibitemShut {NoStop}%
\bibitem [{\citenamefont {{Abeysekara}}\ \emph
  {et~al.}(2017{\natexlab{c}})\citenamefont {{Abeysekara}} \emph
  {et~al.}}]{crab:HAWC}%
  \BibitemOpen
  \bibfield  {author} {\bibinfo {author} {\bibfnamefont {A.~U.}\ \bibnamefont
  {{Abeysekara}}} \emph {et~al.},\ }\bibfield  {title} {\enquote {\bibinfo
  {title} {{Observation of the Crab Nebula with the HAWC Gamma-Ray
  Observatory}},}\ }\href {\doibase 10.3847/1538-4357/aa7555} {\bibfield
  {journal} {\bibinfo  {journal} {\apj}\ }\textbf {\bibinfo {volume} {843}},\
  \bibinfo {eid} {39} (\bibinfo {year} {2017}{\natexlab{c}})},\ \Eprint
  {http://arxiv.org/abs/1701.01778} {arXiv:1701.01778 [astro-ph.HE]}
  \BibitemShut {NoStop}%
\bibitem [{\citenamefont {{Abdo}}\ \emph {et~al.}(2010)\citenamefont {{Abdo}}
  \emph {et~al.}}]{Vela:Fermi10}%
  \BibitemOpen
  \bibfield  {author} {\bibinfo {author} {\bibfnamefont {A.~A.}\ \bibnamefont
  {{Abdo}}} \emph {et~al.},\ }\bibfield  {title} {\enquote {\bibinfo {title}
  {{Fermi Large Area Telescope Observations of the Vela-X Pulsar Wind
  Nebula}},}\ }\href {\doibase 10.1088/0004-637X/713/1/146} {\bibfield
  {journal} {\bibinfo  {journal} {\apj}\ }\textbf {\bibinfo {volume} {713}},\
  \bibinfo {pages} {146--153} (\bibinfo {year} {2010})},\ \Eprint
  {http://arxiv.org/abs/1002.4383} {arXiv:1002.4383 [astro-ph.HE]} \BibitemShut
  {NoStop}%
\bibitem [{\citenamefont {{Abramowski}}\ \emph {et~al.}(2012)\citenamefont
  {{Abramowski}} \emph {et~al.}}]{Vela:HESS12}%
  \BibitemOpen
  \bibfield  {author} {\bibinfo {author} {\bibfnamefont {A.}~\bibnamefont
  {{Abramowski}}} \emph {et~al.},\ }\bibfield  {title} {\enquote {\bibinfo
  {title} {{Probing the extent of the non-thermal emission from the Vela X
  region at TeV energies with H.E.S.S.}}}\ }\href {\doibase
  10.1051/0004-6361/201219919} {\bibfield  {journal} {\bibinfo  {journal}
  {Astron. and Astrophys.}\ }\textbf {\bibinfo {volume} {548}},\ \bibinfo {eid}
  {A38} (\bibinfo {year} {2012})},\ \Eprint {http://arxiv.org/abs/1210.1359}
  {arXiv:1210.1359 [astro-ph.HE]} \BibitemShut {NoStop}%
\bibitem [{\citenamefont {{Grondin}}\ \emph {et~al.}(2013)\citenamefont
  {{Grondin}}, \citenamefont {{Romani}}, \citenamefont {{Lemoine-Goumard}},
  \citenamefont {{Guillemot}}, \citenamefont {{Harding}},\ and\ \citenamefont
  {{Reposeur}}}]{Vela:Grondin13}%
  \BibitemOpen
  \bibfield  {author} {\bibinfo {author} {\bibfnamefont {M.-H.}\ \bibnamefont
  {{Grondin}}}, \bibinfo {author} {\bibfnamefont {R.~W.}\ \bibnamefont
  {{Romani}}}, \bibinfo {author} {\bibfnamefont {M.}~\bibnamefont
  {{Lemoine-Goumard}}}, \bibinfo {author} {\bibfnamefont {L.}~\bibnamefont
  {{Guillemot}}}, \bibinfo {author} {\bibfnamefont {A.~K.}\ \bibnamefont
  {{Harding}}}, \ and\ \bibinfo {author} {\bibfnamefont {T.}~\bibnamefont
  {{Reposeur}}},\ }\bibfield  {title} {\enquote {\bibinfo {title} {{The Vela-X
  Pulsar Wind Nebula Revisited with Four Years of Fermi Large Area Telescope
  Observations}},}\ }\href {\doibase 10.1088/0004-637X/774/2/110} {\bibfield
  {journal} {\bibinfo  {journal} {\apj}\ }\textbf {\bibinfo {volume} {774}},\
  \bibinfo {eid} {110} (\bibinfo {year} {2013})},\ \Eprint
  {http://arxiv.org/abs/1307.5480} {arXiv:1307.5480 [astro-ph.HE]} \BibitemShut
  {NoStop}%
\bibitem [{\citenamefont {{Tibaldo}}\ \emph {et~al.}(2018)\citenamefont
  {{Tibaldo}}, \citenamefont {{Zanin}}, \citenamefont {{Faggioli}},
  \citenamefont {{Ballet}}, \citenamefont {{Grondin}}, \citenamefont
  {{Hinton}},\ and\ \citenamefont {{Lemoine-Goumard}}}]{Vela:Fermi18}%
  \BibitemOpen
  \bibfield  {author} {\bibinfo {author} {\bibfnamefont {L.}~\bibnamefont
  {{Tibaldo}}}, \bibinfo {author} {\bibfnamefont {R.}~\bibnamefont {{Zanin}}},
  \bibinfo {author} {\bibfnamefont {G.}~\bibnamefont {{Faggioli}}}, \bibinfo
  {author} {\bibfnamefont {J.}~\bibnamefont {{Ballet}}}, \bibinfo {author}
  {\bibfnamefont {M.-H.}\ \bibnamefont {{Grondin}}}, \bibinfo {author}
  {\bibfnamefont {J.~A.}\ \bibnamefont {{Hinton}}}, \ and\ \bibinfo {author}
  {\bibfnamefont {M.}~\bibnamefont {{Lemoine-Goumard}}},\ }\bibfield  {title}
  {\enquote {\bibinfo {title} {{Disentangling multiple high-energy emission
  components in the Vela X pulsar wind nebula with the Fermi Large Area
  Telescope}},}\ }\href {\doibase 10.1051/0004-6361/201833356} {\bibfield
  {journal} {\bibinfo  {journal} {Astron. and Astrophys.}\ }\textbf {\bibinfo
  {volume} {617}},\ \bibinfo {eid} {A78} (\bibinfo {year} {2018})},\ \Eprint
  {http://arxiv.org/abs/1806.11499} {arXiv:1806.11499 [astro-ph.HE]}
  \BibitemShut {NoStop}%
\bibitem [{\citenamefont {{Blondin}}\ \emph {et~al.}(2001)\citenamefont
  {{Blondin}}, \citenamefont {{Chevalier}},\ and\ \citenamefont
  {{Frierson}}}]{Vela:Blondin01}%
  \BibitemOpen
  \bibfield  {author} {\bibinfo {author} {\bibfnamefont {J.~M.}\ \bibnamefont
  {{Blondin}}}, \bibinfo {author} {\bibfnamefont {R.~A.}\ \bibnamefont
  {{Chevalier}}}, \ and\ \bibinfo {author} {\bibfnamefont {D.~M.}\ \bibnamefont
  {{Frierson}}},\ }\bibfield  {title} {\enquote {\bibinfo {title} {{Pulsar Wind
  Nebulae in Evolved Supernova Remnants}},}\ }\href {\doibase 10.1086/324042}
  {\bibfield  {journal} {\bibinfo  {journal} {\apj}\ }\textbf {\bibinfo
  {volume} {563}},\ \bibinfo {pages} {806--815} (\bibinfo {year} {2001})},\
  \Eprint {http://arxiv.org/abs/astro-ph/0107076} {astro-ph/0107076}
  \BibitemShut {NoStop}%
\bibitem [{\citenamefont {{de Jager}}\ \emph {et~al.}(2008)\citenamefont {{de
  Jager}}, \citenamefont {{Slane}},\ and\ \citenamefont
  {{LaMassa}}}]{Vela:deJager08}%
  \BibitemOpen
  \bibfield  {author} {\bibinfo {author} {\bibfnamefont {O.~C.}\ \bibnamefont
  {{de Jager}}}, \bibinfo {author} {\bibfnamefont {P.~O.}\ \bibnamefont
  {{Slane}}}, \ and\ \bibinfo {author} {\bibfnamefont {S.}~\bibnamefont
  {{LaMassa}}},\ }\bibfield  {title} {\enquote {\bibinfo {title} {{Probing the
  Radio to X-Ray Connection of the Vela X Pulsar Wind Nebula with Fermi LAT and
  H.E.S.S.}}}\ }\href {\doibase 10.1086/595959} {\bibfield  {journal} {\bibinfo
   {journal} {Astrophys. J. Lett.}\ }\textbf {\bibinfo {volume} {689}},\
  \bibinfo {pages} {L125} (\bibinfo {year} {2008})},\ \Eprint
  {http://arxiv.org/abs/0810.1668} {arXiv:0810.1668} \BibitemShut {NoStop}%
\bibitem [{\citenamefont {{Hinton}}\ \emph {et~al.}(2011)\citenamefont
  {{Hinton}}, \citenamefont {{Funk}}, \citenamefont {{Parsons}},\ and\
  \citenamefont {{Ohm}}}]{Vela:Hinton11}%
  \BibitemOpen
  \bibfield  {author} {\bibinfo {author} {\bibfnamefont {J.~A.}\ \bibnamefont
  {{Hinton}}}, \bibinfo {author} {\bibfnamefont {S.}~\bibnamefont {{Funk}}},
  \bibinfo {author} {\bibfnamefont {R.~D.}\ \bibnamefont {{Parsons}}}, \ and\
  \bibinfo {author} {\bibfnamefont {S.}~\bibnamefont {{Ohm}}},\ }\bibfield
  {title} {\enquote {\bibinfo {title} {{Escape from Vela X}},}\ }\href
  {\doibase 10.1088/2041-8205/743/1/L7} {\bibfield  {journal} {\bibinfo
  {journal} {Astrophys. J. Lett.}\ }\textbf {\bibinfo {volume} {743}},\
  \bibinfo {eid} {L7} (\bibinfo {year} {2011})},\ \Eprint
  {http://arxiv.org/abs/1111.2036} {arXiv:1111.2036 [astro-ph.HE]} \BibitemShut
  {NoStop}%
\bibitem [{\citenamefont {{Slane}}\ \emph {et~al.}(2018)\citenamefont
  {{Slane}}, \citenamefont {{Lovchinsky}}, \citenamefont {{Kolb}},
  \citenamefont {{Snowden}}, \citenamefont {{Temim}}, \citenamefont
  {{Blondin}}, \citenamefont {{Bocchino}}, \citenamefont {{Miceli}},
  \citenamefont {{Chevalier}}, \citenamefont {{Hughes}}, \citenamefont
  {{Patnaude}},\ and\ \citenamefont {{Gaetz}}}]{Vela:Slane18}%
  \BibitemOpen
  \bibfield  {author} {\bibinfo {author} {\bibfnamefont {P.}~\bibnamefont
  {{Slane}}}, \bibinfo {author} {\bibfnamefont {I.}~\bibnamefont
  {{Lovchinsky}}}, \bibinfo {author} {\bibfnamefont {C.}~\bibnamefont
  {{Kolb}}}, \bibinfo {author} {\bibfnamefont {S.~L.}\ \bibnamefont
  {{Snowden}}}, \bibinfo {author} {\bibfnamefont {T.}~\bibnamefont {{Temim}}},
  \bibinfo {author} {\bibfnamefont {J.}~\bibnamefont {{Blondin}}}, \bibinfo
  {author} {\bibfnamefont {F.}~\bibnamefont {{Bocchino}}}, \bibinfo {author}
  {\bibfnamefont {M.}~\bibnamefont {{Miceli}}}, \bibinfo {author}
  {\bibfnamefont {R.~A.}\ \bibnamefont {{Chevalier}}}, \bibinfo {author}
  {\bibfnamefont {J.~P.}\ \bibnamefont {{Hughes}}}, \bibinfo {author}
  {\bibfnamefont {D.~J.}\ \bibnamefont {{Patnaude}}}, \ and\ \bibinfo {author}
  {\bibfnamefont {T.}~\bibnamefont {{Gaetz}}},\ }\bibfield  {title} {\enquote
  {\bibinfo {title} {{Investigating the Structure of Vela X}},}\ }\href
  {\doibase 10.3847/1538-4357/aada12} {\bibfield  {journal} {\bibinfo
  {journal} {\apj}\ }\textbf {\bibinfo {volume} {865}},\ \bibinfo {eid} {86}
  (\bibinfo {year} {2018})},\ \Eprint {http://arxiv.org/abs/1808.03878}
  {arXiv:1808.03878 [astro-ph.HE]} \BibitemShut {NoStop}%
\bibitem [{\citenamefont {{Lattimer}}\ and\ \citenamefont
  {{Prakash}}(2007)}]{NS:review2007}%
  \BibitemOpen
  \bibfield  {author} {\bibinfo {author} {\bibfnamefont {J.~M.}\ \bibnamefont
  {{Lattimer}}}\ and\ \bibinfo {author} {\bibfnamefont {M.}~\bibnamefont
  {{Prakash}}},\ }\bibfield  {title} {\enquote {\bibinfo {title} {{Neutron star
  observations: Prognosis for equation of state constraints}},}\ }\href
  {\doibase 10.1016/j.physrep.2007.02.003} {\bibfield  {journal} {\bibinfo
  {journal} {Phys. Rep.}\ }\textbf {\bibinfo {volume} {442}},\ \bibinfo {pages}
  {109--165} (\bibinfo {year} {2007})},\ \Eprint
  {http://arxiv.org/abs/astro-ph/0612440} {astro-ph/0612440} \BibitemShut
  {NoStop}%
\bibitem [{\citenamefont {{Shapiro}}\ and\ \citenamefont
  {{Teukolsky}}(1983)}]{Shapiro1983}%
  \BibitemOpen
  \bibfield  {author} {\bibinfo {author} {\bibfnamefont {S.~L.}\ \bibnamefont
  {{Shapiro}}}\ and\ \bibinfo {author} {\bibfnamefont {S.~A.}\ \bibnamefont
  {{Teukolsky}}},\ }\href@noop {} {\emph {\bibinfo {title} {Black holes, white
  dwarfs, and neutron stars: The physics of compact objects}}}\ (\bibinfo
  {year} {1983})\BibitemShut {NoStop}%
\bibitem [{\citenamefont {{Yusifov}}\ and\ \citenamefont {{K{\"u}{\c
  c}{\"u}k}}(2004)}]{psr_d:YK04}%
  \BibitemOpen
  \bibfield  {author} {\bibinfo {author} {\bibfnamefont {I.}~\bibnamefont
  {{Yusifov}}}\ and\ \bibinfo {author} {\bibfnamefont {I.}~\bibnamefont
  {{K{\"u}{\c c}{\"u}k}}},\ }\bibfield  {title} {\enquote {\bibinfo {title}
  {{Revisiting the radial distribution of pulsars in the Galaxy}},}\ }\href
  {\doibase 10.1051/0004-6361:20040152} {\bibfield  {journal} {\bibinfo
  {journal} {Astron. and Astrophys.}\ }\textbf {\bibinfo {volume} {422}},\
  \bibinfo {pages} {545--553} (\bibinfo {year} {2004})},\ \Eprint
  {http://arxiv.org/abs/astro-ph/0405559} {astro-ph/0405559} \BibitemShut
  {NoStop}%
\bibitem [{\citenamefont {{Porter}}\ \emph {et~al.}(2017)\citenamefont
  {{Porter}}, \citenamefont {{J{\'o}hannesson}},\ and\ \citenamefont
  {{Moskalenko}}}]{GALPROP2017}%
  \BibitemOpen
  \bibfield  {author} {\bibinfo {author} {\bibfnamefont {T.~A.}\ \bibnamefont
  {{Porter}}}, \bibinfo {author} {\bibfnamefont {G.}~\bibnamefont
  {{J{\'o}hannesson}}}, \ and\ \bibinfo {author} {\bibfnamefont {I.~V.}\
  \bibnamefont {{Moskalenko}}},\ }\bibfield  {title} {\enquote {\bibinfo
  {title} {{High-energy Gamma Rays from the Milky Way: Three-dimensional
  Spatial Models for the Cosmic-Ray and Radiation Field Densities in the
  Interstellar Medium}},}\ }\href {\doibase 10.3847/1538-4357/aa844d}
  {\bibfield  {journal} {\bibinfo  {journal} {\apj}\ }\textbf {\bibinfo
  {volume} {846}},\ \bibinfo {eid} {67} (\bibinfo {year} {2017})},\ \Eprint
  {http://arxiv.org/abs/1708.00816} {arXiv:1708.00816 [astro-ph.HE]}
  \BibitemShut {NoStop}%
\bibitem [{\citenamefont {{Lorimer}}\ \emph {et~al.}(2006)\citenamefont
  {{Lorimer}}, \citenamefont {{Faulkner}}, \citenamefont {{Lyne}},
  \citenamefont {{Manchester}}, \citenamefont {{Kramer}}, \citenamefont
  {{McLaughlin}}, \citenamefont {{Hobbs}}, \citenamefont {{Possenti}},
  \citenamefont {{Stairs}}, \citenamefont {{Camilo}}, \citenamefont {{Burgay}},
  \citenamefont {{D'Amico}}, \citenamefont {{Corongiu}},\ and\ \citenamefont
  {{Crawford}}}]{psr_d:L06}%
  \BibitemOpen
  \bibfield  {author} {\bibinfo {author} {\bibfnamefont {D.~R.}\ \bibnamefont
  {{Lorimer}}}, \bibinfo {author} {\bibfnamefont {A.~J.}\ \bibnamefont
  {{Faulkner}}}, \bibinfo {author} {\bibfnamefont {A.~G.}\ \bibnamefont
  {{Lyne}}}, \bibinfo {author} {\bibfnamefont {R.~N.}\ \bibnamefont
  {{Manchester}}}, \bibinfo {author} {\bibfnamefont {M.}~\bibnamefont
  {{Kramer}}}, \bibinfo {author} {\bibfnamefont {M.~A.}\ \bibnamefont
  {{McLaughlin}}}, \bibinfo {author} {\bibfnamefont {G.}~\bibnamefont
  {{Hobbs}}}, \bibinfo {author} {\bibfnamefont {A.}~\bibnamefont {{Possenti}}},
  \bibinfo {author} {\bibfnamefont {I.~H.}\ \bibnamefont {{Stairs}}}, \bibinfo
  {author} {\bibfnamefont {F.}~\bibnamefont {{Camilo}}}, \bibinfo {author}
  {\bibfnamefont {M.}~\bibnamefont {{Burgay}}}, \bibinfo {author}
  {\bibfnamefont {N.}~\bibnamefont {{D'Amico}}}, \bibinfo {author}
  {\bibfnamefont {A.}~\bibnamefont {{Corongiu}}}, \ and\ \bibinfo {author}
  {\bibfnamefont {F.}~\bibnamefont {{Crawford}}},\ }\bibfield  {title}
  {\enquote {\bibinfo {title} {{The Parkes Multibeam Pulsar Survey - VI.
  Discovery and timing of 142 pulsars and a Galactic population analysis}},}\
  }\href {\doibase 10.1111/j.1365-2966.2006.10887.x} {\bibfield  {journal}
  {\bibinfo  {journal} {Mon. Not. R. Astron. Soc.}\ }\textbf {\bibinfo {volume}
  {372}},\ \bibinfo {pages} {777--800} (\bibinfo {year} {2006})},\ \Eprint
  {http://arxiv.org/abs/astro-ph/0607640} {astro-ph/0607640} \BibitemShut
  {NoStop}%
\bibitem [{\citenamefont {{Haberl}}(2005)}]{INS2005}%
  \BibitemOpen
  \bibfield  {author} {\bibinfo {author} {\bibfnamefont {F.}~\bibnamefont
  {{Haberl}}},\ }\bibfield  {title} {\enquote {\bibinfo {title} {{The
  Magnificent Seven: Nearby Isolated Neutron Stars with strong Magnetic
  Fields}},}\ }in\ \href@noop {} {\emph {\bibinfo {booktitle} {5 years of
  Science with XMM-Newton}}},\ \bibinfo {editor} {edited by\ \bibinfo {editor}
  {\bibfnamefont {U.~G.}\ \bibnamefont {{Briel}}}, \bibinfo {editor}
  {\bibfnamefont {S.}~\bibnamefont {{Sembay}}}, \ and\ \bibinfo {editor}
  {\bibfnamefont {A.}~\bibnamefont {{Read}}}}\ (\bibinfo {year} {2005})\ pp.\
  \bibinfo {pages} {39--44},\ \Eprint {http://arxiv.org/abs/astro-ph/0510480}
  {arXiv:astro-ph/0510480 [astro-ph]} \BibitemShut {NoStop}%
\bibitem [{\citenamefont {{Kaplan}}\ and\ \citenamefont {{van
  Kerkwijk}}(2009)}]{INS2009}%
  \BibitemOpen
  \bibfield  {author} {\bibinfo {author} {\bibfnamefont {D.~L.}\ \bibnamefont
  {{Kaplan}}}\ and\ \bibinfo {author} {\bibfnamefont {M.~H.}\ \bibnamefont
  {{van Kerkwijk}}},\ }\bibfield  {title} {\enquote {\bibinfo {title}
  {{Constraining the Spin-down of the Nearby Isolated Neutron Star RX
  J0806.4-4123, and Implications for the Population of Nearby Neutron
  Stars}},}\ }\href {\doibase 10.1088/0004-637X/705/1/798} {\bibfield
  {journal} {\bibinfo  {journal} {\apj}\ }\textbf {\bibinfo {volume} {705}},\
  \bibinfo {pages} {798--808} (\bibinfo {year} {2009})},\ \Eprint
  {http://arxiv.org/abs/0909.5218} {arXiv:0909.5218 [astro-ph.HE]} \BibitemShut
  {NoStop}%
\bibitem [{\citenamefont {{Tauris}}\ and\ \citenamefont
  {{Manchester}}(1998)}]{psr:beam}%
  \BibitemOpen
  \bibfield  {author} {\bibinfo {author} {\bibfnamefont {T.~M.}\ \bibnamefont
  {{Tauris}}}\ and\ \bibinfo {author} {\bibfnamefont {R.~N.}\ \bibnamefont
  {{Manchester}}},\ }\bibfield  {title} {\enquote {\bibinfo {title} {{On the
  Evolution of Pulsar Beams}},}\ }\href {\doibase
  10.1046/j.1365-8711.1998.01369.x} {\bibfield  {journal} {\bibinfo  {journal}
  {Mon. Not. R. Astron. Soc.}\ }\textbf {\bibinfo {volume} {298}},\ \bibinfo
  {pages} {625--636} (\bibinfo {year} {1998})}\BibitemShut {NoStop}%
\bibitem [{\citenamefont {{Faucher-Gigu{\`e}re}}\ and\ \citenamefont
  {{Kaspi}}(2006)}]{psrP0:FK06}%
  \BibitemOpen
  \bibfield  {author} {\bibinfo {author} {\bibfnamefont {C.-A.}\ \bibnamefont
  {{Faucher-Gigu{\`e}re}}}\ and\ \bibinfo {author} {\bibfnamefont {V.~M.}\
  \bibnamefont {{Kaspi}}},\ }\bibfield  {title} {\enquote {\bibinfo {title}
  {{Birth and Evolution of Isolated Radio Pulsars}},}\ }\href {\doibase
  10.1086/501516} {\bibfield  {journal} {\bibinfo  {journal} {\apj}\ }\textbf
  {\bibinfo {volume} {643}},\ \bibinfo {pages} {332--355} (\bibinfo {year}
  {2006})},\ \Eprint {http://arxiv.org/abs/astro-ph/0512585} {astro-ph/0512585}
  \BibitemShut {NoStop}%
\bibitem [{\citenamefont {{de Jager}}(2008)}]{psrP0:PWN}%
  \BibitemOpen
  \bibfield  {author} {\bibinfo {author} {\bibfnamefont {O.~C.}\ \bibnamefont
  {{de Jager}}},\ }\bibfield  {title} {\enquote {\bibinfo {title} {{Estimating
  the Birth Period of Pulsars through GLAST LAT Observations of Their Wind
  Nebulae}},}\ }\href {\doibase 10.1086/588283} {\bibfield  {journal} {\bibinfo
   {journal} {Astrophys. J. Lett.}\ }\textbf {\bibinfo {volume} {678}},\
  \bibinfo {pages} {L113} (\bibinfo {year} {2008})},\ \Eprint
  {http://arxiv.org/abs/0803.2104} {arXiv:0803.2104} \BibitemShut {NoStop}%
\bibitem [{\citenamefont {{Watters}}\ and\ \citenamefont
  {{Romani}}(2011)}]{psrP0:WR11}%
  \BibitemOpen
  \bibfield  {author} {\bibinfo {author} {\bibfnamefont {K.~P.}\ \bibnamefont
  {{Watters}}}\ and\ \bibinfo {author} {\bibfnamefont {R.~W.}\ \bibnamefont
  {{Romani}}},\ }\bibfield  {title} {\enquote {\bibinfo {title} {{The Galactic
  Population of Young {$\gamma$}-ray Pulsars}},}\ }\href {\doibase
  10.1088/0004-637X/727/2/123} {\bibfield  {journal} {\bibinfo  {journal}
  {\apj}\ }\textbf {\bibinfo {volume} {727}},\ \bibinfo {eid} {123} (\bibinfo
  {year} {2011})},\ \Eprint {http://arxiv.org/abs/1009.5305} {arXiv:1009.5305
  [astro-ph.HE]} \BibitemShut {NoStop}%
\bibitem [{\citenamefont {{Popov}}\ and\ \citenamefont
  {{Turolla}}(2012)}]{psrP0:Popov12}%
  \BibitemOpen
  \bibfield  {author} {\bibinfo {author} {\bibfnamefont {S.~B.}\ \bibnamefont
  {{Popov}}}\ and\ \bibinfo {author} {\bibfnamefont {R.}~\bibnamefont
  {{Turolla}}},\ }\bibfield  {title} {\enquote {\bibinfo {title} {{Initial spin
  periods of neutron stars in supernova remnants}},}\ }\href {\doibase
  10.1007/s10509-012-1100-z} {\bibfield  {journal} {\bibinfo  {journal}
  {Astrophysics and Space Science}\ }\textbf {\bibinfo {volume} {341}},\
  \bibinfo {pages} {457--464} (\bibinfo {year} {2012})},\ \Eprint
  {http://arxiv.org/abs/1204.0632} {arXiv:1204.0632 [astro-ph.HE]} \BibitemShut
  {NoStop}%
\bibitem [{\citenamefont {{Noutsos}}\ \emph {et~al.}(2013)\citenamefont
  {{Noutsos}}, \citenamefont {{Schnitzeler}}, \citenamefont {{Keane}},
  \citenamefont {{Kramer}},\ and\ \citenamefont
  {{Johnston}}}]{psrP0:Noutsos13}%
  \BibitemOpen
  \bibfield  {author} {\bibinfo {author} {\bibfnamefont {A.}~\bibnamefont
  {{Noutsos}}}, \bibinfo {author} {\bibfnamefont {D.~H.~F.~M.}\ \bibnamefont
  {{Schnitzeler}}}, \bibinfo {author} {\bibfnamefont {E.~F.}\ \bibnamefont
  {{Keane}}}, \bibinfo {author} {\bibfnamefont {M.}~\bibnamefont {{Kramer}}}, \
  and\ \bibinfo {author} {\bibfnamefont {S.}~\bibnamefont {{Johnston}}},\
  }\bibfield  {title} {\enquote {\bibinfo {title} {{Pulsar spin-velocity
  alignment: kinematic ages, birth periods and braking indices}},}\ }\href
  {\doibase 10.1093/mnras/stt047} {\bibfield  {journal} {\bibinfo  {journal}
  {Mon. Not. R. Astron. Soc.}\ }\textbf {\bibinfo {volume} {430}},\ \bibinfo
  {pages} {2281--2301} (\bibinfo {year} {2013})},\ \Eprint
  {http://arxiv.org/abs/1301.1265} {arXiv:1301.1265} \BibitemShut {NoStop}%
\bibitem [{\citenamefont {{Igoshev}}\ and\ \citenamefont
  {{Popov}}(2013)}]{psrP0:Igoshev13}%
  \BibitemOpen
  \bibfield  {author} {\bibinfo {author} {\bibfnamefont {A.~P.}\ \bibnamefont
  {{Igoshev}}}\ and\ \bibinfo {author} {\bibfnamefont {S.~B.}\ \bibnamefont
  {{Popov}}},\ }\bibfield  {title} {\enquote {\bibinfo {title} {{Neutron star's
  initial spin period distribution}},}\ }\href {\doibase 10.1093/mnras/stt519}
  {\bibfield  {journal} {\bibinfo  {journal} {Mon. Not. R. Astron. Soc.}\
  }\textbf {\bibinfo {volume} {432}},\ \bibinfo {pages} {967--972} (\bibinfo
  {year} {2013})},\ \Eprint {http://arxiv.org/abs/1303.5258} {arXiv:1303.5258
  [astro-ph.HE]} \BibitemShut {NoStop}%
\bibitem [{\citenamefont {{Cie{\'s}lar}}\ \emph {et~al.}(2018)\citenamefont
  {{Cie{\'s}lar}}, \citenamefont {{Bulik}},\ and\ \citenamefont
  {{Os{\l}owski}}}]{psrP0:2018}%
  \BibitemOpen
  \bibfield  {author} {\bibinfo {author} {\bibfnamefont {M.}~\bibnamefont
  {{Cie{\'s}lar}}}, \bibinfo {author} {\bibfnamefont {T.}~\bibnamefont
  {{Bulik}}}, \ and\ \bibinfo {author} {\bibfnamefont {S.}~\bibnamefont
  {{Os{\l}owski}}},\ }\bibfield  {title} {\enquote {\bibinfo {title} {{Markov
  chain Monte Carlo population synthesis of single radio pulsars in the
  Galaxy}},}\ }\href@noop {} {\bibfield  {journal} {\bibinfo  {journal} {ArXiv
  e-prints}\ } (\bibinfo {year} {2018})},\ \Eprint
  {http://arxiv.org/abs/1803.02397} {arXiv:1803.02397 [astro-ph.IM]}
  \BibitemShut {NoStop}%
\bibitem [{\citenamefont {{Gonthier}}\ \emph {et~al.}(2004)\citenamefont
  {{Gonthier}}, \citenamefont {{Van Guilder}},\ and\ \citenamefont
  {{Harding}}}]{psrP0:G2004}%
  \BibitemOpen
  \bibfield  {author} {\bibinfo {author} {\bibfnamefont {P.~L.}\ \bibnamefont
  {{Gonthier}}}, \bibinfo {author} {\bibfnamefont {R.}~\bibnamefont {{Van
  Guilder}}}, \ and\ \bibinfo {author} {\bibfnamefont {A.~K.}\ \bibnamefont
  {{Harding}}},\ }\bibfield  {title} {\enquote {\bibinfo {title} {{Role of Beam
  Geometry in Population Statistics and Pulse Profiles of Radio and Gamma-Ray
  Pulsars}},}\ }\href {\doibase 10.1086/382070} {\bibfield  {journal} {\bibinfo
   {journal} {\apj}\ }\textbf {\bibinfo {volume} {604}},\ \bibinfo {pages}
  {775--790} (\bibinfo {year} {2004})},\ \Eprint
  {http://arxiv.org/abs/astro-ph/0312565} {astro-ph/0312565} \BibitemShut
  {NoStop}%
\bibitem [{\citenamefont {{Gull{\'o}n}}\ \emph {et~al.}(2014)\citenamefont
  {{Gull{\'o}n}}, \citenamefont {{Miralles}}, \citenamefont {{Vigan{\`o}}},\
  and\ \citenamefont {{Pons}}}]{psrB0:G14}%
  \BibitemOpen
  \bibfield  {author} {\bibinfo {author} {\bibfnamefont {M.}~\bibnamefont
  {{Gull{\'o}n}}}, \bibinfo {author} {\bibfnamefont {J.~A.}\ \bibnamefont
  {{Miralles}}}, \bibinfo {author} {\bibfnamefont {D.}~\bibnamefont
  {{Vigan{\`o}}}}, \ and\ \bibinfo {author} {\bibfnamefont {J.~A.}\
  \bibnamefont {{Pons}}},\ }\bibfield  {title} {\enquote {\bibinfo {title}
  {{Population synthesis of isolated neutron stars with magneto-rotational
  evolution}},}\ }\href {\doibase 10.1093/MNRAS/stu1253} {\bibfield  {journal}
  {\bibinfo  {journal} {Mon. Not. R. Astron. Soc.}\ }\textbf {\bibinfo {volume}
  {443}},\ \bibinfo {pages} {1891--1899} (\bibinfo {year} {2014})},\ \Eprint
  {http://arxiv.org/abs/1406.6794} {arXiv:1406.6794 [astro-ph.HE]} \BibitemShut
  {NoStop}%
\bibitem [{\citenamefont {{Popov}}\ \emph {et~al.}(2010)\citenamefont
  {{Popov}}, \citenamefont {{Pons}}, \citenamefont {{Miralles}}, \citenamefont
  {{Boldin}},\ and\ \citenamefont {{Posselt}}}]{psrB0:P10}%
  \BibitemOpen
  \bibfield  {author} {\bibinfo {author} {\bibfnamefont {S.~B.}\ \bibnamefont
  {{Popov}}}, \bibinfo {author} {\bibfnamefont {J.~A.}\ \bibnamefont {{Pons}}},
  \bibinfo {author} {\bibfnamefont {J.~A.}\ \bibnamefont {{Miralles}}},
  \bibinfo {author} {\bibfnamefont {P.~A.}\ \bibnamefont {{Boldin}}}, \ and\
  \bibinfo {author} {\bibfnamefont {B.}~\bibnamefont {{Posselt}}},\ }\bibfield
  {title} {\enquote {\bibinfo {title} {{Population synthesis studies of
  isolated neutron stars with magnetic field decay}},}\ }\href {\doibase
  10.1111/j.1365-2966.2009.15850.x} {\bibfield  {journal} {\bibinfo  {journal}
  {Mon. Not. R. Astron. Soc.}\ }\textbf {\bibinfo {volume} {401}},\ \bibinfo
  {pages} {2675--2686} (\bibinfo {year} {2010})},\ \Eprint
  {http://arxiv.org/abs/0910.2190} {arXiv:0910.2190 [astro-ph.HE]} \BibitemShut
  {NoStop}%
\bibitem [{\citenamefont {{Blumenthal}}\ and\ \citenamefont
  {{Gould}}(1970)}]{BG70}%
  \BibitemOpen
  \bibfield  {author} {\bibinfo {author} {\bibfnamefont {G.~R.}\ \bibnamefont
  {{Blumenthal}}}\ and\ \bibinfo {author} {\bibfnamefont {R.~J.}\ \bibnamefont
  {{Gould}}},\ }\bibfield  {title} {\enquote {\bibinfo {title}
  {{Bremsstrahlung, Synchrotron Radiation, and Compton Scattering of
  High-Energy Electrons Traversing Dilute Gases}},}\ }\href {\doibase
  10.1103/RevModPhys.42.237} {\bibfield  {journal} {\bibinfo  {journal}
  {Reviews of Modern Physics}\ }\textbf {\bibinfo {volume} {42}},\ \bibinfo
  {pages} {237--271} (\bibinfo {year} {1970})}\BibitemShut {NoStop}%
\bibitem [{\citenamefont {{Chen}}\ and\ \citenamefont
  {{Ruderman}}(1993)}]{deathline1993}%
  \BibitemOpen
  \bibfield  {author} {\bibinfo {author} {\bibfnamefont {Kaiyou}\ \bibnamefont
  {{Chen}}}\ and\ \bibinfo {author} {\bibfnamefont {Malvin}\ \bibnamefont
  {{Ruderman}}},\ }\bibfield  {title} {\enquote {\bibinfo {title} {{Pulsar
  Death Lines and Death Valley}},}\ }\href {\doibase 10.1086/172129} {\bibfield
   {journal} {\bibinfo  {journal} {\apj}\ }\textbf {\bibinfo {volume} {402}},\
  \bibinfo {pages} {264} (\bibinfo {year} {1993})}\BibitemShut {NoStop}%
\bibitem [{\citenamefont {{Hinton}}\ and\ \citenamefont
  {{Hofmann}}(2009)}]{ACT:Hinton2009}%
  \BibitemOpen
  \bibfield  {author} {\bibinfo {author} {\bibfnamefont {J.~A.}\ \bibnamefont
  {{Hinton}}}\ and\ \bibinfo {author} {\bibfnamefont {W.}~\bibnamefont
  {{Hofmann}}},\ }\bibfield  {title} {\enquote {\bibinfo {title}
  {{Teraelectronvolt Astronomy}},}\ }\href {\doibase
  10.1146/annurev-astro-082708-101816} {\bibfield  {journal} {\bibinfo
  {journal} {Ann. Rev. of Astron. and Astrophys.}\ }\textbf {\bibinfo {volume}
  {47}},\ \bibinfo {pages} {523--565} (\bibinfo {year} {2009})},\ \Eprint
  {http://arxiv.org/abs/1006.5210} {arXiv:1006.5210 [astro-ph.HE]} \BibitemShut
  {NoStop}%
\bibitem [{\citenamefont {{Abeysekara}}\ \emph {et~al.}(2013)\citenamefont
  {{Abeysekara}} \emph {et~al.}}]{HAWC2013}%
  \BibitemOpen
  \bibfield  {author} {\bibinfo {author} {\bibfnamefont {A.~U.}\ \bibnamefont
  {{Abeysekara}}} \emph {et~al.},\ }\bibfield  {title} {\enquote {\bibinfo
  {title} {{Sensitivity of the high altitude water Cherenkov detector to
  sources of multi-TeV gamma rays}},}\ }\href {\doibase
  10.1016/j.astropartphys.2013.08.002} {\bibfield  {journal} {\bibinfo
  {journal} {Astroparticle Physics}\ }\textbf {\bibinfo {volume} {50}},\
  \bibinfo {pages} {26--32} (\bibinfo {year} {2013})},\ \Eprint
  {http://arxiv.org/abs/1306.5800} {arXiv:1306.5800 [astro-ph.HE]} \BibitemShut
  {NoStop}%
\bibitem [{\citenamefont {{Malofeev}}\ and\ \citenamefont
  {{Malov}}(1997)}]{Geminga:radio97}%
  \BibitemOpen
  \bibfield  {author} {\bibinfo {author} {\bibfnamefont {V.~M.}\ \bibnamefont
  {{Malofeev}}}\ and\ \bibinfo {author} {\bibfnamefont {O.~I.}\ \bibnamefont
  {{Malov}}},\ }\bibfield  {title} {\enquote {\bibinfo {title} {{Detection of
  Geminga as a radio pulsar}},}\ }\href {\doibase 10.1038/39530} {\bibfield
  {journal} {\bibinfo  {journal} {\nat}\ }\textbf {\bibinfo {volume} {389}},\
  \bibinfo {pages} {697--699} (\bibinfo {year} {1997})}\BibitemShut {NoStop}%
\bibitem [{\citenamefont {{Atkins}}\ \emph {et~al.}(2005)\citenamefont
  {{Atkins}} \emph {et~al.}}]{diffuse:MRGO05}%
  \BibitemOpen
  \bibfield  {author} {\bibinfo {author} {\bibfnamefont {R.}~\bibnamefont
  {{Atkins}}} \emph {et~al.},\ }\bibfield  {title} {\enquote {\bibinfo {title}
  {{Evidence for TeV Gamma-Ray Emission from a Region of the Galactic
  Plane}},}\ }\href {\doibase 10.1103/PhysRevLett.95.251103} {\bibfield
  {journal} {\bibinfo  {journal} {Physical Review Letters}\ }\textbf {\bibinfo
  {volume} {95}},\ \bibinfo {eid} {251103} (\bibinfo {year} {2005})},\ \Eprint
  {http://arxiv.org/abs/astro-ph/0502303} {astro-ph/0502303} \BibitemShut
  {NoStop}%
\bibitem [{\citenamefont {{Bartoli}}\ \emph {et~al.}(2015)\citenamefont
  {{Bartoli}} \emph {et~al.}}]{diffuse:ARGO15}%
  \BibitemOpen
  \bibfield  {author} {\bibinfo {author} {\bibfnamefont {B.}~\bibnamefont
  {{Bartoli}}} \emph {et~al.} (\bibinfo {collaboration} {ARGO-YBJ
  Collaboration}),\ }\bibfield  {title} {\enquote {\bibinfo {title} {{Study of
  the Diffuse Gamma-Ray Emission from the Galactic Plane with ARGO-YBJ}},}\
  }\href {\doibase 10.1088/0004-637X/806/1/20} {\bibfield  {journal} {\bibinfo
  {journal} {\apj}\ }\textbf {\bibinfo {volume} {806}},\ \bibinfo {eid} {20}
  (\bibinfo {year} {2015})},\ \Eprint {http://arxiv.org/abs/1507.06758}
  {arXiv:1507.06758 [astro-ph.IM]} \BibitemShut {NoStop}%
\bibitem [{\citenamefont {{Abdo}}\ \emph {et~al.}(2008)\citenamefont {{Abdo}}
  \emph {et~al.}}]{diffuse:MRGO08}%
  \BibitemOpen
  \bibfield  {author} {\bibinfo {author} {\bibfnamefont {A.~A.}\ \bibnamefont
  {{Abdo}}} \emph {et~al.},\ }\bibfield  {title} {\enquote {\bibinfo {title}
  {{A Measurement of the Spatial Distribution of Diffuse TeV Gamma-Ray Emission
  from the Galactic Plane with Milagro}},}\ }\href {\doibase 10.1086/592213}
  {\bibfield  {journal} {\bibinfo  {journal} {\apj}\ }\textbf {\bibinfo
  {volume} {688}},\ \bibinfo {pages} {1078--1083} (\bibinfo {year} {2008})},\
  \Eprint {http://arxiv.org/abs/0805.0417} {arXiv:0805.0417} \BibitemShut
  {NoStop}%
\bibitem [{\citenamefont {{Prodanovi{\'c}}}\ \emph {et~al.}(2007)\citenamefont
  {{Prodanovi{\'c}}}, \citenamefont {{Fields}},\ and\ \citenamefont
  {{Beacom}}}]{diffuse:TeVexcess07}%
  \BibitemOpen
  \bibfield  {author} {\bibinfo {author} {\bibfnamefont {T.}~\bibnamefont
  {{Prodanovi{\'c}}}}, \bibinfo {author} {\bibfnamefont {B.~D.}\ \bibnamefont
  {{Fields}}}, \ and\ \bibinfo {author} {\bibfnamefont {J.~F.}\ \bibnamefont
  {{Beacom}}},\ }\bibfield  {title} {\enquote {\bibinfo {title} {{Diffuse gamma
  rays from the Galactic Plane: Probing the ''GeV excess" and identifying the
  ''TeV excess"}},}\ }\href {\doibase 10.1016/j.astropartphys.2006.08.007}
  {\bibfield  {journal} {\bibinfo  {journal} {Astroparticle Physics}\ }\textbf
  {\bibinfo {volume} {27}},\ \bibinfo {pages} {10--20} (\bibinfo {year}
  {2007})},\ \Eprint {http://arxiv.org/abs/astro-ph/0603618} {astro-ph/0603618}
  \BibitemShut {NoStop}%
\bibitem [{\citenamefont {{Bykov}}\ \emph {et~al.}(2017)\citenamefont
  {{Bykov}}, \citenamefont {{Amato}}, \citenamefont {{Petrov}}, \citenamefont
  {{Krassilchtchikov}},\ and\ \citenamefont {{Levenfish}}}]{PWN:bowshock2017}%
  \BibitemOpen
  \bibfield  {author} {\bibinfo {author} {\bibfnamefont {A.~M.}\ \bibnamefont
  {{Bykov}}}, \bibinfo {author} {\bibfnamefont {E.}~\bibnamefont {{Amato}}},
  \bibinfo {author} {\bibfnamefont {A.~E.}\ \bibnamefont {{Petrov}}}, \bibinfo
  {author} {\bibfnamefont {A.~M.}\ \bibnamefont {{Krassilchtchikov}}}, \ and\
  \bibinfo {author} {\bibfnamefont {K.~P.}\ \bibnamefont {{Levenfish}}},\
  }\bibfield  {title} {\enquote {\bibinfo {title} {{Pulsar Wind Nebulae with
  Bow Shocks: Non-thermal Radiation and Cosmic Ray Leptons}},}\ }\href
  {\doibase 10.1007/s11214-017-0371-7} {\bibfield  {journal} {\bibinfo
  {journal} {Space Science Rev.}\ }\textbf {\bibinfo {volume} {207}},\ \bibinfo
  {pages} {235--290} (\bibinfo {year} {2017})},\ \Eprint
  {http://arxiv.org/abs/1705.00950} {arXiv:1705.00950 [astro-ph.HE]}
  \BibitemShut {NoStop}%
\bibitem [{\citenamefont {{Uchiyama}}\ \emph {et~al.}(2009)\citenamefont
  {{Uchiyama}}, \citenamefont {{Matsumoto}}, \citenamefont {{Tsuru}},
  \citenamefont {{Koyama}},\ and\ \citenamefont
  {{Bamba}}}]{XrayHalo:Uchiyama2009}%
  \BibitemOpen
  \bibfield  {author} {\bibinfo {author} {\bibfnamefont {H.}~\bibnamefont
  {{Uchiyama}}}, \bibinfo {author} {\bibfnamefont {H.}~\bibnamefont
  {{Matsumoto}}}, \bibinfo {author} {\bibfnamefont {T.~G.}\ \bibnamefont
  {{Tsuru}}}, \bibinfo {author} {\bibfnamefont {K.}~\bibnamefont {{Koyama}}}, \
  and\ \bibinfo {author} {\bibfnamefont {A.}~\bibnamefont {{Bamba}}},\
  }\bibfield  {title} {\enquote {\bibinfo {title} {{Suzaku Observation of HESS
  J1825-137: Discovery of Largely-Extended X-Rays from PSR J1826-1334}},}\
  }\href {\doibase 10.1093/pasj/61.sp1.S189} {\bibfield  {journal} {\bibinfo
  {journal} {Publ. Astron. Soc. Jpn.}\ }\textbf {\bibinfo {volume} {61}},\
  \bibinfo {pages} {S189--S196} (\bibinfo {year} {2009})},\ \Eprint
  {http://arxiv.org/abs/0808.3436} {arXiv:0808.3436} \BibitemShut {NoStop}%
\bibitem [{\citenamefont {{Bamba}}\ \emph {et~al.}(2010)\citenamefont
  {{Bamba}}, \citenamefont {{Anada}}, \citenamefont {{Dotani}}, \citenamefont
  {{Mori}}, \citenamefont {{Yamazaki}}, \citenamefont {{Ebisawa}},\ and\
  \citenamefont {{Vink}}}]{XrayHalo:Bamba2010}%
  \BibitemOpen
  \bibfield  {author} {\bibinfo {author} {\bibfnamefont {A.}~\bibnamefont
  {{Bamba}}}, \bibinfo {author} {\bibfnamefont {T.}~\bibnamefont {{Anada}}},
  \bibinfo {author} {\bibfnamefont {T.}~\bibnamefont {{Dotani}}}, \bibinfo
  {author} {\bibfnamefont {K.}~\bibnamefont {{Mori}}}, \bibinfo {author}
  {\bibfnamefont {R.}~\bibnamefont {{Yamazaki}}}, \bibinfo {author}
  {\bibfnamefont {K.}~\bibnamefont {{Ebisawa}}}, \ and\ \bibinfo {author}
  {\bibfnamefont {J.}~\bibnamefont {{Vink}}},\ }\bibfield  {title} {\enquote
  {\bibinfo {title} {{X-ray Evolution of Pulsar Wind Nebulae}},}\ }\href
  {\doibase 10.1088/2041-8205/719/2/L116} {\bibfield  {journal} {\bibinfo
  {journal} {Astrophys. J. Lett.}\ }\textbf {\bibinfo {volume} {719}},\
  \bibinfo {pages} {L116--L120} (\bibinfo {year} {2010})},\ \Eprint
  {http://arxiv.org/abs/1007.3203} {arXiv:1007.3203 [astro-ph.HE]} \BibitemShut
  {NoStop}%
\bibitem [{\citenamefont {{Joshi}}\ \emph {et~al.}(2017)\citenamefont
  {{Joshi}}, \citenamefont {{Jardin-Blicq}},\ and\ \citenamefont {{HAWC
  Collaboration}}}]{HAWCupgrade2017}%
  \BibitemOpen
  \bibfield  {author} {\bibinfo {author} {\bibfnamefont {V.}~\bibnamefont
  {{Joshi}}}, \bibinfo {author} {\bibfnamefont {A.}~\bibnamefont
  {{Jardin-Blicq}}}, \ and\ \bibinfo {author} {\bibnamefont {{HAWC
  Collaboration}}},\ }\bibfield  {title} {\enquote {\bibinfo {title} {{HAWC
  High Energy Upgrade with a Sparse Outrigger Array}},}\ }\href@noop {}
  {\bibfield  {journal} {\bibinfo  {journal} {International Cosmic Ray
  Conference}\ }\textbf {\bibinfo {volume} {35}},\ \bibinfo {eid} {806}
  (\bibinfo {year} {2017})},\ \Eprint {http://arxiv.org/abs/1708.04032}
  {arXiv:1708.04032 [astro-ph.IM]} \BibitemShut {NoStop}%
\bibitem [{\citenamefont {{Bates}}\ \emph {et~al.}(2014)\citenamefont
  {{Bates}}, \citenamefont {{Lorimer}}, \citenamefont {{Rane}},\ and\
  \citenamefont {{Swiggum}}}]{psrpoppy}%
  \BibitemOpen
  \bibfield  {author} {\bibinfo {author} {\bibfnamefont {S.~D.}\ \bibnamefont
  {{Bates}}}, \bibinfo {author} {\bibfnamefont {D.~R.}\ \bibnamefont
  {{Lorimer}}}, \bibinfo {author} {\bibfnamefont {A.}~\bibnamefont {{Rane}}}, \
  and\ \bibinfo {author} {\bibfnamefont {J.}~\bibnamefont {{Swiggum}}},\
  }\bibfield  {title} {\enquote {\bibinfo {title} {{PSRPOPPy: an open-source
  package for pulsar population simulations}},}\ }\href {\doibase
  10.1093/mnras/stu157} {\bibfield  {journal} {\bibinfo  {journal} {Mon. Not.
  R. Astron. Soc.}\ }\textbf {\bibinfo {volume} {439}},\ \bibinfo {pages}
  {2893--2902} (\bibinfo {year} {2014})},\ \Eprint
  {http://arxiv.org/abs/1311.3427} {arXiv:1311.3427 [astro-ph.IM]} \BibitemShut
  {NoStop}%
\bibitem [{\citenamefont {{Mostafa}}\ and\ \citenamefont {{HAWC
  Collaboration}}(2017)}]{South2017a}%
  \BibitemOpen
  \bibfield  {author} {\bibinfo {author} {\bibfnamefont {M.}~\bibnamefont
  {{Mostafa}}}\ and\ \bibinfo {author} {\bibnamefont {{HAWC Collaboration}}},\
  }\bibfield  {title} {\enquote {\bibinfo {title} {{A future wide field-of-view
  TeV gamma-ray observatory in the Southern Hemisphere}},}\ }in\ \href@noop {}
  {\emph {\bibinfo {booktitle} {APS April Meeting Abstracts}}}\ (\bibinfo
  {year} {2017})\ p.\ \bibinfo {pages} {R4.005}\BibitemShut {NoStop}%
\bibitem [{\citenamefont {{Schoorlemmer}}\ \emph {et~al.}(2017)\citenamefont
  {{Schoorlemmer}}, \citenamefont {{Lopez-Coto}},\ and\ \citenamefont
  {{Hinton}}}]{South2017b}%
  \BibitemOpen
  \bibfield  {author} {\bibinfo {author} {\bibfnamefont {H.}~\bibnamefont
  {{Schoorlemmer}}}, \bibinfo {author} {\bibfnamefont {R.}~\bibnamefont
  {{Lopez-Coto}}}, \ and\ \bibinfo {author} {\bibfnamefont {J.}~\bibnamefont
  {{Hinton}}},\ }\bibfield  {title} {\enquote {\bibinfo {title} {{Baseline
  Design for a Next Generation Wide-Field-of-View Very-High-Energy Gamma Ray
  Observatory}},}\ }\href@noop {} {\bibfield  {journal} {\bibinfo  {journal}
  {International Cosmic Ray Conference}\ }\textbf {\bibinfo {volume} {35}},\
  \bibinfo {eid} {819} (\bibinfo {year} {2017})},\ \Eprint
  {http://arxiv.org/abs/1709.05792} {arXiv:1709.05792 [astro-ph.IM]}
  \BibitemShut {NoStop}%
\bibitem [{\citenamefont {{Mostafa}}\ \emph {et~al.}(2017)\citenamefont
  {{Mostafa}}, \citenamefont {{BenZvi}}, \citenamefont {{Schoorlemmer}},
  \citenamefont {{Sch{\"u}ssler}},\ and\ \citenamefont {{HAWC
  Collaboration}}}]{South2017c}%
  \BibitemOpen
  \bibfield  {author} {\bibinfo {author} {\bibfnamefont {M.}~\bibnamefont
  {{Mostafa}}}, \bibinfo {author} {\bibfnamefont {S.}~\bibnamefont {{BenZvi}}},
  \bibinfo {author} {\bibfnamefont {H.}~\bibnamefont {{Schoorlemmer}}},
  \bibinfo {author} {\bibfnamefont {F.}~\bibnamefont {{Sch{\"u}ssler}}}, \ and\
  \bibinfo {author} {\bibnamefont {{HAWC Collaboration}}},\ }\bibfield  {title}
  {\enquote {\bibinfo {title} {{On the scientific motivation for a wide
  field-of-view TeV gamma-ray observatory in the Southern Hemisphere}},}\
  }\href@noop {} {\bibfield  {journal} {\bibinfo  {journal} {International
  Cosmic Ray Conference}\ }\textbf {\bibinfo {volume} {35}},\ \bibinfo {eid}
  {851} (\bibinfo {year} {2017})}\BibitemShut {NoStop}%
\bibitem [{\citenamefont {{Acharya}}\ \emph {et~al.}(2017)\citenamefont
  {{Acharya}} \emph {et~al.}}]{CTA:17}%
  \BibitemOpen
  \bibfield  {author} {\bibinfo {author} {\bibfnamefont {B.~S.}\ \bibnamefont
  {{Acharya}}} \emph {et~al.} (\bibinfo {collaboration} {The Cherenkov
  Telescope Array Consortium}),\ }\bibfield  {title} {\enquote {\bibinfo
  {title} {{Science with the Cherenkov Telescope Array}},}\ }\href@noop {}
  {\bibfield  {journal} {\bibinfo  {journal} {ArXiv e-prints}\ } (\bibinfo
  {year} {2017})},\ \Eprint {http://arxiv.org/abs/1709.07997} {arXiv:1709.07997
  [astro-ph.IM]} \BibitemShut {NoStop}%
\bibitem [{\citenamefont {{Ridley}}\ and\ \citenamefont
  {{Lorimer}}(2010)}]{LMC:Ridley10}%
  \BibitemOpen
  \bibfield  {author} {\bibinfo {author} {\bibfnamefont {J.~P.}\ \bibnamefont
  {{Ridley}}}\ and\ \bibinfo {author} {\bibfnamefont {D.~R.}\ \bibnamefont
  {{Lorimer}}},\ }\bibfield  {title} {\enquote {\bibinfo {title} {{New limits
  on the population of normal and millisecond pulsars in the Large and Small
  Magellanic Clouds}},}\ }\href {\doibase 10.1111/j.1745-3933.2010.00886.x}
  {\bibfield  {journal} {\bibinfo  {journal} {Mon. Not. R. Astron. Soc.}\
  }\textbf {\bibinfo {volume} {406}},\ \bibinfo {pages} {L80--L84} (\bibinfo
  {year} {2010})},\ \Eprint {http://arxiv.org/abs/1005.4653} {arXiv:1005.4653}
  \BibitemShut {NoStop}%
\bibitem [{\citenamefont {{Israel}}\ \emph {et~al.}(2010)\citenamefont
  {{Israel}}, \citenamefont {{Wall}}, \citenamefont {{Raban}}, \citenamefont
  {{Reach}}, \citenamefont {{Bot}}, \citenamefont {{Oonk}}, \citenamefont
  {{Ysard}},\ and\ \citenamefont {{Bernard}}}]{LMCrad}%
  \BibitemOpen
  \bibfield  {author} {\bibinfo {author} {\bibfnamefont {F.~P.}\ \bibnamefont
  {{Israel}}}, \bibinfo {author} {\bibfnamefont {W.~F.}\ \bibnamefont
  {{Wall}}}, \bibinfo {author} {\bibfnamefont {D.}~\bibnamefont {{Raban}}},
  \bibinfo {author} {\bibfnamefont {W.~T.}\ \bibnamefont {{Reach}}}, \bibinfo
  {author} {\bibfnamefont {C.}~\bibnamefont {{Bot}}}, \bibinfo {author}
  {\bibfnamefont {J.~B.~R.}\ \bibnamefont {{Oonk}}}, \bibinfo {author}
  {\bibfnamefont {N.}~\bibnamefont {{Ysard}}}, \ and\ \bibinfo {author}
  {\bibfnamefont {J.~P.}\ \bibnamefont {{Bernard}}},\ }\bibfield  {title}
  {\enquote {\bibinfo {title} {{submillimeter to centimeter excess emission
  from the Magellanic Clouds. I. Global spectral energy distribution}},}\
  }\href {\doibase 10.1051/0004-6361/201014073} {\bibfield  {journal} {\bibinfo
   {journal} {Astron. and Astrophys.}\ }\textbf {\bibinfo {volume} {519}},\
  \bibinfo {eid} {A67} (\bibinfo {year} {2010})},\ \Eprint
  {http://arxiv.org/abs/1006.2232} {arXiv:1006.2232} \BibitemShut {NoStop}%
\bibitem [{\citenamefont {{Szary}}\ \emph {et~al.}(2014)\citenamefont
  {{Szary}}, \citenamefont {{Zhang}}, \citenamefont {{Melikidze}},
  \citenamefont {{Gil}},\ and\ \citenamefont {{Xu}}}]{psrL:Szary14}%
  \BibitemOpen
  \bibfield  {author} {\bibinfo {author} {\bibfnamefont {A.}~\bibnamefont
  {{Szary}}}, \bibinfo {author} {\bibfnamefont {B.}~\bibnamefont {{Zhang}}},
  \bibinfo {author} {\bibfnamefont {G.~I.}\ \bibnamefont {{Melikidze}}},
  \bibinfo {author} {\bibfnamefont {J.}~\bibnamefont {{Gil}}}, \ and\ \bibinfo
  {author} {\bibfnamefont {R.-X.}\ \bibnamefont {{Xu}}},\ }\bibfield  {title}
  {\enquote {\bibinfo {title} {{Radio Efficiency of Pulsars}},}\ }\href
  {\doibase 10.1088/0004-637X/784/1/59} {\bibfield  {journal} {\bibinfo
  {journal} {\apj}\ }\textbf {\bibinfo {volume} {784}},\ \bibinfo {eid} {59}
  (\bibinfo {year} {2014})},\ \Eprint {http://arxiv.org/abs/1402.0228}
  {arXiv:1402.0228 [astro-ph.HE]} \BibitemShut {NoStop}%
\bibitem [{\citenamefont {{Di Sciascio}}\ and\ \citenamefont {{LHAASO
  Collaboration}}(2016)}]{LHAASO:2016}%
  \BibitemOpen
  \bibfield  {author} {\bibinfo {author} {\bibfnamefont {G.}~\bibnamefont {{Di
  Sciascio}}}\ and\ \bibinfo {author} {\bibnamefont {{LHAASO Collaboration}}},\
  }\bibfield  {title} {\enquote {\bibinfo {title} {{The LHAASO experiment: From
  Gamma-Ray Astronomy to Cosmic Rays}},}\ }\href {\doibase
  10.1016/j.nuclphysbps.2016.10.024} {\bibfield  {journal} {\bibinfo  {journal}
  {Nuclear and Particle Physics Proceedings}\ }\textbf {\bibinfo {volume}
  {279}},\ \bibinfo {pages} {166--173} (\bibinfo {year} {2016})},\ \Eprint
  {http://arxiv.org/abs/1602.07600} {arXiv:1602.07600 [astro-ph.HE]}
  \BibitemShut {NoStop}%
\bibitem [{\citenamefont {Gull^^c3^^b3n}\ \emph {et~al.}(2015)\citenamefont
  {Gull^^c3^^b3n}, \citenamefont {Pons}, \citenamefont {Miralles},
  \citenamefont {Vigan^^c3^^b2}, \citenamefont {Rea},\ and\ \citenamefont
  {Perna}}]{Gullon:2015zca}%
  \BibitemOpen
  \bibfield  {author} {\bibinfo {author} {\bibfnamefont {M.}~\bibnamefont
  {Gull^^c3^^b3n}}, \bibinfo {author} {\bibfnamefont {J.~A.}\ \bibnamefont
  {Pons}}, \bibinfo {author} {\bibfnamefont {J.~A.}\ \bibnamefont {Miralles}},
  \bibinfo {author} {\bibfnamefont {D.}~\bibnamefont {Vigan^^c3^^b2}}, \bibinfo
  {author} {\bibfnamefont {N.}~\bibnamefont {Rea}}, \ and\ \bibinfo {author}
  {\bibfnamefont {R.}~\bibnamefont {Perna}},\ }\bibfield  {title} {\enquote
  {\bibinfo {title} {{Population synthesis of isolated neutron stars with
  magneto-rotational evolution ^^e2^^80^^93 II. From radio-pulsars to
  magnetars}},}\ }\href {\doibase 10.1093/mnras/stv1644} {\bibfield  {journal}
  {\bibinfo  {journal} {Mon. Not. Roy. Astron. Soc.}\ }\textbf {\bibinfo
  {volume} {454}},\ \bibinfo {pages} {615--625} (\bibinfo {year} {2015})},\
  \Eprint {http://arxiv.org/abs/1507.05452} {arXiv:1507.05452 [astro-ph.HE]}
  \BibitemShut {NoStop}%
\end{thebibliography}%

\end{document}